\documentclass[twocolumn]{aastex63}
\usepackage{amsmath}
\usepackage{mathrsfs}
\usepackage{xcolor}
\usepackage[T1]{fontenc}
\usepackage{fourier}
\usepackage{hyperref}
\usepackage[normalem]{ulem}
\usepackage{relsize}
\usepackage{rotating}
\submitjournal{ApJ}

\shorttitle{Persistent radio sources toward fast radio bursts}
\shortauthors{Ibik et al.}

\begin{document}

\title{A search for persistent radio sources toward repeating fast radio bursts discovered by CHIME/FRB}

\correspondingauthor{Adaeze Ibik}
\email{adaeze.ibik@mail.utoronto.ca}

\author[0000-0003-2405-2967]{Adaeze L.~Ibik}
  \affiliation{Dunlap Institute for Astronomy \& Astrophysics, University of Toronto, 50 St.~George Street, Toronto, ON M5S 3H4, Canada}
  \affiliation{David A.~Dunlap Department of Astronomy \& Astrophysics, University of Toronto, 50 St.~George Street, Toronto, ON M5S 3H4, Canada}

\author[0000-0001-7081-0082]{Maria R. Drout}
  \affiliation{David A.~Dunlap Department of Astronomy \& Astrophysics, University of Toronto, 50 St.~George Street, Toronto, ON M5S 3H4, Canada}

\author[0000-0002-3382-9558]{B.~M.~Gaensler}
  \affiliation{Dunlap Institute for Astronomy \& Astrophysics, University of Toronto, 50 St.~George Street, Toronto, ON M5S 3H4, Canada}
  \affiliation{David A.~Dunlap Department of Astronomy \& Astrophysics, University of Toronto, 50 St.~George Street, Toronto, ON M5S 3H4, Canada}
  \affiliation{Department of Astronomy and Astrophysics, University of California Santa Cruz, 1156 High Street, Santa Cruz, CA 95064, USA}

\author[0000-0002-7374-7119]{Paul Scholz}
  \affiliation{Dunlap Institute for Astronomy \& Astrophysics, University of Toronto, 50 St.~George Street, Toronto, ON M5S 3H4, Canada}
  \affiliation{Department of Physics and Astronomy, York University, 4700 Keele Street, Toronto, Ontario, ON MJ3 1P3, Canada}
  
\author[0000-0002-5519-9550]{Navin Sridhar}
  \affiliation{Department of Astronomy, Columbia University, New York, NY 10027, USA}
  \affiliation{Cahill Center for Astronomy and Astrophysics, MC 249-17 California Institute of Technology, Pasadena CA 91125, USA}

\author[0000-0001-8405-2649]{Ben Margalit}
  \affiliation{School of Physics and Astronomy, University of Minnesota, Minneapolis, MN 55455, USA}
  
\author[0000-0001-6812-7938]{T.~E.~Clarke}
  \affiliation{Naval Research Laboratory, Code 7213, 4555 Overlook Ave SW, Washington DC, 20375, D.C., United States}
  
  \author[0000-0002-4119-9963]{Casey J.~Law}
  \affiliation{Cahill Center for Astronomy and Astrophysics, MC 249-17 California Institute of Technology, Pasadena CA 91125, USA}
  \affiliation{Owens Valley Radio Observatory, California Institute of Technology, Big Pine CA 93513, USA}
    
\author[0000-0003-2548-2926]{Shriharsh P.~Tendulkar}
  \affiliation{Department of Astronomy and Astrophysics, Tata Institute of Fundamental Research, Mumbai, 400005, India}
  \affiliation{National Centre for Radio Astrophysics, Post Bag 3, Ganeshkhind, Pune, 411007, India}
  \affiliation{CIFAR Azrieli Global Scholars Program, CIFAR, Toronto, Canada}

\author[0000-0002-2551-7554]{Daniele Michilli}
  \affiliation{MIT Kavli Institute for Astrophysics and Space Research, Massachusetts Institute of Technology, 77 Massachusetts Ave, Cambridge, MA 02139, USA}
  \affiliation{Department of Physics, Massachusetts Institute of Technology, 77 Massachusetts Ave, Cambridge, MA 02139, USA}
  
    \author[0000-0003-0307-9984]{Tarraneh Eftekhari}
  \affiliation{Center for Interdisciplinary Exploration and Research in Astrophysics (CIERA) and Department of Physics and Astronomy, Northwestern University, Evanston, IL 60208, USA}

  \author[0000-0002-3615-3514]{Mohit Bhardwaj}
\affiliation{Department of Physics, Carnegie Mellon University, 5000 Forbes Avenue, Pittsburgh, 15213, PA, USA}

\author[0000-0003-4052-7838]{Sarah Burke-Spolaor}
  \affiliation{Department of Physics and Astronomy, West Virginia University, P.O. Box 6315, Morgantown, WV 26506, USA}
  \affiliation{Center for Gravitational Waves and Cosmology, West Virginia University, Chestnut Ridge Research Building, Morgantown, WV 26505, USA}
  
\author[0000-0002-2878-1502]{Shami Chatterjee}
  \affiliation{Department of Astronomy and Cornell Center for Astrophysics and Planetary Science, Cornell University, Ithaca NY 14853, USA}

\author[0000-0001-6422-8125]{Amanda M.~Cook}
  \affiliation{Dunlap Institute for Astronomy \& Astrophysics, University of Toronto, 50 St.~George Street, Toronto, ON M5S 3H4, Canada}
  \affiliation{David A.~Dunlap Department of Astronomy \& Astrophysics, University of Toronto, 50 St.~George Street, Toronto, ON M5S 3H4, Canada}

\author[0000-0003-2317-1446]{Jason W.T. Hessels}
 \affiliation{Department of Physics, McGill University, 3600 rue University, Montr\'eal, QC H3A 2T8, Canada}
  \affiliation{Trottier Space Institute, McGill University, 3550 rue University, Montr\'eal, QC H3A 2A7, Canada}
  \affiliation{Anton Pannekoek Institute for Astronomy, University of Amsterdam, Science Park 904, 1098 XH Amsterdam, The Netherlands}
  \affiliation{ASTRON, Netherlands Institute for Radio Astronomy, Oude Hoogeveensedijk 4, 7991 PD Dwingeloo, The Netherlands}
  
  \author[0000-0001-6664-8668]{Franz Kirsten}
  \affiliation{ASTRON, Netherlands Institute for Radio Astronomy, Oude Hoogeveensedijk 4, 7991 PD Dwingeloo, The Netherlands}
   \affiliation{Department of Space, Earth and Environment, Chalmers University of Technology, Onsala Space Observatory, 439 92, Onsala, Sweden}
   
\author[0000-0003-3457-4670]{Ronniy C.~Joseph}
  \affiliation{Department of Physics, McGill University, 3600 rue University, Montr\'eal, QC H3A 2T8, Canada}
  \affiliation{Trottier Space Institute, McGill University, 3550 rue University, Montr\'eal, QC H3A 2A7, Canada}

    \author[0000-0001-9345-0307]{Victoria M. Kaspi}
\affiliation{Department of Physics, McGill University, 3600 rue University, Montr\'eal, QC H3A 2T8, Canada}
\affiliation{Trottier Space Institute, McGill University, 3550 rue University, Montr\'eal, QC H3A 2A7, Canada}

\author[0000-0002-5857-4264]{Mattias Lazda}
  \affiliation{Dunlap Institute for Astronomy \& Astrophysics, University of Toronto, 50 St.~George Street, Toronto, ON M5S 3H4, Canada}
  \affiliation{David A.~Dunlap Department of Astronomy \& Astrophysics, University of Toronto, 50 St.~George Street, Toronto, ON M5S 3H4, Canada}
  
\author[0000-0002-4279-6946]{Kiyoshi W.~Masui}
  \affiliation{MIT Kavli Institute for Astrophysics and Space Research, Massachusetts Institute of Technology, 77 Massachusetts Ave, Cambridge, MA 02139, USA}
  \affiliation{Department of Physics, Massachusetts Institute of Technology, 77 Massachusetts Ave, Cambridge, MA 02139, USA}
  
\author[0000-0003-0510-0740]{Kenzie Nimmo}
  \affiliation{MIT Kavli Institute for Astrophysics and Space Research, Massachusetts Institute of Technology, 77 Massachusetts Ave, Cambridge, MA 02139, USA}

\author[0000-0002-8897-1973]{Ayush Pandhi}
  \affiliation{David A.~Dunlap Department of Astronomy \& Astrophysics, University of Toronto, 50 St.~George Street, Toronto, ON M5S 3H4, Canada}
  \affiliation{Dunlap Institute for Astronomy \& Astrophysics, University of Toronto, 50 St.~George Street, Toronto, ON M5S 3H4, Canada}
  
\author[0000-0002-8912-0732]{Aaron B.~Pearlman}
  \affiliation{Department of Physics, McGill University, 3600 rue University, Montr\'eal, QC H3A 2T8, Canada}
  \affiliation{Trottier Space Institute, McGill University, 3550 rue University, Montr\'eal, QC H3A 2A7, Canada}
  
\author[0000-0002-4795-6973]{Ziggy Pleunis}
  \affiliation{Dunlap Institute for Astronomy \& Astrophysics, University of Toronto, 50 St.~George Street, Toronto, ON M5S 3H4, Canada}
  \affiliation{Anton Pannekoek Institute for Astronomy, University of Amsterdam, Science Park 904, 1098 XH Amsterdam, The Netherlands}
  \affiliation{ASTRON, Netherlands Institute for Radio Astronomy, Oude Hoogeveensedijk 4, 7991 PD Dwingeloo, The Netherlands}
  
\author[0000-0001-7694-6650]{Masoud Rafiei-Ravandi}
  \affiliation{Department of Physics, McGill University, 3600 rue University, Montr\'eal, QC H3A 2T8, Canada}
  \affiliation{Trottier Space Institute, McGill University, 3550 rue University, Montr\'eal, QC H3A 2A7, Canada}

 \author[0000-0002-6823-2073]{Kaitlyn Shin}
  \affiliation{MIT Kavli Institute for Astrophysics and Space Research, Massachusetts Institute of Technology, 77 Massachusetts Ave, Cambridge, MA 02139, USA}
  \affiliation{Department of Physics, Massachusetts Institute of Technology, 77 Massachusetts Ave, Cambridge, MA 02139, USA}
  
\author[0000-0002-2088-3125]{Kendrick M.~Smith}
\affiliation{Perimeter Institute for Theoretical Physics, 31 Caroline Street N, Waterloo, ON N25 2YL, Canada}

\begin{abstract}
The identification of persistent radio sources (PRSs) coincident with two repeating fast radio bursts (FRBs) supports FRB theories requiring a compact central engine. However, deep non-detections in other cases highlight the diversity of repeating FRBs and their local environments. Here, we perform a systematic search for radio sources towards 37 CHIME/FRB repeaters using their arcminute localizations and a combination of archival surveys and targeted observations. Through multi-wavelength analysis of individual radio sources, we identify two (20181030A-S1 and 20190417A-S1) for which we disfavor an origin of either star formation or an active galactic nucleus in their host galaxies and thus consider them candidate PRSs. We do not find any associated PRSs for the majority of the repeating FRBs in our sample. For 8 FRB fields with Very Large Array imaging, we provide deep limits on the presence of PRSs that are 2--4 orders of magnitude fainter than the PRS associated with FRB\,20121102A. Using Very Large Array Sky Survey imaging of all 37 fields, we constrain the rate of luminous ($\gtrsim$10$^{40}$ erg s$^{-1}$) PRSs associated with repeating FRBs to be low. Within the context of FRB-PRS models, we find that 20181030A-S1 and 20190417A-S1 can be reasonably explained within the context of magnetar, hypernebulae, gamma-ray burst afterglow, or supernova ejecta models---although we note that both sources follow the radio luminosity versus rotation measure relationship predicted in the nebula model framework. Future observations will be required to both further characterize and confirm the association of these PRS candidates with the FRBs.

\end{abstract}


\section{Introduction} \label{sec:intro}

More than 750 short-duration energetic radio transients, known as fast radio bursts \citep[FRBs;][]{Lorimer2007}, have been reported to date \citep[e.g.][]{chime-amiri2021, Nimmo2023}. There are two apparent populations of FRBs: ``one-off'' FRBs and ``FRB repeaters'', the latter of which have shown more than one burst. While only accounting for $<$10\% of known FRBs \cite{chime-amiri2021, CHIME2023, Michilli2022}, repeaters are particularly interesting since they provide more opportunities for detailed follow-up studies of the FRB.  While we do not yet know whether repeaters and one-off FRBs are two separate classes \citep[e.g.][]{Pleunis2021}, the existence of repeaters demonstrates that the progenitors of at least some FRBs are not cataclysmic (\citealt{Spitler2016}; for a review of FRB models, see \citealt{Platts_2019}). 

Despite multi-wavelength search efforts \citep{Andreoni_2020, Kilpatrick_2021, Nunez2021,Yan2024}, no prompt counterparts to FRBs have been found\footnote{The FRB-like burst of Galactic magnetar SGR 1935+2154 did show a hard X-ray burst counterpart \citep{CHIME-Anderson2020, Bochenek2020}}. However, long-lived counterparts have been seen in radio emission at frequencies between 100\,MHz and 22\,GHz. In particular, two well-localized repeating FRBs, FRB\,20121102A \citep{Chatterjee2017} and FRB\,20190520B \citep{Niu2021}, have each been associated with a long-lived ``persistent radio source'' (PRS). The PRSs that are coincident with FRB\,20121102A and FRB\,20190520B are similar: both are compact on milli-arcsecond (mas) scales \citep{Marcote2017, Bhandari2023}, have non-thermal (negative) spectral indices at a brightness temperature of $T > 5\times10^{7}$\,K for FRB\,20121102A,  luminosities of $L_\mathrm{radio}$ $\approx$ 10$^{29}$\,erg $s^{-1}$Hz$^{-1}$ and are spatially offset from the nuclei of their dwarf host galaxies by $<$10\,pc. Both FRBs associated with these PRSs exhibit high and highly variable Faraday rotation measures (RMs) \citep{Michilli2018extreme,Niu2021} suggesting a highly magnetized origin \citep[e.g.][]{Margalit2018}.

PRSs are thus defined as long-lived compact radio sources that are brighter than the local star formation in the galaxy and are clearly not  active galactic nuclei (AGN). It has been proposed that the luminosity of a PRS is correlated with the RM of the FRB if the RM primarily arises from the persistent emission region \citep{Yang2020}. Recently, a potential PRS with a positive spectral index ($\alpha \sim 0.97\pm 0.54$) was reported for FRB\,20201124A \citep{Bruni2023}. While the PRS luminosity correlates with the FRB's RM, similar to the other two confirmed PRSs, its lower luminosity, and other characteristics suggest a possible origin from star formation on a subarcsecond scale \citep{Nimmo2022,Dong-dust2024}. The two confirmed PRSs have radio flux densities higher than that expected by star formation activities in their host galaxies. Their fluxes have also been observed to be variable over short time scales ($\lesssim1$ year) \citep[e.g.][]{Rhodes2023,Zhang2023}. However, the time variability observed for the PRS of FRB\,20121102A has been suggested to be a result of refractive scintillation as opposed to intrinsic \citep{Chen2023}.

It is possible that a single central engine could both produce FRBs and power the PRS. This general idea has been detailed in multiple models: the magnetar model \citep{Beloborodov2017,Margalit2018,margalit2019,Bhandari2023}, gamma-ray burst afterglows \citep{ Murase2016, Beloborodov2017, Metzger2017}, ultra-luminous X-ray binaries \citep{Sridhar_2021,
Sridhar&Metzger_22, Sridhar+24}, and synchrotron heating \citep{Yang2016,Li2020} among others \citep{Platts_2019}\footnote{https://frbtheorycat.org/index.php/Main$\_$Page}. The PRS of FRB\,20121102A exhibits analogous characteristics to the radio emission observed from a previous luminous supernova (SLSN; \citealt{Eftekhari2019}). This indicates a possible connection between the two classes of transients, with the SLSN being one of the proposed FRB progenitor channels. It is also possible for a PRS to be produced by the massive accreting black hole of the host galaxy \citep{Zhang_2020}.

If all repeating FRBs are associated with PRSs, this would have implications for their progenitor channels and emission mechanisms. In contrast, robust non-detections of PRSs in a sample of FRB repeaters \citep[as recorded by e.g.][]{Marcote2020, kirsten2021, Nimmo2022} could imply (i) a different progenitor channel for some events, (ii) that the physical conditions implied by the PRS in FRB\,20121102A are not \emph{required} for the production of repeating FRBs or (iii) that PRSs have shorter lifetimes such that the emission is no longer detectable at the time of observations. It is also possible that PRSs could be associated with some one-off FRBs and not only with repeaters \citep{Law2022}. However, given that the 2 FRBs with PRSs and the one with a candidate PRS are all repeaters, these have shown that at a minimum PRSs are an important phenomenon associated with repeating FRB nature. While the first two discovery PRSs are remarkably similar, if confirmed, including FRB\,20201124A brings diversity to the properties of the PRS population. This diversity inspires the need to identify and study more PRSs to understand their connection with FRBs and shed light on the emission mechanisms, progenitor channels, and population variation of FRBs \citep{Vohl2023, Dong2024}.

Typically, robust multi-wavelength associations require  $\lesssim$ 1\,arcsecond localizations for FRBs, and hence the detection of a burst with an interferometer \citep{EftekhariandBerger2017, Eftekhari2018}. However, \cite{Eftekhari2018} demonstrate that due to the lower density of radio sources in the sky compared to faint optical sources \citep{Gordon2021, Driver2016}, robust FRB-PRS associations can be made for coarser localizations ($\sim$20 arcseconds). In addition, they argue that for larger localization regions that preclude firm associations, constraints can still be placed on possible radio associations as a function of luminosity. 

Arcsecond localizations are not available for most FRBs discovered by the Canadian Hydrogen Intensity Mapping Experiment (CHIME) telescope. Localizations with CHIME's best-recorded precision ($\sim$1\,arcminute) can be obtained using the channelized raw voltages of the FRB signals from the telescope feeds (i.e.\ the baseband data) following the techniques described by \cite{Michilli_2021}.

Although these baseband positions exceed the threshold of $\lesssim$20" found by \citep{Eftekhari2018} to robustly associate a radio source with an FRB based on probability of chance alignment arguments, it is still possible to search for radio sources within the CHIME/FRB localization regions. One can then analyze the properties of these radio sources along with their host galaxies to identify any that may originate from a source other than star formation or an AGN in their host galaxy. While follow-up observations would be necessary to confirm an FRB association with any radio sources of interest, such studies still have the potential to provide insights into the possible prevalence of PRSs or the depth of any non-detections. 

\begin{deluxetable*}{llcccccccc}
\tabletypesize{\small}
\tablecaption{Summary of Archival Radio surveys employed in this study \label{tab:catalogue details}}
\tablehead{\colhead{Survey} & \colhead{Telescope} & \colhead{Observation} & \colhead{Frequency} & \colhead{Angular} & \colhead{Sky} & \colhead{Sensitivity$^a$} & \colhead{N$_r^b$} & \colhead{$\mu$$^c$} & \colhead{Reference} \\
\colhead{} & \colhead{} & \colhead{Dates} & \colhead{} & \colhead{Resolution} & \colhead{Coverage} & \colhead{} & \colhead{} & \colhead{} & \colhead{} \\
\colhead{} & \colhead{} & \colhead{} & \colhead{(\,GHz)} & \colhead{(\,$^{\prime\prime}$)} & \colhead{(\,deg$^{2}$)} & \colhead{(\,mJy/beam)} & \colhead{} & \colhead{} & \colhead{}
  }
\startdata
VLASS & VLA & 2017$-$2024 & 2$-$4 & 2.5 & 33,885 &  0.12 & 3 & 3.9$\pm$2.5
     & \cite{McConnell_2020}\\
     FIRST & VLA & 1993$-$2004 & 1.5 & 5 & 10,575 & 
     0.15 & 2 & 3.9$\pm$2.5
     & \cite{Becker1995} \\
     NVSS & VLA & 1993$-$1996 & 1.4 & 45 & 23,264 & 
     2.50 & 4 & 5.1$\pm$2.7
     & \cite{Condon1998}\\
     RACS & ASKAP & 2019$-$2020 & 0.887$-$1.655 & 15 & 36,656 &
     0.25 & 1 & 2.4$\pm$2.1
     & \cite{McConnell_2020}\\
     TGSS ADR1 & GMRT & 2010$-$2012 & 0.150 & 25 & 36,900 & 
     3.50 & 2 & 0.7$\pm$1.4
     & \cite{Intema2017} \\
    LoTSS &  LOFAR & 2014$-$2021  & 0.10$-$0.24 & 6 & 5635 & 
    0.08 & 8 & 34.6$\pm$6.4 
    & \cite{Shimwell2022} 
\enddata
\tablenotetext{a}{1$\sigma$ root-mean-square sensitivity.}
\tablenotetext{b}{N$_\mathrm{r}$ is the number of radio sources found for each catalog in the FRB localization regions. Compared to $\mu$, No catalog shows any excess radio source in the field (See \S ~\ref{subsec:global-implication} for a discussion).}
\tablenotetext{c}{$\mu$ is the expected number of radio sources given the total area of CHIME/FRB fields searched that overlaps with each of these catalogs. The errors are propagated using \cite{Gehrels1986A} Poisson probability error function for 1-$\sigma$ values (see \S ~\ref{subsec:pcc-global} for details).}

\end{deluxetable*}

This type of study serves as a precursor for classifying radio sources as PRSs for upcoming telescopes like the CHIME/FRB Outrigger project \citep{Leung2021,Mena-Parra2022,Cassanelli2022,Lanman2024}, the fast radio transient-detection program at MeerKAT \citep[MeerTRAP;][]{Rajwade2022meerkat}, the Deep Synoptic Array \citep[DSA-110;][]{Ravi2022DS110} and the Bustling Universe Radio Survey Telescope in Taiwan \citep[BURSTT;][]{BURSTT2022} which will improve the number of precisely localized FRBs and allow for more robust multi-wavelength association.  Meanwhile, by leveraging a substantial sample of repeating FRBs discovered by CHIME/FRB, significant constraints can be placed on the presence of FRB\,20121102A-like PRSs using CHIME/FRB repeaters and arcsecond radio continuum surveys. 

In this work, we searched for and studied radio sources found within the localization regions of the 37 recently published CHIME/FRB repeaters for which $\sim$arcmin baseband localizations were available \citep{Bhardwaj_2021, BhardwajR42021, CHIME2023, Michilli2022}. We report the data used for the search, relevant observations carried out, and the result of the search in \S ~\ref{sec:data-and-observations}. We then describe the multi-wavelength data used and the diagnostic analysis carried out for each of the radio sources in \S ~\ref{sec:multiwavelength-data} and \S ~\ref{sec:multi-wavelength-analysis} respectively. 
The summary of the search result and candidate potential PRSs are presented in \S ~\ref{sec:discussion}. We then describe the implication of the radio sources in the context of existing FRB-PRS models in \S ~\ref{subsec:comparison}, followed by the conclusion in \S ~\ref{sec:conclusion}. Detailed results of individual radio sources can be found in Appendix ~\ref{sec:individual-radio-sources}.
In all cases, we convert redshifts to luminosity distances assuming a flat $\Lambda_{\mathrm{\textit{CDM}}}$ cosmology with $H_{\mathrm{0}}$ $=$ 67.7 km\,s$^{-1}$ Mpc$^{-1}$, $\Omega_{\mathrm{m}}$ = 0.31, and $\Omega _{\mathrm{\wedge}}$ = 0.68 \citep{planck2018vi}.

\section{A Search for PRS Candidates in the Localization Region of CHIME/FRB Repeaters} \label{sec:data-and-observations}

Here we describe the process used to search for radio sources in the fields of CHIME/FRB repeaters. Specifically, after describing the FRB fields searched in \S ~\ref{subsec:sample-selection}, we describe the radio catalogs (\S ~\ref{subsec:radio-catalogs}) and targeted VLA observation images searched (\S ~\ref{subsec:realfast-obs} and \S ~\ref{subsec:vlite}) and present the initial result of the searches (\S ~\ref{sec:archival-result} and \S ~\ref{sec:vla-result}). The probability statistics of the archival and deep VLA results are described in \S ~\ref{subsec:pcc-global} and \S ~\ref{subsec:pcc-fr}.

\subsection{FRB Sample} \label{subsec:sample-selection}
Of the 52 repeating FRBs presented by \cite{CHIME2023, Michilli2022}  which represents almost all known repeaters, we have selected objects that have $\lesssim$2 arcminute localizations. 
This resulted in 37 FRB regions searched for PRSs including FRB\,20181030A, and FRB\,20200120E which were initially presented by \citeauthor{Bhardwaj_2021} (\citeyear{Bhardwaj_2021,BhardwajR42021}). While for a majority of these FRBs, the baseband localization region is on order 1$-$2 arcmin from the CHIME baseband positions, we note that 2 FRBs in our sample (FRB\,20180916B; \citealt{Marcote2020} and FRB\,20200120E; \citealt{Kirsten_2020}) have milliarcsecond positions available.

\subsection{Radio Data} \label{sec:radio-data}
\subsubsection{Radio survey catalog search} \label{subsec:radio-catalogs}
 
We first searched for the presence of candidate PRSs within the 90\% confidence level error regions of the 37 FRBs using archival radio catalogs from the VLA Sky Survey \citep[VLASS;][]{Lacy_2020}, the NRAO VLA Sky Survey \citep[NVSS;][]{Condon1998}, the Faint Images of the Radio Sky at Twenty-Centimeters \cite[FIRST;][]{Becker1995} survey, the TIFR-GMRT Sky Survey \cite[TGSS;][]{Intema2017},  the Rapid ASKAP Continuum Survey \cite[RACS;][]{McConnell_2020} and the high-resolution component of the LOw-Frequency ARray LOFAR Two-metre Sky Survey \cite[LoTSS;][]{Shimwell2022}. For each survey, we list their frequency, angular resolution, sky coverage, and sensitivity in Table ~\ref{tab:catalogue details}.

\subsubsection{Deep VLA Observation} \label{subsec:realfast-obs}
 
In addition, we have carried out observations of FRB repeater fields with the Karl G. Jansky Very Large Array (VLA) interferometer in different array configurations of the VLA, all in the 1$-$2\,GHz-band, through program numbers 18B-405 (PI: Casey Law), 19B-223, 19A-331, 20B-280, 20A-469, 21B-176, and 21A-387 (PI: Shriharsh Tendulkar). In the 1$-$2\,GHz-band, the full width at half power of the primary beam is 28 arcmin, which is 14 times larger than the typical CHIME FRB localization region. Details of the observations are given in Table ~\ref{tab:radio-obs}. No FRBs were detected in the commensal realfast \citep{Law2018} observations during these programs.

The Common Astronomy Software Applications package \citep[CASA;][]{casa2007} was used to perform the data reduction, flux measurement calibration, and imaging of the data. Specifically, we used the Python-based CASA pipeline tool \texttt{pwkit} released by \cite{peterwilliams2017}. We flagged for radio frequency interference (RFI) using the automatic AOFlagger \citep{Offringa2010}. The bandpass and phase calibration were carried out using the specified calibrators as listed in Table ~\ref{tab:radio-obs}.  After data reduction, we imaged the total intensity component (Stokes I) of the source visibilities, setting the cell size so there would be 4$-$5 pixels across the width of the beam. All calibrated data were imaged using the CLEAN algorithm and primary beam correction was carried out. For cases where there is more than one observation of a particular field, we combine the calibrated visibilities in the uv-plane to produce a single higher-sensitivity image shown in Figure ~\ref{fig:VLA-images}. 
There are some observations taken during the time when the VLA telescopes are moving from A to D configuration. We note that as a result of this movement, the data quality from this set-up could be poor and could affect the final image. 

\texttt{AEGEAN} \citep{Hancock2012,Hancock2018} was used to find sources and measure fluxes and associated uncertainties within each image. The summary of all measured radio flux densities and upper limits are presented in Appendix~\ref{sec:individual-radio-sources} Table ~\ref{tab:radio-archive}, and discussed in Section~\ref{sec:vla-result} below. 

\begin{deluxetable*}{llccccccc}
\tabletypesize{\small}
\tablecaption{Summary of Deep Radio Observations with VLA at 1.5 GHz for listed FRB fields\label{tab:radio-obs}}
\tablehead{\colhead{FRB} & \colhead{Configuration} & \colhead{Date} & \colhead{bandpass calibrator} & \colhead{phase calibrator} & \colhead{RMS Sensivity$^a$} & \colhead{Ns$^b$} & \colhead{Integration$^c$}\\
\colhead{} & \colhead{} & \colhead{} & \colhead{} & \colhead{} & \colhead{(\,$\mu$Jy/beam)} & \colhead{}  & \colhead{Time (hr)}}
\startdata  
     FRB\,20180814A & B & March, June, July, 2019  & 0542+498=3C147 & J0410+7656 & 6.5 & 2 & 16\\
     & A, A$\rightarrow$D$^f$ & 22 Oct, 2020, Feb$-$March, 2021  & 0542+498=3C147 & J0410+7656 & 5.4 & 4 & 11.5 \\
      & combined & combined  & 0542+498=3C147 & J0410+7656 & 3.5 & 5 & 27.5$^{d}$ \\  
     FRB\,20180916B & B & 7$-$22 June, 2019 & 0542+498=3C147 & J0217+7349 & 6 & 0 & 12\\
      FRB\,20181030A & A$\rightarrow$D$^f$ & 2$-$12 March, 2021 & 1331+305=3C286 & J1035+5628 & 10 & 1 & 4\\
      FRB\,20190117A & BnA & 2 Nov, 2020 & 0137+331=3C48 & J2139+1423 & 16 & 1 & 1\\
     FRB\,20190208A & A & 5$-$7 Feb, 2021 & 0137+331=3C48  & J1852+4855  & 5 & 1 & 10 \\
     FRB\,20190303A$^{e}$ & B, A$\rightarrow$D$^f$ & Nov, Dec, 2019, Jan 2020 &  1331+305=3C286  & J1313+5458 & 5 & 2 & 10\\
      & A & Feb, 2021 &  1331+305=3C286  & J1352+3126 & 5 & 2 & 15.5\\
     FRB\,20190417A & A & 25$-$27 Feb, 2021 & 1331+305=3C286 & J1944+5448 & 8.5 & 2 & 4\\
     FRB\,20200120E & A & 29 Dec, 2020, 13$-$18 Jan, 2021 & 0542+498=3C147 & J1048+7143 & 9 & 0 & 5   
\enddata
\tablecomments{The results are from combined observations for all fields except for FRB\,20190117A which has a single observation.}
    $^a$ 1-$\sigma$ root-mean-square error of the image. \\
    $^b$ Number of sources within the FRB error region.\\
    $^c$ The total on-source time including integration and overhead.\\
    $^{d}$ Noise level obtained from the combination of the visibilities of all images from 2019, 2020, and 2021 in all arrays to enhance image resolution.\\
    $^{e}$ Images were combined separately as shown in this table because of the morphology of the extended sources found in this field.\\
$^f$ A$\rightarrow$D configuration is an observational setup at the time when the VLA telescopes are moving from A to D configuration. As a result of this movement, the data quality from this set-up can be poor most of the time, hence we advise caution when interpreting results from such set-ups.
\end{deluxetable*}

\subsubsection{Commensal VLITE Observations} \label{subsec:vlite} 

Finally, we obtained simultaneous data for our VLA observations using the VLA Low-band Ionosphere and Transient Experiment (VLITE) which is a commensal low-frequency system on the VLA that operates in parallel with nearly all observing programs above 1\,GHz \citep{Clarke}. VLITE records a parallel data stream during regular programs on a subset of up to 18 VLA antennas with a bandwidth of 64 MHz centered at 352 MHz. The  VLITE data are correlated with a custom DiFX-based correlator \citep{2007PASP..119..318D} and processed through a dedicated calibration and imaging pipeline. The pipeline uses a combination of Obit \citep{obit} and AIPS \citep{aips} for standard reductions including removing RFI solving for the delay, complex gain, and bandpass. VLITE uses the \citet{pb2017} flux density scale and an additional calibration uncertainty of 15\% has been added to the measurement errors. 

We used PyBDSF \citep{pybdsf} to measure fluxes and associated measurement errors from the VLITE images. We present the measurements in Appendix~\ref{sec:individual-radio-sources} Table~\ref{tab:radio-archive} where non-detections within the error region are reported as 5$\sigma$ upper limits.

 \subsection{Archival Search Result } \label{sec:archival-result}
Out of 37 FRB fields searched 13 unique radio sources in the archival surveys were found in all the regions combined. Specifically, there are 8 LoTSS radio sources within the localization regions for 7 CHIME/FRB repeaters, 4 sources in NVSS, 2 in FIRST, 2 in TGSS, 1 in RACS, and 3 in VLASS.

Some of these radio sources are from the same location but seen at different epochs and frequencies as represented by the catalogs. In addition, we obtained upper limits on the flux densities for the potential existence of radio sources at the positions of the remaining 34 FRBs where there were no VLASS sources even though they are within the VLASS coverage. We adopt VLASS for this purpose, as it was the most sensitive catalog searched at GHz frequencies. A comprehensive investigation of the properties of these sources is presented in \S ~\ref{sec:multi-wavelength-analysis} and their flux density limits and frequencies are reported in Appendix~\ref{sec:individual-radio-sources} Table ~\ref{tab:radio-archive}.

\begin{figure*}
\centering
\includegraphics[width=\textwidth]{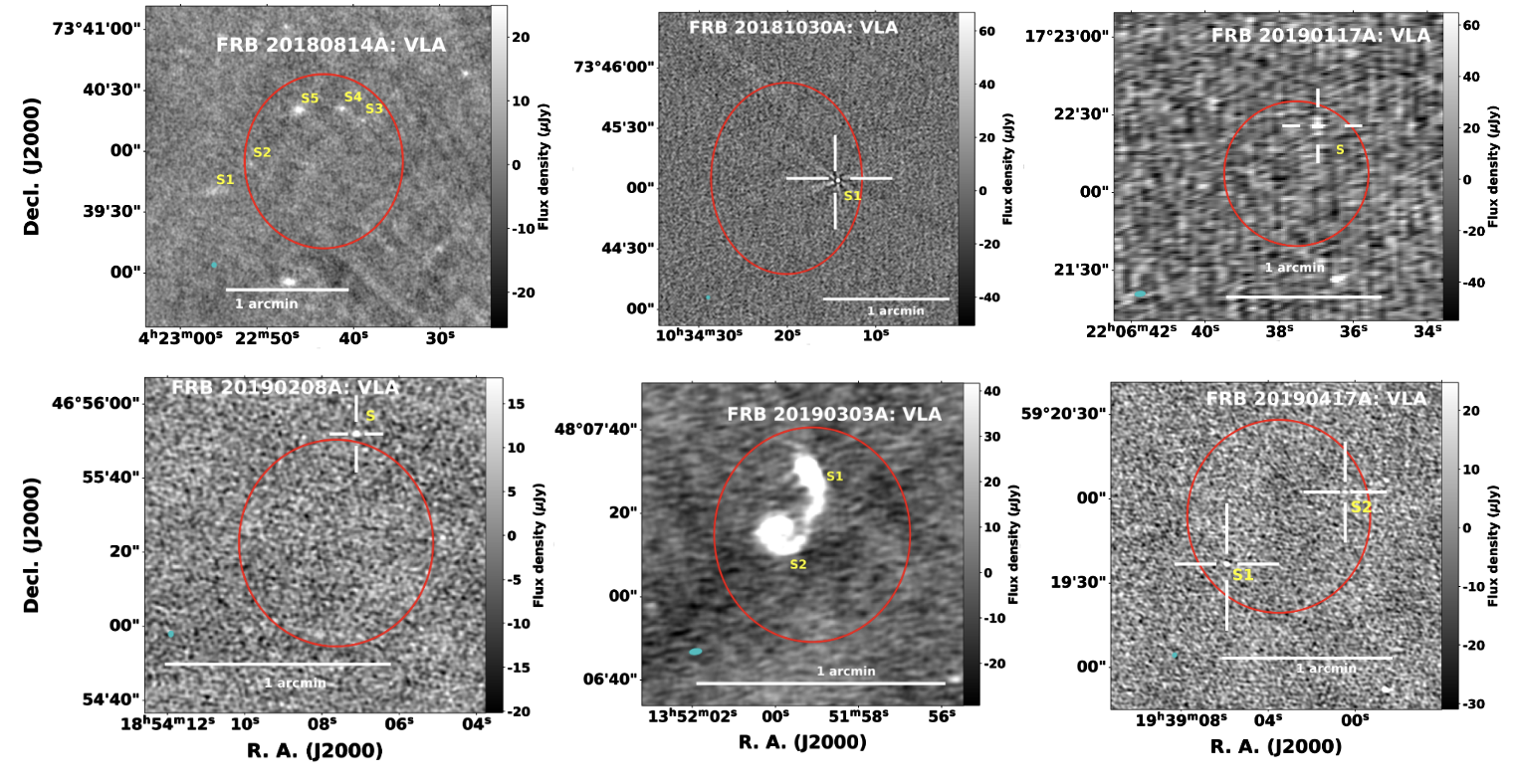}
\caption{VLA images capturing the 6 FRB fields featuring detected radio sources (unresolved, resolved, and extended). The 90\% confidence level baseband error regions of the FRBs are depicted as red ellipses. For some fields, the white crosshair marks the location of the sources, while others have the sources labeled with 'S' corresponding to the number of radio sources in the field. The beam of each image is represented by a small cyan-colored ellipse at the lower left corner of each image.} \label{fig:VLA-images}
\end{figure*}

\subsection{Deep VLA results} \label{sec:vla-result}
Here we report the sources found in the field of the FRBs observed in deep VLA observations. Out of 8 FRB fields imaged with deep-targeted VLA observations, there are 6 fields with one or more radio sources within the CHIME baseband error region. Specifically, we found 5 sources for FRB\,20180814A, 1 source for FRB\,20181030A, 1 source for FRB\,20190208A, 1 source for FRB\,20190117A, 2 sources for FRB\,20190417A, and 2 sources for FRB\,20190303A for a total of 12 radio sources from the deep VLA images. In Figure~\ref{fig:VLA-images}, we show the deep VLA images for these six fields with both the location of detected radio sources and the CHIME baseband localization region indicated. In addition, we compute the 5-$\sigma$ upper limit based on the root-mean-square (r.m.s.) noise within the localization region of the FRB in the final image. These would represent limits on the presence of a PRS in the event that none of the radio sources identified are associated with the FRB.

\begin{figure*}
\centering
\includegraphics[width=.9\textwidth]{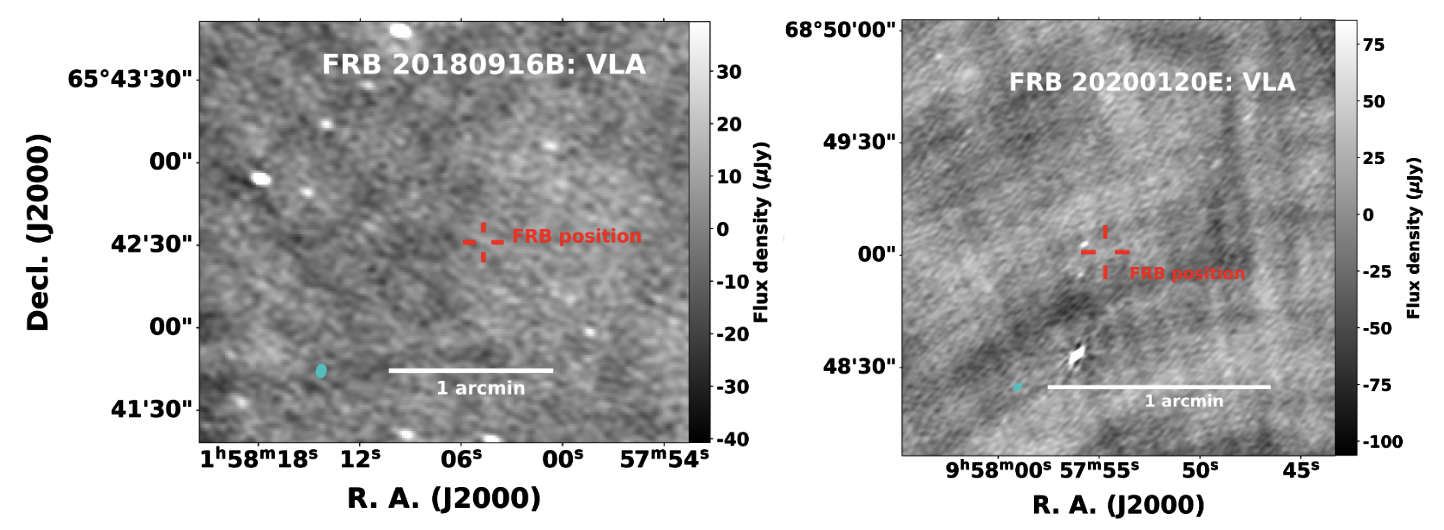}
\caption{VLA images--FRB\,20180916B and FRB\,20200120E. There is no radio source at the location of these FRBs. The red crosshairs indicate the subarcsecond position of the FRBs but are much larger than the representative of the positional uncertainties. } \label{fig:VLA-upp}
\end{figure*}

The other two FRBs (FRB\,20180916B and FRB\,20200120E) have sub-arcsecond positions from Very Long Baseline Interferometry (VLBI) that were used for the search instead. There is no radio source in the VLA images at the location of either FRB, hence we measure 3-$\sigma$ flux density upper limits at these locations.  In Figure ~\ref{fig:VLA-upp}, we show the radio non-detection in the deep VLA images for these two fields with the subarcsecond location of their FRBs indicated. A summary of measured radio flux densities and limits is given in Appendix~\ref{sec:individual-radio-sources} Table ~\ref{tab:radio-archive}.

\subsection{Expected Number of radio sources, $\mu$} \label{subsec:pcc-global}
While we will perform a detailed analysis to assess the potential origin of each radio source below, we can also compare the number of radio sources identified within the CHIME localization regions to expectations based on the source counts in each catalog searched. Doing so will allow us to assess whether there is any evidence for a (statistical) excess of sources compared to background levels, potentially due to association with the FRBs.

To do this, we used the total number of sources found for each catalog, the total sky area covered by the entire survey, and the total CHIME/FRB area that overlaps with the catalog to estimate the average number of sources ($\mu$) expected from searching the entire CHIME/FRB region\footnote{The expected number of radio sources, $\mu$ is estimated as $\mu = \frac{A_\mathrm{b}}{A_\mathrm{cat}} \times N$, where $N$ is the total number of sources from each catalog, $A_\mathrm{cat}$ is the area covered by a specific survey catalog and $A_\mathrm{b}$ is the total area of the 90\% confidence level of CHIME/FRB baseband regions searched.}. This value could be used to obtain the probability of finding the observed number of radio sources for each catalog. The errors on the expectation number are propagated using \cite{Gehrels1986A} Poisson probability error function for 1-$\sigma$ values. This analysis was done for all the catalogs searched and the results are presented in Table ~\ref{tab:catalogue details}. All catalog results have fewer or an equal number of sources compared to the expected number. Thus, we do not find evidence of an excess of sources in the CHIME localization regions at the depths covered by these archival surveys. While in most surveys the number of sources found is consistent within errors to the expected number we note that the number of LoTSS sources identified is a factor of $\sim$4 lower than expected. This may mean that the $\sim$0.043 deg$^{2}$ covered by the CHIME repeater baseband regions searched in this study are not representative of the general source density found in the 5600\,deg$^{2}$ covered by LoTSS DR2 \citep{Shimwell2022}.

\subsection{Individual Chance Coincidence Probability, P$_\mathrm{cc,rad}$ of associating each radio source to its FRB} \label{subsec:pcc-fr}

In addition to searching for statistical excesses of sources, we can also estimate the probability of chance coincidence between each radio source and its CHIME FRB localization region, P$_\mathrm{cc,rad}$, based on the size of the FRB uncertainty region and the flux density of radio source following \citealt{Eftekhari2018} (see Table ~\ref{tab:probabilities} for a brief description). We assumed a Poisson distribution of radio sources across the sky by calculating the chance coincident probability as $P_\mathrm{cc,rad} = 1 - exp\,(-\pi R_\mathrm{FRB}^{2}n(>S_\nu)$), where $R_\mathrm{FRB}$ is the 90\% error radius of the FRB, $S_\nu$ is the flux density of the radio source and n($>S_\nu$) is the number density of radio sources as bright or brighter than the flux density ($>S_\nu$) of the radio source found. For sources that were found from the deep VLA images and other archival sources at 1.4 and 1.5 GHz, we used the MeerKAT DEEP2 \citep{Matthews2021} source count dataset. For VLASS sources, we used the VLASS source count described by \cite{Gordon2021} and for the other low-frequency radio sources, we assumed a spectral index of $-$0.71 according to \cite{Gordon2021} to estimate their 1.4\,GHz MeerKAT equivalent. 

Given the size of the CHIME/FRB uncertainty region, we expect that the chance probability of associating the radio sources to the FRB (P$_\mathrm{cc,rad}$) will not be low, however, for completeness's sake, we report these values for the sources in Appendix~\ref{sec:individual-radio-sources} Table ~\ref{tab:radio-archive}. To accommodate effects resulting from multiple fields being searched, we also computed the adjusted chance coincidences following the application of the Bonferroni correction, also known as the 'look-else-where' effect \citep{gs14}. Consistent with the findings of \cite{Eftekhari2018}, we find that for localization regions of the size provided by CHIME baseband data, it is not typically possible to associate an individual radio source to a CHIME FRB based on purely statistical arguments. Finally, due to the difficulty with the accurate determination of flux densities of blended sources, we did not estimate P$_\mathrm{cc,rad}$ values for any such sources.

\begin{deluxetable*}{lllccccc}
\tabletypesize{\small}
\tablecaption{Description of Probabilities and Expectation values used \label{tab:probabilities}} 
\tablehead{\colhead{Symbol} & \colhead{Name} & \colhead{Description} & \colhead{Sections} & \colhead{Tables}  \\
\colhead{} & \colhead{}  & \colhead{} & \colhead{}  & \colhead{}}  
\startdata
$\mu$ & Expected number of radio sources & Probability of finding the observed number of radio sources &  \S ~\ref{subsec:pcc-global} & Table ~\ref{tab:catalogue details} \\
 & &  within the entire CHIME/FRB regions searched for each catalog &  & \\ \hline
P$_\mathrm{cc,rad}$ & Chance Probability of Association & Chance probability that a radio source &  \S ~\ref{subsec:pcc-fr}, ~\ref{sec:individual-radio-sources} & Table ~\ref{tab:radio-archive} \\ 
& between a radio source and its FRB & is found within its FRB's uncertainty region &   & \\ \hline
P$_\mathrm{cc,gal}$ & Chance Probability of Association  & Chance coincident probability of associating a radio source  &  \S ~\ref{subsec:pcc-rg}, ~\ref{sec:individual-radio-sources} & Table ~\ref{tab:radio-opt} \\
 & between a radio source and its host galaxy & to its plausible host galaxy. & &  & \\
 & & The  value corrected for the Bonferroni effect is named  P$_\mathrm{cc,RG}$& &  & 
\enddata
\end{deluxetable*}

\section{Multiwavelength Data for PRS Candidates}\label{sec:multiwavelength-data}

In total, we identified 25 radio sources within the CHIME localization regions of 37 repeating FRBs from both archival and targeted observations. As shown in \S ~\ref{subsec:pcc-global} and \S ~\ref{subsec:pcc-fr}, in many cases the $\sim$arcminute size of these localization regions precludes making a robust association with the FRB from statistical arguments. We will therefore rely on multiwavelength analysis to investigate the possible origin of each radio source and identify sources of particular interest as PRS candidates. Here we describe the multiwavelength data that will be used in this analysis.

\subsection{Optical and Infrared Photometry}

\subsubsection{Cross-match with optical catalogs}\label{subsec:cross-match}
We first searched for optical counterparts to the radio sources within 2 arcseconds of each radio source position using the Sloan Digital Sky Survey (SDSS; \citealt{Ahn2012}), The Pan-STARRS catalog (PS1) \citep{chambers2016pan} and Dark Energy Spectroscopy Instrument (DESI), Data Release 9 (DR9) \citep{dsl+19} Legacy Imaging Survey photometric catalogs. We use PanSTARRS/DESI catalog flags to verify that none of these optical sources are classified as probable stars. Out of 25 radio sources (including those found in archival surveys and the targeted VLA observations), we found optical sources that are spatially consistent with 17 of them. In Table ~\ref{tab:radio-opt}, we list relevant properties of these optical sources that are provided by optical catalogs and will be used in the analysis in the following sections. This includes position, apparent r-band magnitudes, and half-light radius.

We did have a unique case of 20181030A-S where the radio source physically overlaps a spiral arm of the nearby (D$_L$ $\sim$20 Mpc; \citealt{Theureau2007}) galaxy, NGC 3252. While the radio source is located $\sim$1 arcminute from the galaxy nucleus, by visual inspection and considering the physical size of the galaxy, we infer that the radio source may be associated with the galaxy.

\subsubsection{Gemini Photometry} \label{subsec:Gemini-Photometry}
In addition, as part of ongoing CHIME/FRB follow-up efforts, we requested deep Gemini imaging of the field of FRB\,20190417A. The original motivation was the high dispersion measure (DM) of the FRB \citep[1378.2$\pm$2\,pc\,cm$^{-3}$;][]{Fonseca2020}. 
Given that higher DM can correspond to higher redshifts \citep{Macquart_2020}, this would allow us to search for fainter optical sources than previously identified by other surveys such as PanSTARRS and DESI.

Gemini North imaging of the field of FRB\,20190417A was carried out in the r-, i-, and z-bands on 29th July 2022 (for r-band) and 19th September 2022 (for i- and z-bands). Total exposure time was 6 $\times$ 250s for the i- and r-bands and 10 $\times$ 150s for z-band. The data were reduced using the standard packages of \texttt{gmos} and \texttt{gemini} routine in \texttt{PYRAF} \citep{pyraf2012}. This included overscan correction, flat fielding, image alignment, mosaicing, and combination. The different exposure images were combined to produce a single image. 

We used \texttt{PHOTUTILS} \citep{larry_bradley_2023} and \texttt{SExtractor} \citep{Bertin1996} source-finding software to carry out background subtraction and estimate the properties of the source associated with the radio source. Specifically, we measured the apparent magnitude by calibrating it against standard stars from PanSTARRS and then used the \texttt{iraf} routine to estimate the full-width half maximum of the optical source to measure the half-light radius. 

\subsubsection{Infrared Photometry}\label{subsec:infrared-photometry}
For radio sources within the coverage, we retrieved photometric data for each of the radio sources from the Wide-field Infrared Survey Explorer (WISE) \citep{Wright2010} where available.

\subsection{Optical Spectroscopy} \label{sec:Optical-Spectroscopy}

For radio sources where we find an optical counterpart, we compile their spectroscopic redshifts and emission line fluxes when possible. These will be to estimate other galaxy properties in the sections below.

\subsubsection{Archival/Published Spectroscopy} \label{ssubsec: archival-spec}

Redshift information was taken from the literature for some radio sources where the spectrum of their host galaxy is already published. Specifically, this was available for 20190303A-S1, 20190303A-S2, 20180814A-S \citep{Michilli2022}, as well as 20200223B-S and 20191106C-S \citep{Ibik2023}. Redshift information for these sources is listed in Appendix~\ref{sec:individual-radio-sources} Table ~\ref{tab:radio-derived}. 

\begin{figure*}[t!]
\centering
\includegraphics[width=.85\textwidth]{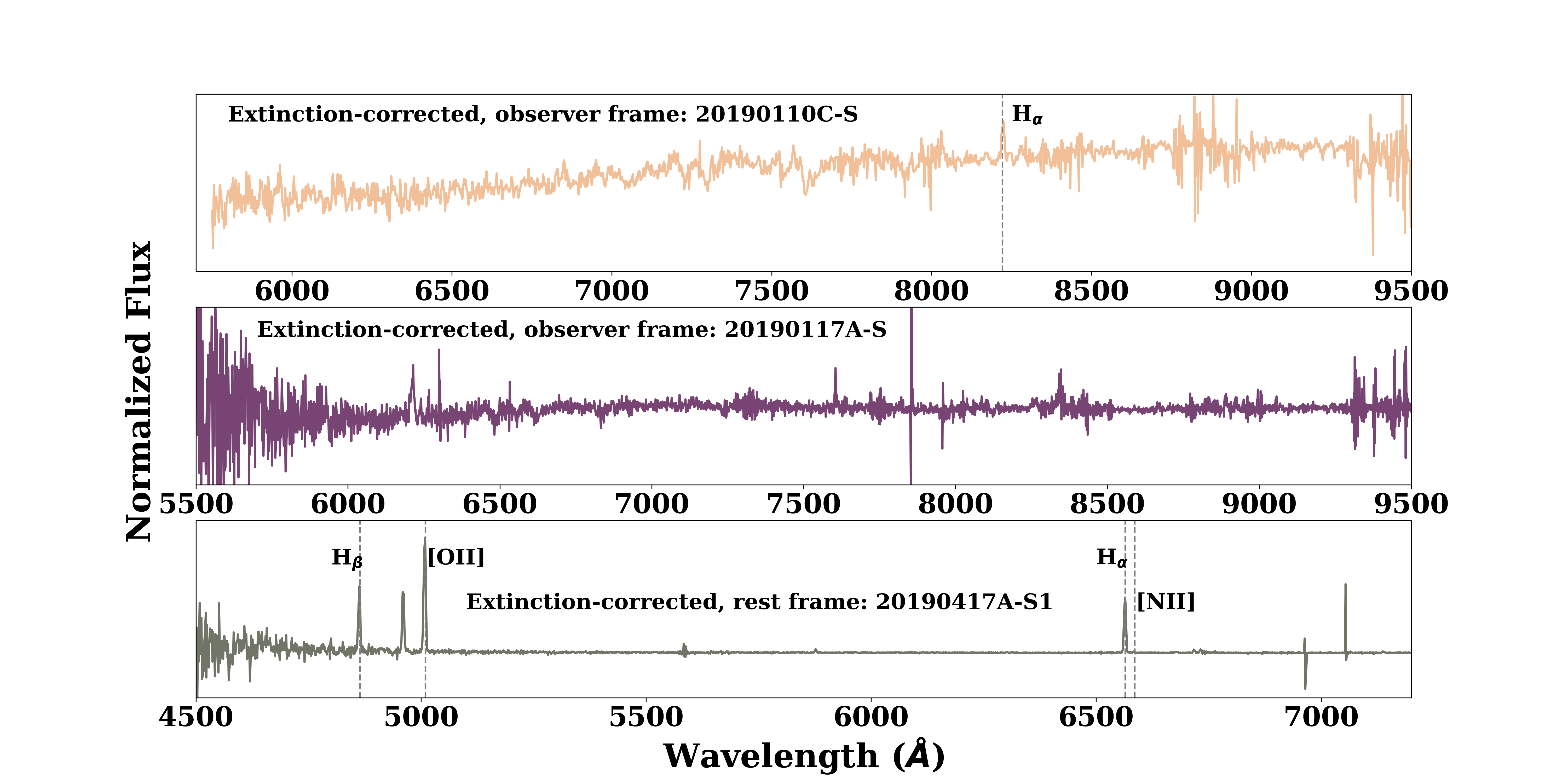}
\caption{Gemini North spectra for the plausible host galaxies of radio sources 20190110C-S (top panel), 20190117A-S (middle panel), and 20190417A-S1 (bottom panel). In the top panel, the spectrum exhibits a bright emission line, possibly H$_{\mathrm{\alpha}}$, which is also clearly visible in the 2-dimensional spectra. 
In the middle panel, there are no prominent emission or absorption lines, making it unsuitable for estimating the redshift of the likely host galaxy of 20190117A-S. In the bottom panel, the spectrum of the likely host galaxy of 20190417A-S1 displays distinct emission line features, indicated by dashed gray lines. These lines were used to derive a redshift of $z_\mathrm{spec} = 0.12817(2)$ and other properties of the galaxy as discussed in \S ~\ref{subsec:R18}.} \label{fig:gemini-spectra}
\end{figure*}

\subsubsection{Gemini North Spectroscopy} \label{subsec: gemini-spec}

In addition, we obtained observations of the likely host galaxies of 20190110C-S, 20190117A-S, and 20190417A-S1 with the Gemini Multi-Object Spectrograph (GMOS; \citealt{GMOS2004}). The data were obtained on 2022 August 20 and 2022 September 20 for exposure times of (5 $\times$ 1200s), (7 $\times$ 1200s), and (6 $\times$ 1060s) respectively. We used the $1.0^{\prime\prime}$ long-slit, R400 grating and OG515 ($>$ 520\,nm) blocking filter, at central wavelengths of 720/730\,nm.

The data were reduced using the standard \texttt{gmos} and \texttt{gemini} packages in \texttt{PYRAF} \citep{pyraf2012} and also with \texttt{PypeIt} \citep{Prochaska2020}. This included overscan correction, flat fielding, sky subtraction, wavelength calibration, and extraction. The flux calibration was done using a standard star observed with the same setup on a different night. Chip gaps and cosmic rays were removed from individual exposures and the different exposure spectra were combined to produce a final spectrum. All three GMOS spectra are shown in Figure~\ref{fig:gemini-spectra}.

For the spectrum of the plausible host galaxy of 20190117A-S (middle panel of Figure ~\ref{fig:gemini-spectra}), there were no prominent emission or absorption features seen. We recorded the continuum-level flux caused by the emission. However, due to the lack of emission lines, we could not estimate the redshift of the galaxy.

For the spectrum of the plausible host galaxy of 20190110C-S (top panel of Figure ~\ref{fig:gemini-spectra}), we found only one obvious spectroscopic emission feature located between 8000 and 8500 \AA. (This feature was also visible in the 2-dimensional spectra). If this feature corresponds to H$_{\alpha}$ then the resulting redshift would be $z_\mathrm{spec} = 0.24$. 

For the spectrum of the plausible host galaxy of 20190417A-S1 (bottom panel of Figure ~\ref{fig:gemini-spectra}), we found late-type galaxy emission features. To estimate the redshift of the galaxy from the spectrum, we measured the observed central wavelengths of the H$_{\mathrm{\alpha}}$, H$_{\mathrm{\beta}}$, [NII], and SII emission lines. We recorded a redshift of $z_\mathrm{spec} =$ 0.12817(2). The spectrum was then corrected to the rest-frame and corrected for Milky Way extinction using the \cite{Cardelli1989} extinction law and an E(B-V)$_{\rm{MW}} = 0.0729$ mag from \citet{Schlegel1998A}. We find no evidence for significant additional internal extinction when calculating the Balmer decrement. The line fluxes of H$_{\mathrm{\alpha}}$, H$_{\mathrm{\beta}}$, [OIII]($\lambda 5006\AA$), and [NII]($\lambda 6582\AA$) were then measured by fitting a Gaussian profile to the rest-frame spectrum. The process was carried out in an automated manner and repeated 100 times to estimate the error in the line fluxes due to uncertainty at the continuum level.

\begin{deluxetable*}{llcccccccccc}
\tabletypesize{\small}
\tablecaption{Radio-to-optical association results \label{tab:radio-opt}}
\tablehead{\colhead{Radio} & \colhead{Radio} & \colhead{RA$_\mathrm{gal}$} & \colhead{DEC$_\mathrm{gal}$} & \colhead{$z$$^{h}$} & \colhead{m$_{r}$} & \colhead{R$_\mathrm{50}$} & \colhead{optical} & \colhead{offset} & \colhead{RO ratio$^{g}$} & \colhead{P$_\mathrm{cc,gal}^{a}$} & \colhead{P$_\mathrm{cc,RG}^{a}$} \\
\colhead{source} & \colhead{catalog} & \colhead{(J2000)} & \colhead{(J2000)} & \colhead{} & \colhead{(mag)} & \colhead{(")}  & \colhead{catalog} & \colhead{(")} & \colhead{} }
\startdata 
20181030A-S1 & VLA & 10:34:14.25 &	$+$73:45:53.9 & 0.00385 & 12.88 & 31.72$^{f}$ & (NGC 3252) & 57.6$\pm$0.1$^{c}$ & $-$ & 0.0015 & 0.0532 \\
20190110C-S & VLASS & 16:37:17.82	& $+$41:26:34.1 & $<$0.22 & 22.32 & 0.66$^{f}$ & SDSS/DESI & 0.24$\pm$0.07 & 2.3 & 0.0252 & 0.6106 \\
20200619A-S & VLASS & 18:10:17.35 &	$+$55:37:15.5 & $<$0.45 & 20.40 & 0.68$^{e}$ & PS1/DESI & 0.44$\pm$0.13 & 1.9 & 0.0054 & 0.1803 \\
20190208A-S & VLA &  18:54:07.11 & $+$46:55:51.9 & $<$0.68 & 20.44 & 0.22$^{e}$ & PS1/DESI & 0.04$\pm$0.03 & 0.5 & 0.0005 & 0.0195 \\
20190117A-S & VLA  & 22:06:36.96 & $+$17:22:25.7 & $<$0.46 & 23.48 & 0.1$^{be}$ & DESI & 0.08$\pm$0.07 & 2.2 & 0.0020 & 0.0705 \\
20190417A-S1 & VLA  & 19:39:05.82	& $+$59:19:36.7 & 0.128 & 21.47$^{d}$ & 0.3$^{de}$ & DESI/Gemini & 0.56$\pm$0.06$^{d}$ & 1.3 & 0.0049 & 0.1659 \\
\enddata
\tablecomments{This table displays exclusively unresolved radio sources linked with either extended/resolved or unresolved galaxies.}
$^{a}$ $P_\mathrm{cc,gal}$ is the chance coincident probability of finding an optical source of the given magnitude and half-light radius while $P_\mathrm{cc,RG}$ is the value corrected for `look-elsewhere-effect' (see \S ~\ref{subsec:pcc-rg} and Table ~\ref{tab:probabilities}). \\
$^{b}$ This source is classified by DESI to be a point source with a low probability of being a star, hence there is no measurement of its half-light radius. Given the nature of the optical source, we adopted 0.1" for the calculation of the chance coincident calculation. \\
$^{c}$ The radio source is at the edge of a closeby NGC galaxy, hence the reason for the large offset.\\
$^{d}$ The optical source information was obtained from Gemini photometry (see \S ~\ref{subsec:Gemini-Photometry} for details).\\
$^{e}$ Unresolved optical counterpart.\\
$^{f}$ Extended optical counterpart. \\
$^{g}$ RO ratio is the logarithm of the radio-to-optical flux ratios according to RO ratio $= log_\mathrm{10}$ (S$_\mathrm{1.4GHz}$/F$_\mathrm{opt}$), where values $<$1.4 implies star formation-related radio emission (see \S ~\ref{subsec:RO-ratio}).\\
$^{h}$ Redshifts limits are the FRB z$_{\rm{max}}$ while the others are z$_{\rm{spec}}$ of the optical counterpart of the radio source. Details on the redshifts of the sources are presented in Appendix Table ~\ref{tab:radio-derived}.
\end{deluxetable*}

\section{Multiwavelength Analysis of PRS Candidates} \label{sec:multi-wavelength-analysis}

\begin{deluxetable}{lccc}
\tabletypesize{\small}
\tablecaption{Summary of the properties of known PRSs \label{tab:known-prs}}
\tablehead{\colhead{PRS property} & \colhead{FRB\,20121102A$^{a}$} & \colhead{FRB\,20190520B$^{a}$} & \colhead{FRB\,20201124A$^{*}$}}
\startdata
     \emph{Size} (pc): & $<$0.7 & $<$9 & $<$700\\
     \emph{z$_{\rm{spec}}$}: &  0.19273  & 0.24 & 0.098\\
     \emph{$\nu$} (GHz): &  1.6 & 1.7 & 15\\
     \emph{L$_{\nu}$} (ergs\,s$^{-1}$\,Hz$^{-1}$): &  2.8$\times$10$^{29}$\, & 3.0$\times$10$^{29}$ & 5.3$\times$10$^{27}$  \\
     \emph{Spectral index, $\alpha$}: &  $-$0.4$\pm$0.5 & $-$0.41$\pm$0.04  & 1.00$\pm$0.43\\
      \emph{offset} : &  $<$12\,mas ($<$40\,pc)  & 20\,mas (80\,pc) & 0.1$\arcsec$\\
       \emph{DM} (pc\,cm$^{-3}$): &  558.0  & 1204.7 & 413 \\
        \emph{RM} (rad\,m$^{-2}$): &  1.46$\times$10$^{5}$  & 2$\times$10$^{5}$  & $-$889.5\\
\enddata
Note:\\
$^{a}$ Details taken from \cite{Chatterjee2017}, \cite{Tendulkar_2017} and \cite{Marcote2017}.\\
$^{b}$ Details taken from \cite{Niu2021} and \cite{Bhandari2023} \\
$^{*}$ This is a candidate PRS and information was taken from \cite{Bruni2023}.
\end{deluxetable}

Here, we present a detailed multi-wavelength analysis of each of the radio sources identified in the CHIME/FRB localization regions. Our primary goal is to assess their potential nature with the aim of identifying promising PRS candidates. Throughout, we will use the properties of the previously published PRSs as shown in Table \ref{tab:known-prs} as a guide, while also focusing on properties that can be used to assess whether the radio sources are simply consistent with expectations for either: (i) star formation in their host galaxies or (ii) AGN.

\subsection{Size of the Radio Source} \label{subsec:size}

The size of a radio source is a key property for deciding its nature. Based on the two confirmed examples to date (FRB\,20121102A and FRB\,20190520B), PRSs are expected to be compact ($<$ 10 pc; \citealt{Marcote2017, Bhandari2023}). Given the frequency of our observations and the typical beam size of the VLA, we, therefore, expect similar PRSs to be unresolved in our data, even for relatively nearby events \footnote{For an example, at a redshift of $z_\mathrm{spec} = 0.1$, beam size of $1.3^{\prime\prime}$---representative for the VLA in A configuration and 1.5 GHz---would place a relatively conservative limit on the size of the radio-emitting source of $<$2.4 kpc}. 

We therefore characterize the extent of each radio source using measurements of their semi-major axis, semi-minor axis, and position angle. We classify sources into one of three categories: (i) unresolved, corresponding to sources where the ratio of the area of the radio source to the beam size is $\lesssim$1.2, (ii) resolved sources, which are radio sources whose size exceeds that of their beam and can be fit by a single 2D Gaussian, or (iii) extended sources, which we identify as sources that can be fit by more than one Gaussian component, including both multi-component and complex sources. We adopt a threshold of $\lesssim$1.2 for our definition of unresolved sources because it can be difficult to define the edge of some sources with respect to that of the beam. The categories assigned to each detected radio source are listed in Appendix~\ref{sec:individual-radio-sources} Table ~\ref{tab:radio-archive}.

Of the 25 radio sources identified, 10 are unresolved by our definition. Moving forward, we only consider these as potential PRS candidates. These candidates include, specifically: 20200619A-S, 20190110C-S, 20181030A-S1, 20190208A-S, 20190117A-S, and 20190417A-S1, which have likely optical host associations, as well as 20180814A-S2, 20180814A-S3, 20181119A-S, and 20191114A-S, which lack a coincident optical association.

\subsection{Redshift Cut-off and Luminosity} \label{subsec:redshift-cut}

In order for a radio source to be a viable PRS candidate, it must not be located at a redshift higher than the maximum allowed for the FRB based on its DM. We take values of DM and z$_{\mathrm{max}}$ already published by \cite{CHIME2023} and \cite{Michilli2022}. For cases where they are not available, we calculate z$_{\mathrm{max}}$ using the \texttt{FRUITBAT} software \citep{Batten2019} by subtracting the DM contribution expected for the disk of the Milky Way from the NE2001 model \citep{ne2001}, DM$_\mathrm{MW}$(NE2001), but do not attempt to correct for contributions from either the MW halo (DM$_\mathrm{halo}$) or the FRB host galaxy (DM$_\mathrm{host}$) to be conservative. In Appendix~\ref{sec:individual-radio-sources} Table ~\ref{tab:radio-derived}, we list z$_{\mathrm{max}}$ for all FRBs in our sample.

Of the 10 unresolved radio sources described above, we have spectroscopic redshifts for the underlying galaxies for three of them. Of these, the probable hosts of 20181030A-S1 and 20190417A-S1 have lower z$_{\mathrm{spec}}$ than the z$_{\mathrm{max}}$ for their respective FRBs, and thus remain viable. For 20190110C-S, the tentative redshift of z$_{\mathrm{spec}}=0.24$ (which was based on a single emission line) is slightly higher than the z$_{\mathrm{max}}=0.22$ of the FRB. However, because there is uncertainty in this redshift estimate, we chose to still consider this source in the analysis in the following sections. Of the remaining unresolved radio sources, 4 have no detected optical counterparts, and 3 others have no redshift information. As a result, we do not rule out any of the 10 unresolved sources as PRS candidates due to their redshifts.

In addition, for each radio source detected within the vicinity of each FRB, we utilize the z$_{\mathrm{spec}}$ of its host galaxy to compute the luminosity and other relevant properties of the radio source. In instances where a spectroscopic redshift is unavailable, we employ the z${_\mathrm{max}}$ of the FRB, estimated from the DM, to derive an upper limit for the luminosity of the radio source. The observed spectral luminosities are listed in Appendix~\ref{sec:individual-radio-sources} Table ~\ref{tab:radio-derived}.

\subsection{Variability and spectral shape } \label{subsec:variability}

The current limited sample size makes it difficult to determine the time-dependent properties and spectral shapes of PRSs. While the two known PRSs exhibit short-term variability ($<$1 year; \citealt{Chatterjee2017, Niu2021}), they also appear to vary at longer timescales \citep[e.g. $>$1 year for FRB\,20121102A;][]{Rhodes2023}. However, the observed time variability seen for the PRS associated with FRB\,20121102A \citep{Chatterjee2017, Marcote2017} may be attributed to refractive scintillation rather than intrinsic source variability \citep{Chen2023}. The two known PRSs exhibit non-thermal spectral shapes, consistent with synchrotron emission, with FRB\,20121102A displaying a  broken power-law spectral shape.  However, we note that the PRS candidate FRB\,20201124A shows a positive spectral index. Due to the significant uncertainty surrounding their variability and spectral characteristics, we refrain from using these traits to define PRSs. Instead, we present our observations to contribute to a better understanding of these sources as the sample size grows in the future.  In this subsection, we investigate the variability and spectral indices of the identified radio sources. 

For radio sources (among the 10 unresolved sources) with more than 2 detections at the same frequency but different times, we generate a light curve and calculate the variability of the source, $V_s$, using the $t-$statistical methods outlined by \cite{Mooley2016} as $V_s = |\Delta S/\sigma_{S}| \geq 5$ where $\Delta S = S_{2} - S_{1}$ and $\sigma_{S}$ is the standard deviation between the two fluxes used. We adopt $V_s$ values $>$ 5 (at 5-$\sigma$) as variable sources. Due to limited data availability, we are unable to make definitive statements regarding variability for most of the radio sources. Using VLASS epochs 1 and 2 for two radio sources, namely 20200619A-S and 20190110C-S, we record $V_s$ values of 0.35 and 0.36 respectively, indicating the absence of detected variability within a 2-year timescale. However, we recorded $V_s = 16$ within 10 days of VLA observations at 1.5 GHz for 20181030A-S1, which could be due to refractive scintillation.
This source is particularly interesting given its short-scale variability behavior which is similar to that seen for the PRS of FRB\,20121102 and FRB\,20190520B at similar frequencies.

\begin{figure}[t]
\centering
\includegraphics[width=.45\textwidth]{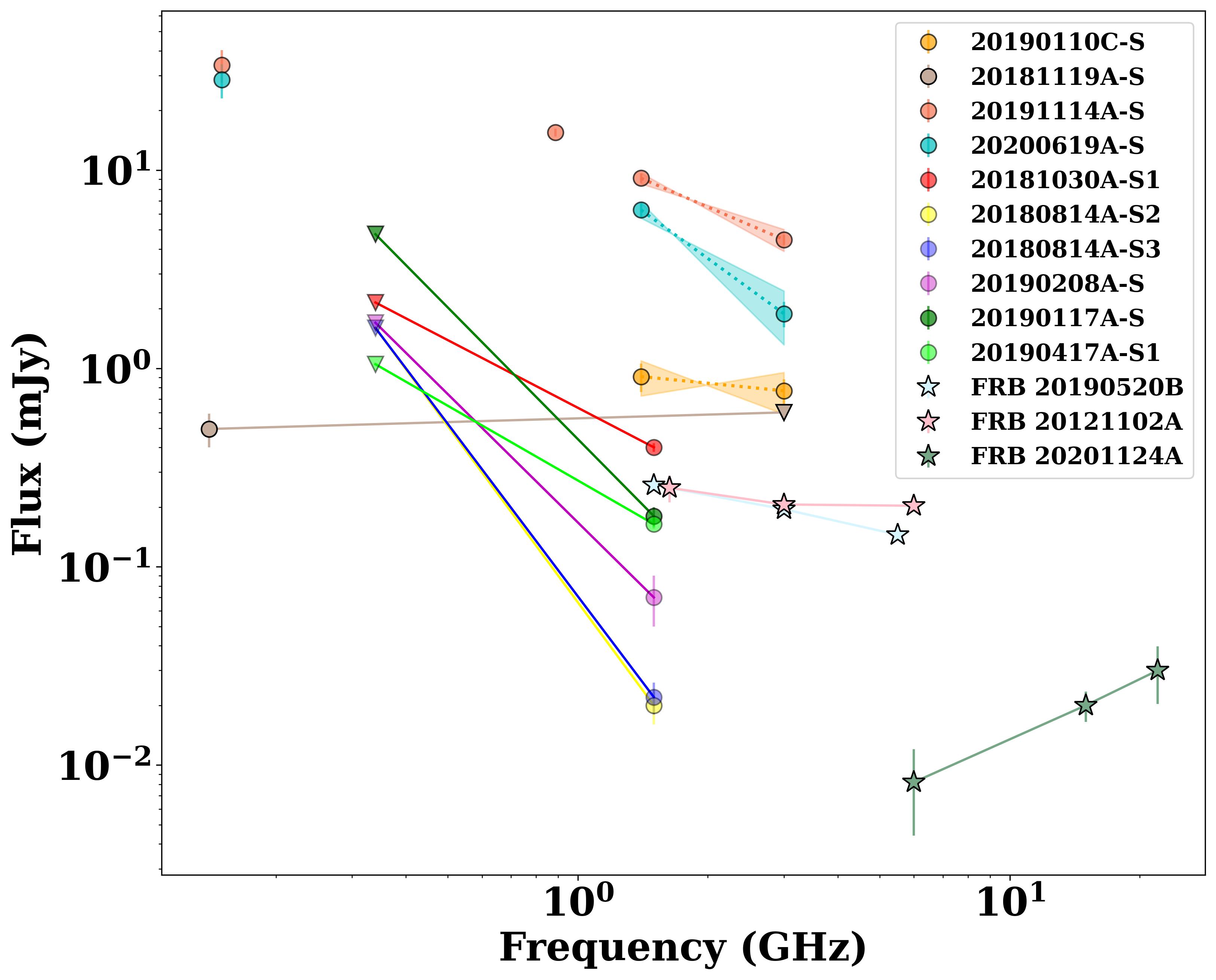}
\caption{Spectral Energy Distribution (SED) of the unresolved radio sources shown as colored circles while the colored stars are known PRSs (FRB\,20121102A, FRB\,20190520B) and a potential third PRS. In general, most of the sources have non-positive spectral shapes similar to that seen for FRB\,20121102A and FRB\,20190520B. 20181119A-S (brown) is the only source with a positive spectra shape among the radio sources but similar to that of the potential PRS of FRB\,20201124A. Solid lines connect contemporaneous flux periods.} 
\label{fig:sed}
\end{figure}

For the 10 unresolved radio sources, we estimate their spectral indices via multiple methods: (i) for the six sources identified in targeted VLA imaging, we use the first and second Taylor terms of the Stokes's images, $I_{0}$ and $I_{1}$ alongside their image residuals to estimate the in-band spectral indices and their uncertainties according to $\alpha$ $=$ $I_{1}/I_{0}$, (ii) for the six sources identified in targeted VLA imaging we also use upper limits from contemporaneous VLITE observations to places limits on $\alpha$, and (iii) for the four sources identified in archival surveys, we use detections in multiple surveys/frequencies to estimate the spectral indices. For methods (i) and (ii) the results are consistent with errors, except for one source (20190417A-S2). For method (iii) we caution that these observations were taken at different times (in some cases more than 10 years apart), and thus should be treated with caution.
Results are listed in Appendix~\ref{sec:individual-radio-sources} Table~\ref{tab:radio-derived} and shown in Figure~\ref{fig:sed}, where contemporaneous observations are connected by solid lines.  All these radio sources seem to exhibit a non-thermal spectral shape except for 20181119A-S which has a positive spectral shape similar to the spectral shape of the low luminosity PRS of FRB\,20201124A ($\alpha \sim 0.9$) \citep{Bruni2023}. The non-thermal shape of the other radio sources is consistent with the spectral indices recorded for FRB\,20121102A ($\alpha = -0.27\pm0.24$ \citep{Marcote2017}) and FRB190820B ($\alpha = -0.41\pm0.04$ \citep{Niu2021}).

\subsection{Offset between Radio source and Optical sources} \label{subsec:offset}

For the 6 radio sources that are both unresolved and have nearby optical associations, we calculate the offset between the radio and optical sources. These offsets will be used both to comment on the possibility that the radio source(s) may be due to AGN in their host galaxies and to calculate the probability of chance association between the radio and optical sources in \S ~\ref{subsec:pcc-rg} below.

In Table ~\ref{tab:radio-opt}, we list the angular offsets with their associated uncertainties for all 6 radio sources. Also listed are the calculated half-light radii for each of the optical sources. In four cases, the optical sources are unresolved, while in two cases, they are resolved. In the latter case, host-galaxy normalized offsets can also be estimated. For the two sources with spectroscopic redshifts---20181030A-S1 and 20190417A-S1---the angular offsets correspond to physical offsets of 5.59$\pm$0.01 kpc and 1.68$\pm$0.18 kpc, respectively.

For one of the 6 radio sources (20181030A-S1), the high offset rules out an AGN in the underlying optical galaxy (although the possibility that it is a background source, unassociated with the optical galaxy, will be discussed below). The remaining 5 sources all have offsets from the underlying optical sources of $\lesssim$$0.5\prime\prime$. Given the uncertainties in these offsets, as well as the potential for some systematic offsets between the astrometry in radio and optical images, these sources may be consistent with the nuclei of their hosts.  While AGNs are expected in the centers of galaxies, we note that it is still possible for a PRS to be found near the center of its host, similar to what was seen for the PRS of FRB\,20121102A\footnote{The host galaxy of FRB\,20121102A is a dwarf galaxy. As a result, the nucleus of the galaxy may not be well defined to clearly determine an accurate offset.} (offset $=$ $0.2\prime\prime$; \citealt{Tendulkar_2017}). We therefore do not explicitly eliminate any PRS candidates based on this metric, but will consider this along with other properties below.

\subsection{Chance Probability of Association between the radio source and its plausible host galaxy, P$_\mathrm{cc,gal}$} \label{subsec:pcc-rg}

As described above, 6 of the 10 unresolved radio sources identified in our search physically overlap with an optical source.  In instances where we fail to identify optical counterparts, this may be attributed to limitations in the magnitude depth of the searched public archival catalogs, potentially rendering them unable to detect high-redshift sources, or it may be due to the absence of an optical counterpart for the radio sources themselves. However, even for cases where we identify potential optical counterparts, it is necessary to assess the chance coincident probability, P$_\mathrm{cc,gal}$ of associating the radio source to its optical counterpart that is a galaxy of a given brightness, $m_r$ (see Table ~\ref{tab:probabilities} for a quick description).

We assumed a Poisson distribution of galaxies across the sky and calculated the probability of a chance coincidence occurring within a radius, $R$ using P$_\mathrm{cc,gal}$ = 1 $-$ $exp\,(-\pi R^{2}\sigma(\leq m_r))$. The r-band galaxy number count reported by \cite{Driver_2016} was used to calculate the projected areal number density of galaxies brighter than the host galaxy r-band magnitude, $\sigma$($\leq$ m$_r$).  To estimate $R$, we applied the prescription described by 
 \cite{Bloom2002} given as $R =$ max[2$R_\mathrm{radio}$,$\sqrt{R_\mathrm{0}^{2} + 4R_\mathrm{hlg}^{2}}$], where $R_\mathrm{radio}$ is the error radius of the radio source, $R_\mathrm{0}$ is the offset between the radio source and its host galaxy, and $R_\mathrm{hlg}$ is the half-light radius of the host galaxy. The result of this calculation gives P$_\mathrm{cc,gal}$ and is reported for each radio source with associated optical counterparts in Table ~\ref{tab:radio-opt}. 
 
While this formalism for the probability of chance alignment is common in the field of transients, we also computed adjusted chance coincidences using the Bonferroni correction---also known as the 'look-elsewhere' effect. This accounts for the fact that we simultaneously searched 37 different CHIME/FRB localization regions for radio sources, and thus may expect to find even relatively low probability alignments with some frequency. This correction is represented by the equation $P_\mathrm{cc,RG} = 1 - (1 - P_\mathrm{cc,gal})^{n}$, where $n$ is the number of fields examined ($n=37$ in our case). These adjusted values are also listed in Table~\ref{tab:radio-opt}.

Examining Table~\ref{tab:radio-opt}, we see that 5 of the sources have P$_\mathrm{cc,gal}$ $\lesssim$ 0.005, indicating a strong likelihood of association with their optical counterparts. The remaining source has P$_\mathrm{cc,gal}$$\sim$ 0.025. While the look-elsewhere effect naturally raises the possibility that we would identify an individual source with P$_\mathrm{cc,gal}$ $<$ 0.01 (which corresponds to 1\% probability) by chance (see the final column in Table~\ref{tab:radio-opt}), it is unlikely for all six sources to be so. We therefore proceed with analysis in the sections below that assume the optical and radio sources are associated.

\subsection{Radio-to-Optical ratio } \label{subsec:RO-ratio}
Given that some of our sources are close to the centers of their likely host galaxies based on their projected offsets we need an additional diagnostic method to check whether the radio source is consistent with expectations for either star formation or AGN emission. Fortunately, the radio-to-optical flux ratio (RO ratio) of the source can provide just such a diagnostic metric. 

We calculate the ratio as RO ratio $=$ Log$_\mathrm{10}(S_\mathrm{1.4GHz}$/F$_\mathrm{opt})$ where S$_\mathrm{1.4GHz}$ is the flux density in Jansky of the radio source at 1.4\,GHz and F$_\mathrm{opt}$ is the optical flux density in Jansky obtained from the AB r-band magnitude of the galaxy according to the equation F$_\mathrm{opt}$ $=$ 3631$\times$10$^{-0.4m_{r}}$ \citep[e.g.][]{Machalski1999, Afonso2005, Seymour2008}. Using this formalism, \cite{Eftekhari2021} adopt a rough threshold where RO ratio $<$ 1.4 is consistent with star formation and RO ratio $>$ 1.4 is consistent with AGN activities or other radio emissions such as pulsar wind nebula, hypernebula, SN remnants, GRB afterglows, etc.

Using this same threshold, among the 5 unresolved radio sources with probable host galaxy associations and offset $< 1\prime\prime$, we find that 20190208A-S (RO ratio $\sim$0.5) is likely due to star formation. When the RO ratios of the remaining 4 radio sources---20190110C-S ($\sim$2.3), 20200619A-S ($\sim$1.9), 20190117A-S ($\sim$2.2), and 20190417A-S1 ($\sim$1.3)---are compared with that of the PRS of FRB\,20121102A (RO ratio $=$ 2.9 using 250 $\mu$Jy at 1.63\,GHz) and FRB\,20190520B (RO ratio $=$ 1.7 using 258 $\mu$Jy at 1.5\,GHz), their RO ratios are consistent with either AGN or PRSs.

We note that we do not calculate an RO ratio for the 6th source that has a large offset but overlaps the spiral arm of a nearby galaxy (20181030A-S) because using the PanSTARRS template images we do not identify a specific optical counterpart at the location of the radio source.

\subsection{Infrared WISE Diagnostic analysis of the radio source} \label{subsec:WISE}
In order to further assess if there is any evidence for either AGN activity or obscured star formation in the galaxies associated with our PRS candidates, we examine their infrared colors from the Wide-Field Infrared Survey Explorer (WISE; \citealt{Wright2010}). In the case where there is a WISE detection, we take the WISE band values from the archives and over-plot the color ratios on the WISE diagnostic plot of \cite{Wright2010}. The WISE diagnostic plot allows us to classify the infrared emission from our radio sources into different types of emission. From the plot, the parameter space overlapping with starburst, low-ionization nuclear emission-line regions (LINERs) galaxies, ultra/luminous infrared galaxies (LIRGs/ULIRGs), Quasi-stellar objects (QSOs), Seyferts, and obscured AGN are regarded as dusty regions \citep[e.g.][]{Reddy2006,Wright2010,2012WISE,Reddy2012,Jarett2017WISE}.

Comparing the ratios, we conclude that one of our radio sources (20200619A-S) along with the PRS associated with FRB\,20190520B has inferred colors consistent with star-forming galaxies without much dust contribution. 20190110C-S is consistent with an AGN, Seyfert, or dusty galaxies while 20190117A-S and the PRS of FRB\,20201124A are consistent with a dusty star-forming environment. There is no detected WISE sources at the locations of 20190208A-S, 20181030A-S1, or 20190417A-S1 nor the PRS associated with FRB\,20121102A.

\begin{figure}[t]
\centering  
\includegraphics[width=.45\textwidth]{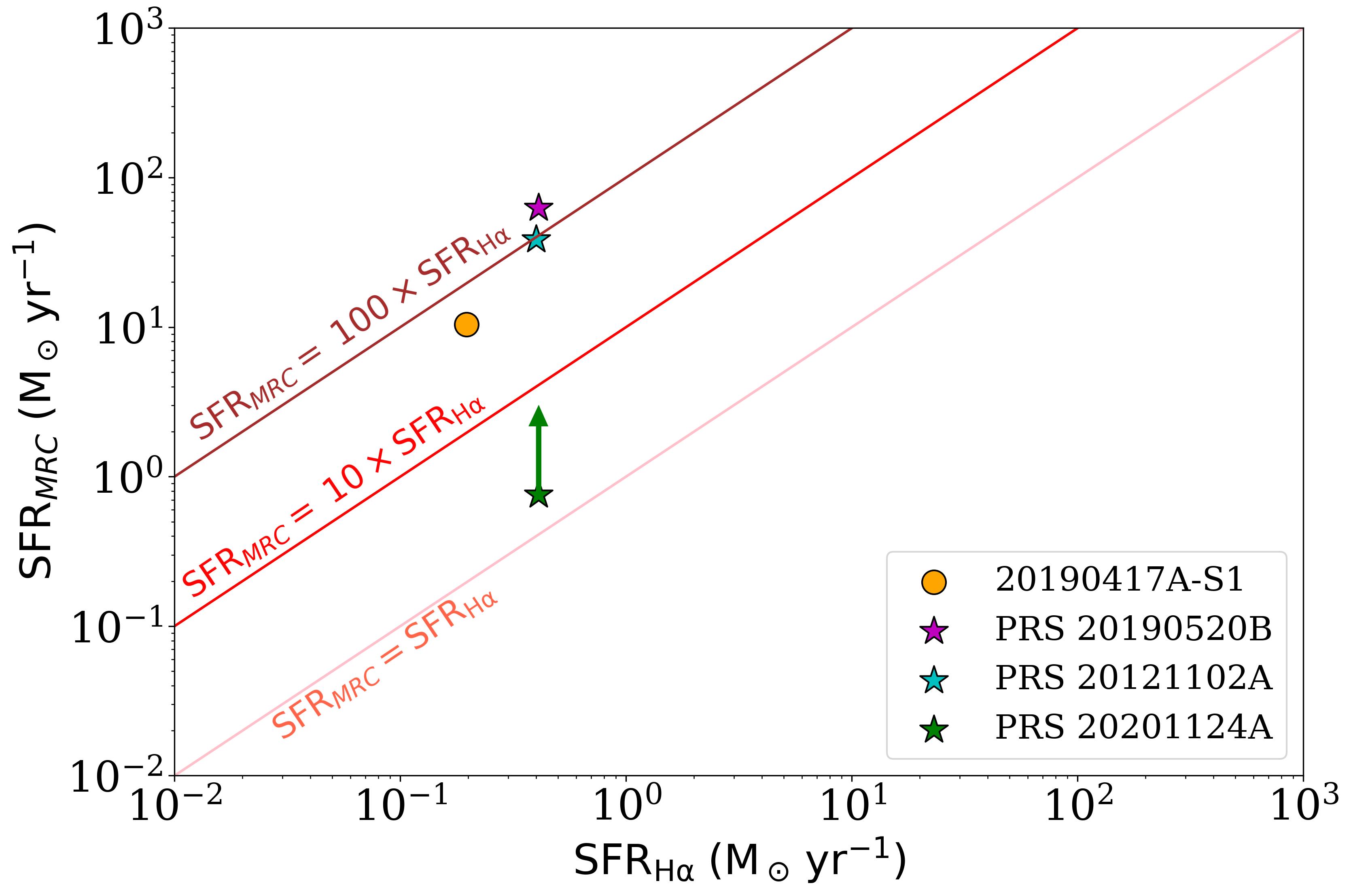} 
\caption{ A plot of mid-radio continuum star formation rate, SFR$_\mathrm{MRC}$ versus $H_\mathrm{\alpha}$ estimated star formation rate in the context of that for FRB\,20121102A (cyan), FRB\,20190520B (magenta) and FRB\,20201124A (dark green) with the SFR$_\mathrm{MRC}$ of the first two estimated from their 1.5 GHz PRS luminosity. The SFR$_\mathrm{MRC}$ plotted for FRB\,20201124A is from the 6GHz flux density and represents the lowest possible SFR$_\mathrm{MRC}$ as indicated by the gray upward arrow. The location of the orange circle (20190417A-S1) indicates that there is an additional cause of radio emission responsible for the observed emission. } 
\label{fig:sfr-opt-rad}
\end{figure}

\subsection{Star Formation Rates} \label{subsec:SFR}

It is possible to further assess whether the detected radio sources are consistent with, or in excess of, expectations for star formation from their host galaxies by comparing the star formation rate inferred from the radio flux to that from other metrics, such as H$_\mathrm{\alpha}$ luminosity. Out of the 6 unresolved radio sources with host associations, only one (20190417A-S1) has  H$_\mathrm{\alpha}$ flux measurement required for this analysis. In particular, we note that while we obtained spectra for both 20190117A-S and 20190110C-S, no emission lines were present in the former and only one line was visible in the latter (leading to uncertainty in its identification). In addition, we have the global H$_\mathrm{\alpha}$ flux available for the galaxy that overlaps with 20181030A-S1 (located in one of the spiral arms) but we do not have a measurement of H$_\mathrm{\alpha}$ flux at the location of the radio source, as would be necessary to assess if it is consistent with an unresolved knot of star formation. While there is a tentative H$_\mathrm{\alpha}$ detection for 20190110C-S, we do not consider it for this analysis because of the uncertainty in z$_{\mathrm{spec}}$.

To complete this assessment, we use the relationship of \cite{Murphy2011} to convert H$_\mathrm{\alpha}$ luminosity to a star formation rate (SFR$_\mathrm{H_\alpha}$) and the equations from \cite{tsk+17} to assess implications for the radio star formation rate. The equations from \cite{tsk+17} include contributions to the radio flux from both thermal and non-thermal sources, and we use them to compute a mid-continuum radio-inferred star formation rate (SFR$_\mathrm{MRC}$). For these calculations, we assume an electron temperature of $10^{4}$ K, a non-thermal spectral index of $\alpha = -0.8$, and a ratio of thermal to total star formation of 0.1 at 1.5 GHz (see Table 6 of \citealt{tsk+17}). We use a Monte Carlo method to propagate the uncertainty in the radio flux density to SFR$_\mathrm{MRC}$. We simulate 1000 flux density values, drawing from a Gaussian probability distribution with the same 1-sigma uncertainty as for the radio source. We then use these to calculate 1000 SFR$_\mathrm{MRC}$ values and record their mean and standard deviation.

\begin{figure*}[t]
\centering
\includegraphics[width=.9\textwidth]{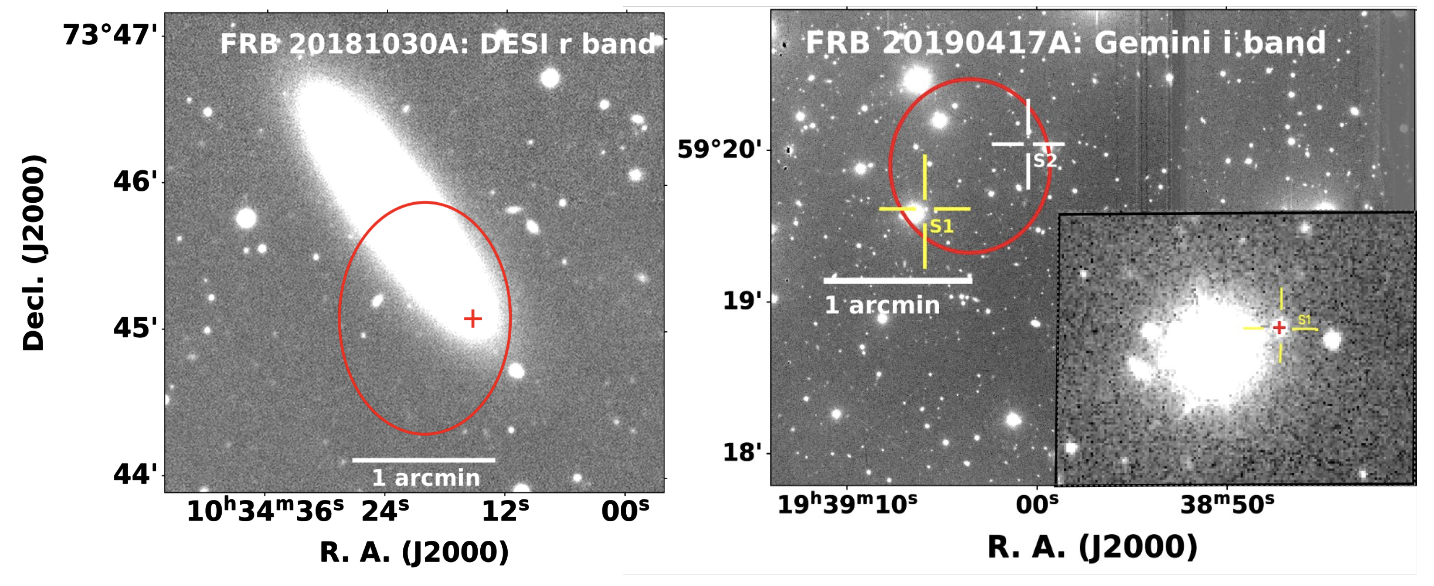}
\caption{Optical images showing associations of the two potential radio sources from our sample. \emph{Left:} DESI r-band field of FRB\,20181030A showing the host galaxy of the FRB and the radio source---20181030A-S1.\emph{Right:} Gemini North deep image of the field of FRB\,20190417A. The yellow crosshair shows the location of the likely host of the radio source 20190417A-S1, which is the faint optical source near a bright star. The red `+' sign indicates the location of the radio source for both fields. } \label{fig:optical-images}
\end{figure*}

Based on this analysis, the observed radio flux (SFR$_\mathrm{MRC} = 10.41\pm0.53$ M$_\mathrm{\odot}\,yr^{-1}$) of 20190417A-S1 is significantly higher than what would be expected based on the optical star formation rate (SFR$_\mathrm{H_\alpha} = 0.20$ M$_\mathrm{\odot}\,yr^{-1}$).  This is evident in Figure ~\ref{fig:sfr-opt-rad} where we plot the SFR$_\mathrm{H_\alpha}$ versus SFR$_\mathrm{MRC}$ showing that either a significant fraction of the star formation in the host of 20190417A-S1 is obscured, or that another emission source is responsible for the radio emission. Unfortunately, no WISE color information is available for the host of 20190417A-S1. Also shown are the two previously reported PRSs and the potential PRS of FRB\,20201124A. As can be seen, 20190417A-S1 (orange) falls in a similar location to both FRB\,20121102A (cyan) and FRB\,20190520B (magenta). 

The SFR$_\mathrm{H_\alpha} = 0.033$ M$_\mathrm{\odot}\,yr^{-1}$ that we have for 20181030A-S1 is not from the local neighborhood of the potential PRS, hence we could not use it for this analysis. However, we calculate SFR$_\mathrm{MRC} = 0.02$ M$_\mathrm{\odot}\,yr^{-1}$ for 20181030A-S1 and compare it with typical values of SFR$_\mathrm{H_\alpha}$ $\sim$ 10$^{-4}$ M$_\mathrm{\odot}\,yr^{-1}$ for individual HII regions from \cite{Crowther2013}.  SFR$_\mathrm{MRC}$ is over 100 times greater than SFR$_\mathrm{H_\alpha}$ values for most HII regions indicating that either there is an extreme star-forming region in this exact location of NGC-3252, or that another emission source is responsible for the radio emission. The fact that  SFR$_\mathrm{MRC}$ is greater than SFR$_\mathrm{H_\alpha}$ for each of these sources suggests that there is additional radio emission beyond what can be attributed to unobscured star formation (which would be visible in H$_{\alpha}$). This excess radio emission could, in principle, be due to multiple source including (i) a PRS, (ii) an AGN, or (iii) obscured star formation. In section \ref{subsec:summary-prs}, we will consider this information with other metrics described above (e.g., WISE IR diagnostics, optical-to-radio radio) to assess the most promising PRS candidates within our sample.

\section{Summary of PRS search and candidates}\label{sec:discussion}  

Out of a sample of 37 CHIME/FRB fields searched, we identified a total of 25 radio sources: 13 archival sources (including an NVSS source near FRB\,20181030A that has been reported by \citealt{BhardwajR42021}), and 12 radio sources from the deep VLA observations. Appendix~\ref{sec:individual-radio-sources}, Table ~\ref{tab:radio-archive} and ~\ref{tab:radio-derived} show the properties of all the radio sources. Out of 25 radio sources, there are 10 unresolved sources while the rest are resolved/extended sources.

In \S ~\ref{subsec:summary-prs} below, we summarize the result of the multi-wavelength analysis for the 10 unresolved radio sources and conclude which are the most promising PRS candidates. We then place deep limits on the presence of a PRS for all the FRBs observed with VLA in \S ~\ref{subsec:deep-limits} and discuss the global implication for the radio search result in \S ~\ref{subsec:global-implication}.

\subsection{Summary of Potential PRS candidates} \label{subsec:summary-prs}

Our ability to characterize each of the 10 unresolved radio sources identified varies based on the quantity of multi-wavelength data available, in particular likely optical host associations and redshifts. Here we discuss individual radio sources, progressing with those from most to least information about their putative hosts. 

Two of the unresolved radio sources have positions that overlap with optical galaxies for which we have secure redshifts that are lower than z$_{max}$ for the FRBs. 20181030A-S1 overlaps with a spiral arm of the nearby NGC\,3252 (which was identified by \cite{BhardwajR42021} as the most probable host for this FRB), while 20190417A-S1 is offset $\sim$0.5$\arcsec$ from an unresolved optical galaxy at z$_{\rm{spec}} = 0.128$. Both sources show similarities to confirmed PRSs associated with FRB\,20121102A and FRB\,20190520B. Both have non-thermal spectral indices (\S~\ref{subsec:variability}) and, if associated with their underlying galaxies, have luminosities of 3.1$\times$10$^{35}$ and 1.1$\times$10$^{38}$ erg s$^{-1}$, respectively. The latter is similar to previously identified PRSs, while the former is multiple orders of magnitude less luminous. 

In terms of other possible origins, both 20181030A-S1 and 20190417A-S1 have radio luminosities in excess of expectations for star formation within their hosts (local in the case of 20181030A-S1; \S~\ref{subsec:SFR}). In both cases, we also disfavor that they are AGN in the underlying optical galaxy. For 20181030A-S1, this is due to its significant offset from the nucleus of NGC 3252. For 20190417A-S1, the optical spectrum obtained shows narrow emission lines with ratios indicative of a star-forming galaxy on a BPT diagram (see Appendix \S~\ref{subsec:R18}). Its position is also offset from the galaxy center at $>$3$\sigma$, although we caution that this does not consider possible astrometric offsets between optical and radio images. While the radio sources could be background AGN, unassociated with the identified optical galaxies, we found a probability of chance alignment of $<0.05\%$ for both (\S~\ref{subsec:pcc-rg}). In the case of 20181030A-S1, this low probability is driven primarily by the rarity of m$_r$ $\sim$ 13 mag galaxies, although we note that background AGN have been identified in other nearby galaxies \citep[e.g.][]{Massey2019}. While we do not identify a specific counterpart to the radio sources in archival optical, UV, or X-ray catalogs, we cannot formally rule out this possibility. However, despite this, we consider both 20181030A-S1 and 20190417A-S1 promising PRS candidates. 

Four additional unresolved radio sources identified in our search have associated optical counterparts, all of which also
have spectral indices consistent with non-thermal emission. For two of these sources, we obtained optical spectroscopy of the underlying optical sources (\S~\ref{sec:Optical-Spectroscopy}). The spectrum of the optical source underlying 20190110C-S showed a single emission line, which if interpreted as H$_{\alpha}$ places it above z$_{\rm{max}}$ for the FRB (\S~\ref{subsec:redshift-cut}). While this is tentative, most other reasonable interpretations for the line would place the source at even higher redshifts. In addition, the colors of the optical source are relatively red ($g-r$ color of $\sim$4) and the WISE diagnostic colors overlap with AGN/Seyfert galaxies (\S~\ref{subsec:WISE}). We therefore consider it unlikely to be a PRS. The radio source 20190117A-S has a position consistent with the centroid of its (unresolved) optical counterpart. However, an optical spectrum does not reveal any prominent emission lines indicative of AGN activity, and WISE infrared colors are more consistent with dusty galaxies. Thus, while the RO ratio is in excess of typical expectations for star formation (\S~\ref{subsec:RO-ratio}), future analysis examining the possibility of obscured star formation would be necessary to assess the viability of this source as a PRS.

For the final two sources with optical associations, we have not yet obtained any optical spectroscopy. 20190208A-S has a position consistent with the centroid of its (unresolved) optical counterpart. In addition, its RO ratio is consistent with expectations for star formation (\S~\ref{subsec:RO-ratio}). In contrast, 20200619A-S has a RO ratio in excess of expectations for star formation, but its WISE infrared colors are consistent with star-forming galaxies \emph{without} significant AGN activity (\S~\ref{subsec:WISE}). In addition, its position is nominally offset from the centroid of its (unresolved) optical counterpart at more than 3$\sigma$, but still as a probability of chance alignment of only $\sim$0.5\%. This source is therefore promising, although we consider it more cautiously than 20181030A-S1 and 20190417A-S1 as a PRS candidate until optical spectroscopy can be obtained.

Finally, there are 4 radio sources with no known host optical host associations. Two have relatively high z$_{\rm{max}}$ values (z$_{\rm{max}} = $ 0.43 and 0.52 for 20191119A-S and 20191114A-S, respectively), and it is possible that their host galaxies are simply below the detection threshold of the optical surveys searched. At these redshifts, with a survey depth of m$\sim$24 mag, we would only be sensitive to galaxies brighter than M$\lesssim-18$ mag. The final two sources are located in the field of FRB\,20180814A. This FRB has a relatively low value of z$_{\rm{max}} = 0.091$ and has a most probable host galaxy PanSTARRS-DR1 J042256.01+733940.7, as assessed by \cite{Michilli2022}. These two radio sources are offset by 22$\arcsec$ and 80$\arcsec$ from this galaxy (which has a visible diameter in PanSTARRS of $\lesssim$10$\arcsec$). It is therefore probable that they are not associated with the FRB. 

In conclusion: 20181030A-S1 and 20190417A-S1 are the most promising PRS candidates identified in this search to date, although we cannot fully rule out the possibility that they are (background) AGN. Additional follow-up to further explore the potential nature of these targets is ongoing. Figure ~\ref{fig:optical-images} shows the DESI and Gemini images for 20181030A-S1 and 20190417A-S1 fields respectively with the red cross indicating the position of the radio source. In addition, While some radio sources identified (e.g. 20190110C-S, 20180814A-S2, 20180814A-S3) are unlikely to be PRSs, others (e.g. 20200619A-S) warrant further investigation to assess their viability.

 \begin{deluxetable}{lcccc}
\centering
\tabletypesize{\small}
\tablecaption{Deep limits on PRS \label{tab:deep-limit}} 
\tablehead{\colhead{FRB} & \colhead{Upper limit F$_{\nu}$} & \colhead{$\sigma$-level$^{d}$} & \colhead{Upper limit L$_{\nu}^{a}$} & \colhead{z$_{\mathrm{spec}}$$^{e}$}\\
\colhead{} & \colhead{($\mu$Jy)}  & \colhead{} & \colhead{(erg\,s$^{-1}$\,Hz$^{-1}$)} &  \colhead{Ref.}}  
\startdata
FRB\,20180814A$^{c}$ & 17.5 & 5$\sigma$ & $<$2.1$\times$10$^{27}$ & 1 \\
FRB\,20180916B$^{b}$ & 18.0 & 3$\sigma$ & $<$4.8$\times$10$^{26}$ & 2 \\
FRB\,20181030A$^{c}$ & 50.0 & 5$\sigma$ & $<$2.4$\times 10^{25}$ & 3 \\
FRB\,20190117A & 80.0 & 5$\sigma$ & $-$ & $-$ \\
FRB\,20190208A & 25.0 & 5$\sigma$ & $-$ & $-$ \\
FRB\,20190303A$^{c}$ & 25.0 & 5$\sigma$ & $<$2.6$\times$10$^{27}$ & 1 \\
FRB\,20190417A & 42.5 & 5$\sigma$ & $-$ & $-$ \\
FRB\,20200120E$^{b}$ & 27.0 & 3$\sigma$ & $<$4.2$\times$10$^{23}$ & 4 
\enddata
Note \\
$^{a}$ The upper limit luminosities are estimated using the FRB redshifts where the FRB host is known and its z$_{\mathrm{spec}}$ is available. All measurements are at 1.5 GHz.\\
$^{b}$ These FRBs have sub-arcsecond localizations and the PRS limits here are comparable to previously published values in the literature.
\\
$^{c}$ These FRBs have probable host galaxies identified using their CHIME/FRB baseband localizations.\\
$^{d}$ We measured 5$\sigma$ level flux density limit for FRBs with baseband localization regions and 3$\sigma$ for the ones with subarcseconds localizations.\\
$^{e}$ References for the z$_{\mathrm{spec}}$ of the published host galaxies of the FRBs. \\
1: \citealt{Michilli2022}, 2: \citealt{Marcote2020}, 3: \citealt{BhardwajR42021}, 4: \citealt{Bhardwaj_2021} and \citealt{kirsten2021}
\end{deluxetable}

\begin{figure*}
\centering
\includegraphics[width=.9\textwidth]{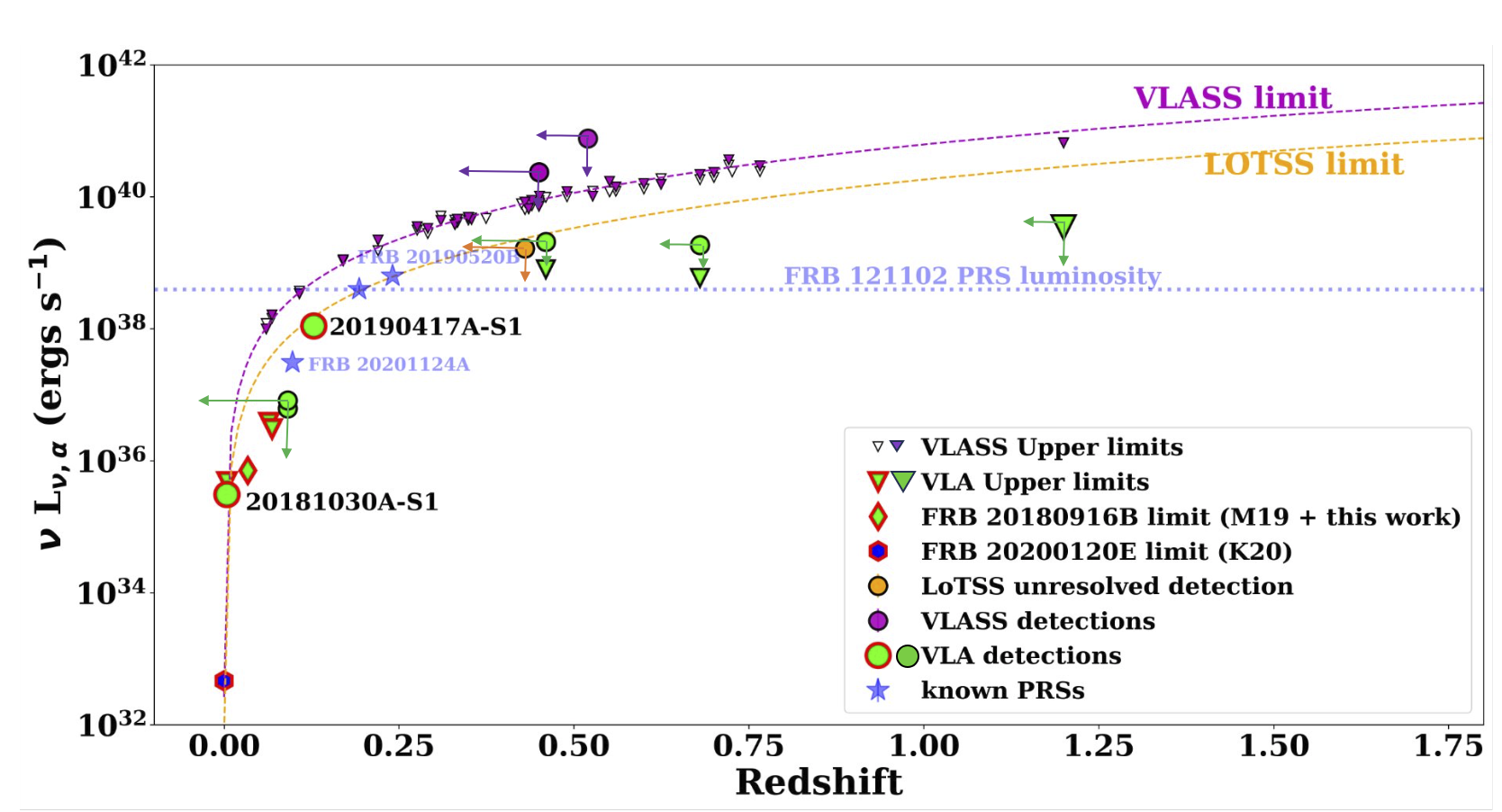}
\caption{Luminosity versus redshift plot showing all unresolved radio sources as circle markers from this work. The detections and deep limits obtained from VLA observations are plotted as light green inverted triangles and colored circles. The markers with red-colored marker edges are estimated with $z_{\rm{spec}}$ while others are estimated using $z_{\rm{max}}$. The luminosities shown in circles for these detections except the ones with red marker edges are considered upper limits since they were estimated using $z_{\rm{max}}$. In addition to the PRS of FRB\,20121102A, 20190520B, and a potential PRS of FRB\,20201124A (all light blue stars), we also show limits for FRB\,20180916B (diamonds), and FRB\,20200120E (hexagons). M19 and K20 represent \citet{Marcote_2019} and \citet{Kirsten_2020}, respectively. The blue dotted line is the luminosity of the PRS of FRB\,20121102A. Upper limits on PRSs from VLASS (small purple and white inverted triangles for first and second epochs) are also plotted and 2 VLASS detections are shown as purple circles. The purple and orange dashed lines are the VLASS and LoTSS 5$\sigma$ sensitivity limit as a function of redshift.  The two potential PRSs from this work are shown as bigger circles with black-lettered labels: 20181030A-1 and 20190417A-1. 3-$\sigma$ and 5-$\sigma$ luminosity upper limits on the presence of a PRS are shown as inverted light green triangles for FRBs fields observed with VLA for FRBs having subarcsecond and baseband localizations respectively. } 
\label{fig:vlasslim}
\end{figure*}

\subsection{Deep limits on the presence of PRS}\label{subsec:deep-limits}

While we identify a number of promising PRS candidates, future follow-up will be needed to confirm any association with the FRB. In Table ~\ref{tab:deep-limit} we therefore also quote deep limits on the flux density of any PRS, which are valid in the situation that none of the radio sources identified are linked to the FRB. We quote limits only for the fields targeted by our VLA observations described in Section~\ref{subsec:realfast-obs} as these are deeper than archival surveys such as VLASS. 

When computing the flux density upper limits, we take 5 times the measured RMS level for the six FRBs with baseband localization regions and 3 times the measured RMS level for the two FRBs with known subarcsecond localizations.  Five of the eight FRBs with targeted VLA imaging also have known or most probable host galaxies presented in the literature \citep{Michilli2022,Marcote2020,BhardwajR42021,Bhardwaj_2021,kirsten2021}. For these events, we also quote an upper limit on the spectral radio luminosity using z$_{\mathrm{spec}}$ for these galaxies. We refrain from quoting a spectral luminosity limit for the other events as they will not be significant. 

Broadly, the luminosity limits presented in Table~\ref{tab:deep-limit} are 2$-$4 orders of magnitude fainter than those of the two known PRSs. For the two events with subarcsecond localizations, previous limits on the presence of a PRS were available in the literature. We found the same luminosity limit for the presence of a PRS associated with FRB\,20180916B as \citet{Marcote2020}, but a less stringent limit for FRB\,20200120E than \citet[][who found L$_{\rm{PRS}} <$ 3.1$\times$10$^{23}$\, erg\,s$^{-1}$\,Hz$^{-1}$]{kirsten2021}.

\subsection{Constraints on PRS prevalence as a function of luminosity and redshift from the global search} \label{subsec:global-implication}

In Figure~\ref{fig:vlasslim} we summarize the key global results from our search for potential PRSs in the localization region of CHIME/FRB repeaters using both archival radio surveys (37 FRB regions searched) and targeted VLA observations (a subset of 8 FRB regions searched). In particular, we plot redshift versus radio luminosity for the VLASS, VLA, and LoTSS radio detections and upper limits described above as well as the previously published PRSs (FRB\,20121102A, FRB\,20190520B, and FRB\,20201124A). For reference, we also plot lines representing the VLASS and LoTSS sensitivity as a function of redshift. 

When placing objects on this plot, we adopt a spectroscopic redshift whenever it is known and otherwise use $z_\mathrm{max}$ for the FRB. Points for which spectroscopic redshifts were used are outlined in red. In cases where $z_\mathrm{max}$ was used, these points represent upper limits on both the redshift and therefore \emph{also} the luminosity of the radio detection/upper limit. To emphasize this, we also label these points with both horizontal and vertical arrows. (In contrast, points that are upper limits on the luminosity because they are derived from radio non-detections as either downward-facing triangles, diamonds, or hexagons as labeled in the legend). The radio luminosities on this plot are calculated as $\nu L_{\nu, \alpha}$ where we have taken both the observed frequency and adopted redshift into account. In particular, for all observations, we calculate the spectral luminosity at a rest-frame frequency of 1.5 GHz using the equation:
$L_{\nu, \alpha} = \frac{4\pi D^{2}_{L}S_{\nu}}{(1+z)^{1+\alpha}} \left( \frac{1.5~\rm{GHz}}{\nu~ \rm{GHz}}\right)^{\alpha}$, where $D_{L}$ is the luminosity distance, $S_{\nu}$ is the observed flux density, $\nu$ is the observed frequency, z is the source redshift and $\alpha$ is the spectral index. We take $\alpha = - 0.4$ (the spectral index estimated for the PRS of FRB\,20121102A).

As noted above, VLASS was the deepest archival survey searched for all 37 FRB regions at GHz frequencies (where the three known PRSs were first identified). However, as can be seen in Figure~\ref{fig:vlasslim}, VLASS is only sensitive to the luminosities of previously identified PRSs at low redshift ($z\lesssim0.11$). As only 4 of the 37 CHIME FRBs searched have $z_\mathrm{max}$ values below this, we can make only weak statements on the prevalence of PRSs with similar luminosities to that of FRB121102 based on this global search\footnote{Since zero of these have unresolved VLASS radio sources in the field of view, we would infer that limit of $<$25\% of repeating FRBs with $z_\mathrm{max} < 0.11$ show PRSs with similar luminosity to FRB121102}. Instead, deeper targeted searches, such as those presented here for a subset of 8 FRBs, will be required to make firmer statistical statements on prevalence.

However, more broadly, we can probe the possibility of more luminous FRBs than have been identified to date with all-sky surveys such as VLASS. In particular, VLASS would be sensitive to PRSs with luminosities of $\sim$10$^{40}$ erg s$^{-1}$ (a factor of $\sim$25 times more luminous than the PRS associated with FRB\,20121102A) out to a redshift of $z=0.5$. As only one of the 22 CHIME repeaters with $z_\mathrm{max} < 0.5$ that were searched as part of this study even had an unresolved VLASS source in its localization region, we can conclude that such luminous PRSs must be rare ($<$5\%). This is consistent with our finding that there is no evidence for a statistical excess of radio sources in the CHIME localization regions within these archival surveys (see Section~\ref{subsec:pcc-global} and Table~\ref{tab:catalogue details}).

As shown in Figure~\ref{fig:vlasslim}, the global LoTSS sensitivity in $\nu L_{\nu}$ is approximately a factor of 3.5 deeper than that of VLASS. This is due to a combination of (i) the slightly lower average survey rms (0.08 mJy/beam versus 0.12 mJy/beam, see Table~\ref{tab:catalogue details}) and (ii) the lower observing frequency coupled with the assumed negative spectral index. Given these assumptions, we would infer that LoTSS would be sensitive to PRSs with similar luminosities to FRB\,20121102A out to a redshift of $z=0.19$ (while zero of five FRB regions searched with $z_\mathrm{max}$ less than this value have an unresolved LoTSS source). In addition, based on these calculations, LoTSS would be sensitive to PRSs with radio luminosities of $\sim$10$^{40}$ erg s$^{-1}$ out to a larger redshift of $z=0.8$. Given that only one of the 35 FRB regions searched with $z_\mathrm{max} < 0.8$ had an unresolved LoTSS source, this would further decrease the estimated prevalence of such luminous PRSs to $<$3\%.

We caution that these LoTSS constraints are dependent both on the assumed spectral index of the PRS as well as the possibility of spectral breaks at lower frequencies (which were not accounted for when placing objects on Figure~\ref{fig:vlasslim}). In particular, we note that neither of the PRSs associated with FRB\,20121102A or FRB\,20190520B is detected in LoTSS (while predictions based on the spectral index would place them close to the 5$\sigma$ sensitivity limit of the survey). However, these results are still consistent with the conclusion that luminous PRSs are rare.

\begin{figure*}[t]
\centering
\includegraphics[width=.45\textwidth]{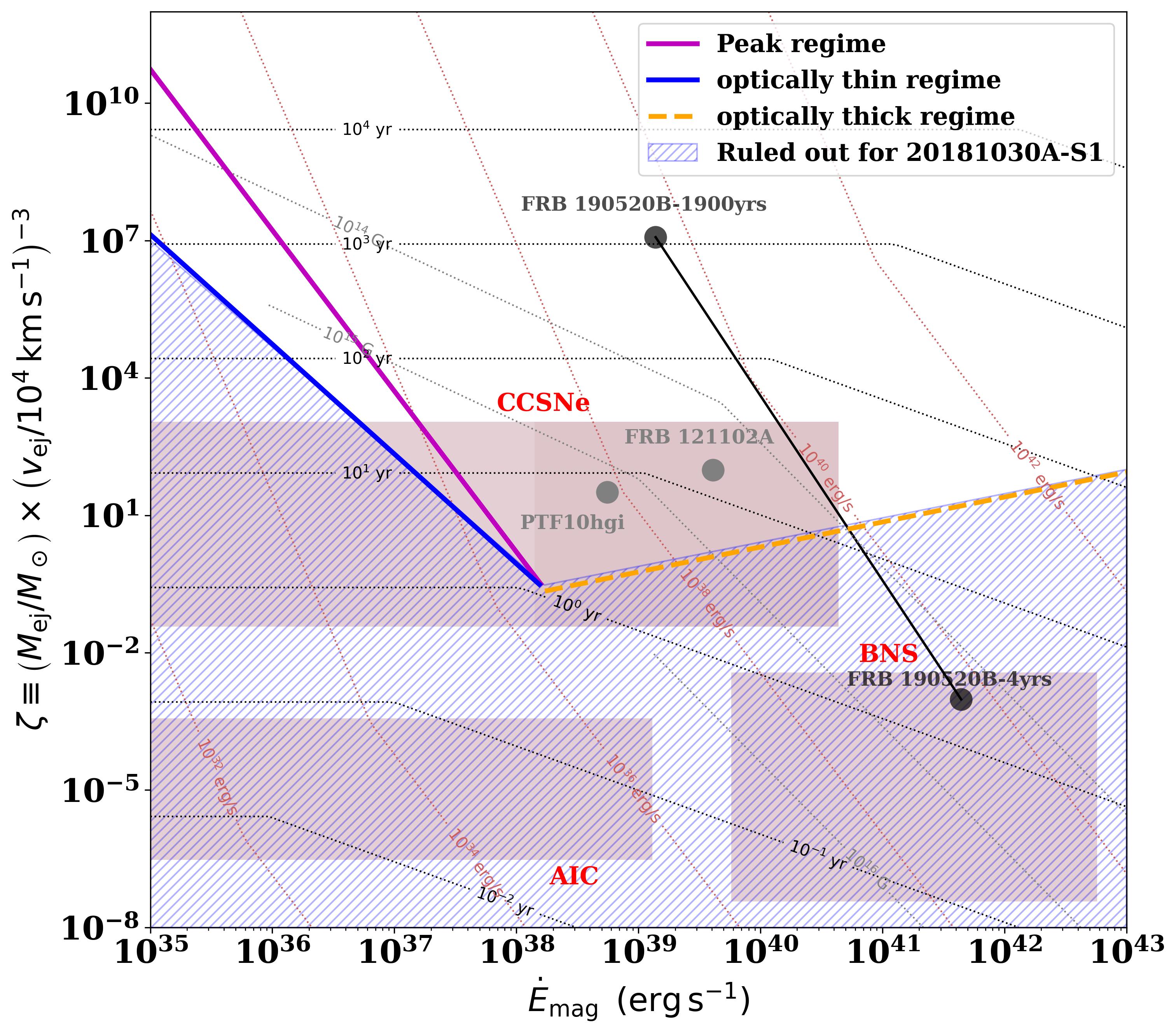}
\includegraphics[width=.45\textwidth]{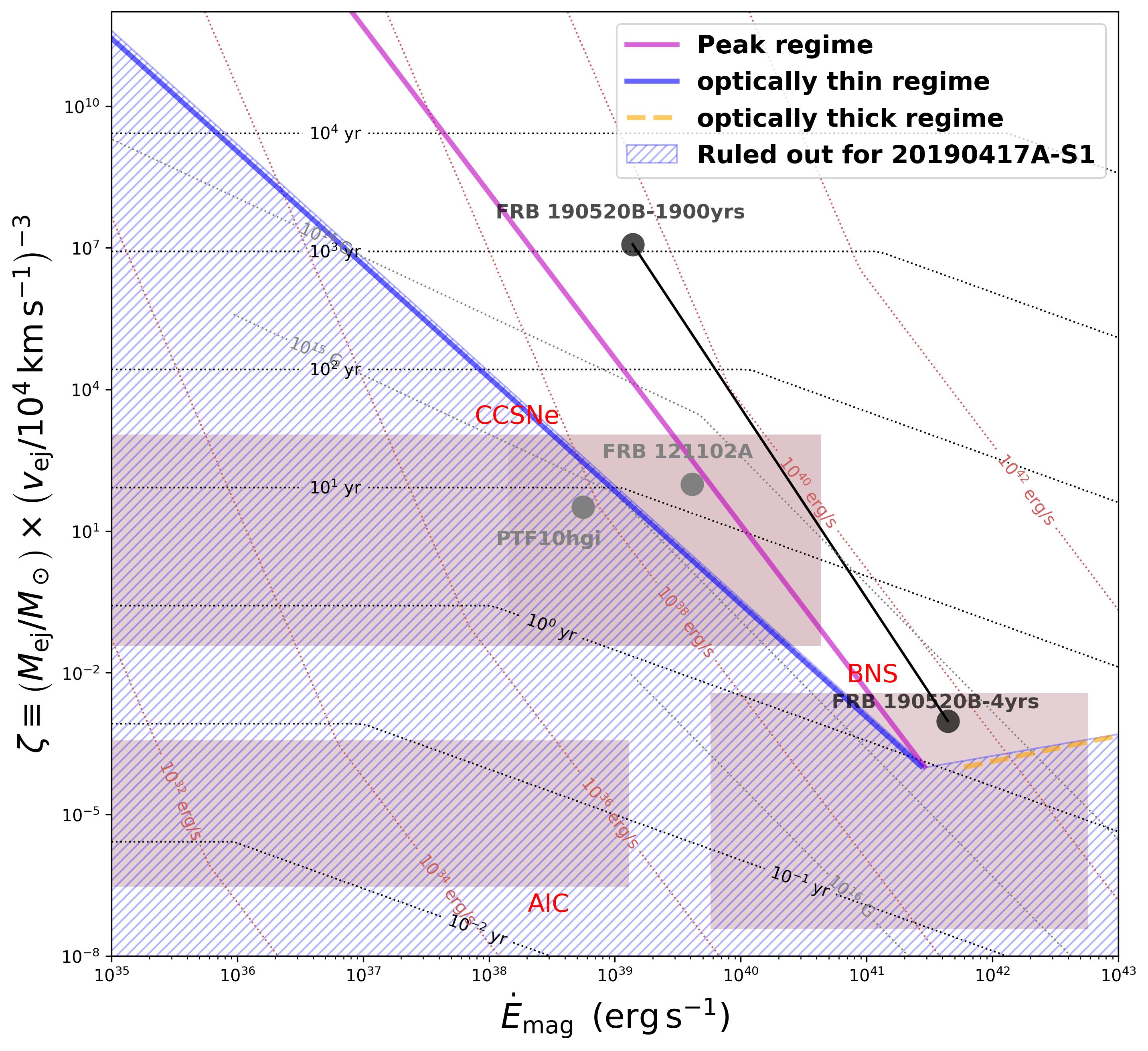}
\caption{Characteristics of the nebular radio emission emitted by magnetars originating from diverse progenitor pathways such as a superluminous supernova (SLSNe), a long gamma-ray bursts (LGRB), a binary neutron star (BNS) merger, or accretion-induced collapse (AIC) of a white dwarf for 20181030A-S1 (\emph{Left}) and 20190417A-S1 (\emph{Right}). For the different regimes predicted by our model, the blue-shaded regions correspond to the parameter spaces ruled out for 20181030A-S1 and 20190417A-S1 respectively. The allowed ejecta density parameter space for 20181030A-S1 is consistent with magnetars born through core-collapse i.e. SLSNe and LGRB while that of 20190417A-S1 is consistent with core-collapse SN and binary neutron star mergers. The properties of the PRS of FRB\,20121102A \citep{margalit2019} and FRB\,20190520B \citep{Bhandari2023} are also overplotted and consistent with the unshaded regions of the parameter space of 20181030A-S1 and 20190417A-S1.} 
\label{fig:mag-bns-aic}
\end{figure*}

 \section{Comparison between potential PRSs and FRB-PRS models } \label{subsec:comparison}

Various models have been proposed to explain the PRS seen at the location of FRBs. Here, we examine the implication of a subset of these models in the context of the two radio sources (20181030A-S1 and 20190417A-S1) classified as potential PRSs by our analysis (\S~\ref{subsec:magnetar}--\ref{subsec:summary-of-models}; see \S~\ref{subsec:summary-prs} for a summary of key properties of these candidates), followed by a discussion of model implications for cases where a PRS was not detected (\S~\ref{subsec:implication-nondetect}). We emphasize that this analysis inherently assumes each radio source is associated with the FRB, which has yet to be confirmed. In addition, we only consider models where we have observable parameters for the radio sources that are sufficient to make an inference. Specifically, we consider four models: (i) a pulsar wind or ion-electron wind nebula, (ii) a hypernebula, (iii) an off-axis jet GRB afterglow, and (iv) supernova ejecta interaction.

For the pulsar wind or ion-electron wind nebula model (\S ~\ref{subsec:magnetar}), both the FRB and the PRS emission are powered by the magnetar's rotational or magnetic energy. For the hypernebula model (\S ~\ref{subsubsec:hypernebula}), the FRB is emitted along the jet funnel due to magnetized shocks and reconnection events far away from the accreting engine while the PRS is powered by the interaction of the shock with the circumstellar materials released by the hyper accretion of the system. Within the off-axis jet GRB and the supernova models, the FRB emission could come through the dissipation of rotational or magnetic energy from the compact object created in the explosion, while the PRS is attributed to the afterglow of a long-duration gamma-ray burst (\S ~\ref{subsec:grb-afterglow}), or the shock interaction between the supernova ejecta and the surrounding circumstellar medium (\S ~\ref{subsec:sn-ejecta}).

For all models, we need the age of the radio-emitting region. Although this is unknown for the two objects considered here, we can set a lower limit based on the time of the discovery of the FRB by CHIME. For 20181030A-S1, we use the FRB discovery date (2018 October 30) to establish a lower age limit of $t_\mathrm{age} >$ 2.36 years (864 days) at the time of the last radio detection (2021 March 6). For 20190417A-S1, we use the FRB discovery date to estimate a lower age limit of $t_\mathrm{age} >$ 1.9 years (682 days) at the radio observation date (2021 February 27).
 
 \subsection{PWN/Magnetar Ion-Electron wind nebula }\label{subsec:magnetar}

We consider the case of an ion-electron wind nebula as the cause of the PRS according to \cite{Margalit2018}. We place constraints on the magnetic energy ($E_\mathrm{B}$), the age ($t_\mathrm{age}$), and the size ($R\mathrm{n}$) of the nebula using the luminosity and the upper limit measurement on the size of the radio source from VLA imaging at 1.5 GHz. In addition, we use the models of \cite{margalit2019} to constrain the allowed ejecta properties as a function of energy injection rates, and comment on their consistency with expectations for various magnetar progenitor channels (core-collapse supernovae, binary neutron star mergers, and accretion-induced collapse of WDs).

We first consider the case of 20181030A-S1. We combine the rotation measure (RM) value of FRB\,20181030A of $\sim$36.6$\pm$0.2\,rad\,m$^{-2}$\citep{Mckinven2023} and the luminosity of the radio source (L$_\mathrm{\nu} =$ 2.1 $\times$ 10$^{26}$\,ergs\,s$^{-1}$\,Hz$^{-1}$) to estimate the magnetic energy using Equation 22 of \cite{Margalit2018}. We assume that the magnetization of the injected outflow, $\sigma$, is $\sim 1$, and that the mean energy per particle in a proton-electron composition, $\chi$, is 0.2 GeV. This yields estimates for the magnetic energy of $E_\mathrm{B} \sim 2.9 \times 10^{46}$\,ergs\,s$^{-1}$\,Hz$^{-1}$ and magnetic field strength of B$\sim 2.26 \times 10^{14}$ G. We then calculate an upper limit on the age of the nebulae of $t_\mathrm{age} <$ 3011 years using $t_\mathrm{age} \sim E_\mathrm{B}/\nu L_\mathrm{\nu}$ ($\nu$ is the 1.5 GHz frequency of the radio observations). 

To estimate the upper limit on the true size of the nebula, we use the beam size of the combined VLA image of the radio source which is 1.1$\prime\prime$ for the semi-major axis. At the angular diameter distance of the galaxy ($D_\mathrm{A} \sim 17$\,Mpc), we obtained an upper limit on the size of the nebula to be $<$92\,pc. \cite{Mckinven2023} reported a high linear polarization fraction and varying polarization angle from the RM analysis for this FRB. This means that the FRB is likely coming from the neighborhood of its central engine if powered by a magnetar \citep{Mckinven2023}, similar to the `nearby' or magnetospheric models of FRB \citep{Kumar2020,Kumar2017,Kumar2024,Murase2016}. 

Next, we can estimate the allowed parameter space for the ejecta density parametrized by $ \zeta = (M_\mathrm{ej}/M_\mathrm{\odot}) \times (v_\mathrm{ej}/10^{4}$ km s$^{-1}) ^{-3}$ assuming a range of energy injection rates, $\dot{E}_\mathrm{mag}$, where $M_\mathrm{ej}$ is the ejecta mass and $v_\mathrm{ej}$ is the ejecta velocity of the event. To do this, we solved the analytic equations for synchrotron radiation assuming the detected 1.5 GHz radio emission was in one of three regimes: (i) in the optically thick regime (ii) at peak frequency, and (ii) in the optically thin regime. These are represented in Equations 19, 23, and 21 of \cite{margalit2019} respectively. The left panel of Figure ~\ref{fig:mag-bns-aic} shows the result of this analysis using the spectral luminosity of 20181030A-S1. When we consider the lower age limit constraint of $>3$ years (based on the time between the FRB discovery the the radio observations), the remaining allowed parameter space is shown as an unshaded region. We also note that the lower peak regime and the optically thick regime are ruled out by the measured in-band spectral index of $\sim -0.7$ (\S~\ref{subsec:variability}). 

\begin{figure*}[t]
\centering
\includegraphics[width=\textwidth]{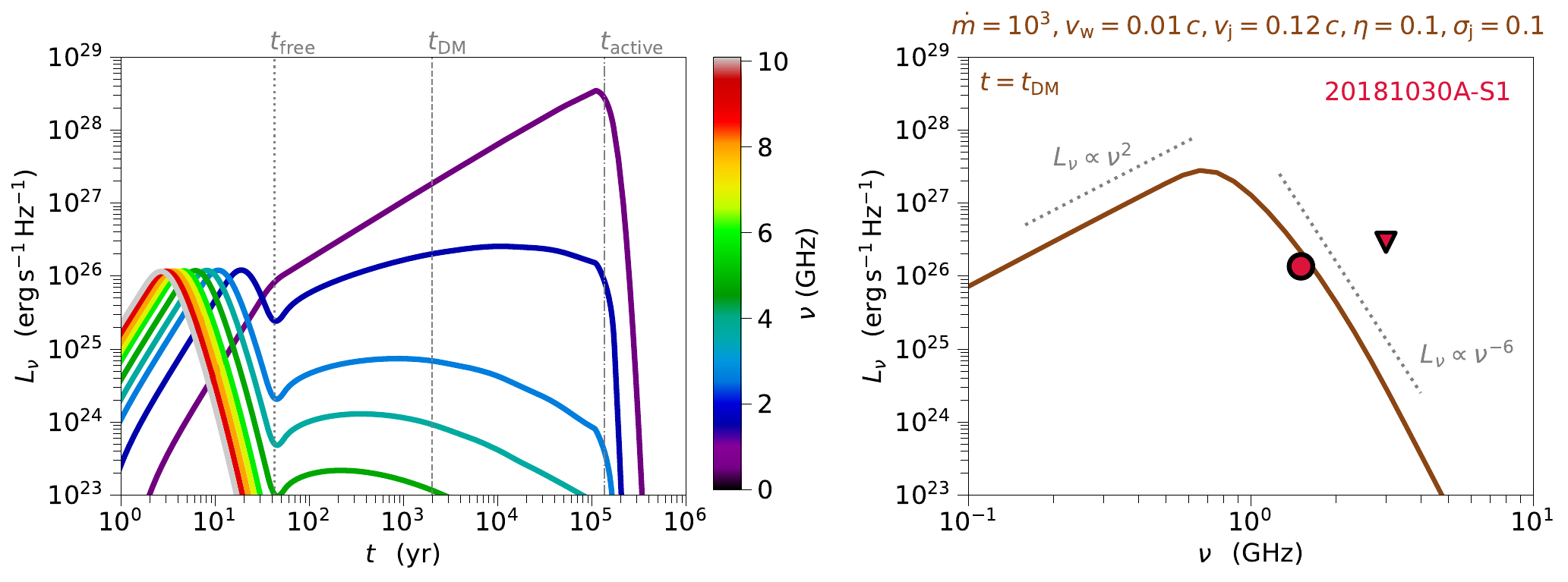}
\caption{Radio synchrotron emission from an accretion-powered hypernebula surrounding the FRB\,20181030A. Left panel: Light curves of the expanding hypernebula in different bands (color-coded). Vertical grey dotted, dashed, and dash-dotted lines denote the free-expansion timescale of the hypernebula (Eq.~\ref{eq:t_free}), the moment during the evolution when the contribution of the hypernebula material to the DM matches the observed $DM_{\rm host}$, and the active duration of the central accreting engine $t_{\rm active}$, respectively (see Sec.~\ref{subsubsec:hypernebula} for more details on the system's parameters). Right panel: Spectral energy distribution at $t=t_{\rm DM}\sim2000$\,yr. The red markers are the observed spectrum (downward-facing triangle is an upper limit) of the proposed PRS associated with FRB\,20181030A.} 
\label{fig:hypernebula}
\end{figure*}

According to the allowable parameter range, the binary neutron star and accretion-induced collapse models are ruled out as a potential source of the magnetar if 20181030A-S1 is associated with the FRB. For comparison, the parameter estimation for the PRS of FRB\,20190520B from \cite{Bhandari2023} is also displayed, where the two ends of the line correspond its lower and upper age limits (with constraints from the lower limit aligning with the parameter space estimated for magnetar formed by a binary neutron star merger). Alongside FRB\,20121102A, the PRS candidate identified here for FRB\,20181030A suggests consistency with magnetars formed in SLSNe/LGRB---ruling out binary neutron star mergers and accretion-induced collapse of a white dwarf. 

We next follow the same procedure for the radio source 20190417A-S1. We note that the high and varying RM of this FRB (4681 $-$ 4429 rad\,m$^{-2}$; \citealt{Feng2022, Mckinven2023}) includes it among other repeaters (including FRB\,20121102A \citep{Chatterjee2017} and FRB\,20190520B \citep{Niu2021}) that are probably residing in a dynamic magnetoionic environments makes this radio source an object of particular interest. We obtain estimates of $E_\mathrm{B} \sim 2.11 \times 10^{50}$\,ergs\,s$^{-1}$\,Hz$^{-1}$, B$\sim 1.9 \times 10^{16}$ G, and upper limit on $t_\mathrm{age} <$ 56,186 years.  We then use the beam size of the combined VLA image of the radio source which is 1.3" for the semi-major axis at an angular diameter distance of $D_\mathrm{A} \sim 475.5$\,Mpc to obtain an upper limit on the size of the nebula to be $<$3.2\,kpc ($<$9.75$\times 10^{21}$\,cm). When considering the allowed ejecta parameters as a function of energy injection (see the right panel of Figure ~\ref{fig:mag-bns-aic}), our analysis yields a range of parameter spaces consistent with core-collapse supernova and binary neutron star mergers. In the future, accurate measurements of the sizes of the radio sources via the VLBI method and a well-defined SED could yield tighter constraints for both 20181030A-S1 and 20190417A-S1.

\subsection{Hypernebula model} \label{subsubsec:hypernebula}

Another paradigm proposed to self-consistently explain the observed properties of FRBs and their associated PRS is a hypernebula that is inflated by the baryon-rich outflows ejected by hyper-accreting compact objects \citep{Sridhar&Metzger_22, Sridhar+24}. Recently, this model was employed to explain the observed properties of FRBs 20210117A \citep{Bhandari+23a}, 20201124A \citep{Dong-dust2024}, and 20190520B \citep{Bhandari2023}. In this picture, the FRB is emitted along the jet funnel due to magnetized shocks and reconnection events far away from the accreting engine \citep{Sridhar_2021}. We investigate the possibility of this model here in light of our observations.

We first consider 20181030A-S1. We take the maximum isotropic-equivalent luminosity of FRB\,20181030A to be $10^{40}\,{\rm erg\,s^{-1}}$ \citep{CHIME_19a}; this requires an accretion rate of $\dot{M}_\bullet \gtrsim10^{3} \dot{M}_{\rm Edd}$ for the FRB to be accretion-jet powered \citep{Sridhar_2021}, where $\dot{M}_{\rm Edd}\sim1.3\times10^{39}\,{\rm erg\,s^{-1}}$ is the Eddington mass transfer rate for an accreting $10\,M_{\odot}$ black hole (or a requirement of $\dot{M}_\bullet \gtrsim10^{4} \dot{M}_{\rm Edd}$ for the FRB to be powered by a neutron star). The large-angled, slower disk winds (with speeds $v_{\rm w}\sim0.01\,c$, and mass-loss rate $\dot{M}_{\rm w}\sim\dot{M}_\bullet$) powered by the hyper-accreting disk will drive a forward shock into the circumstellar medium (with an assumed density $n\approx10\,{\rm cm}^{-3}$). The compact object also powers a faster wind/jet along the spin axis (with speeds $v_{\rm j}\sim0.12\,c$) that drives the termination shock upon interaction with the slower disk winds. 

Following \cite{Sridhar&Metzger_22}, we calculate the observable properties of the hypernebula due to these interactions for the following assumed physical parameters: jet magnetization parameter (ratio of the magnetic enthalpy density to the plasma enthalpy density) $\sigma_{\rm j}=0.1$, ratio of the wind luminosity to the jet luminosity $\eta=0.1$, the fraction of the shock power that goes into heating the electrons $\varepsilon_{\rm e}=0.5$, the mass of the accreting compact object $M_\bullet=10\,M_\odot$ (assuming a black hole), and mass of the companion accretor star $M_\star=30\,M_\odot$: this sets the active lifetime of the system to be $t_{\rm active}\sim M_\star/\dot{M}_\bullet$. The radio synchrotron light curves from the hypernebula at different bands are provided in the left panel of Fig.~\ref{fig:hypernebula}.

The free expansion timescale of the outflowing winds (before they start to decelerate) is,
\begin{equation} \label{eq:t_free}
    t_{\rm free} \approx 42\,{\rm yr}\left(\frac{L_{\rm w,39}}{n_1}\right)^{1/2}\left(\frac{v_{\rm w}}{0.01\,c}\right)^{-2.5}.
\end{equation}
Here, we adopt the short-hand notation, $Y_{x} \equiv Y/10^x$ for quantities in cgs units. The contribution of the expanding shell to the dispersion measure (assuming the shell remains ionized) bridging the free expansion and deceleration phases is,
\begin{equation}\label{eq:DM}
    \text{DM}_{\rm sh} \simeq \frac{M_{\rm sh}}{4\pi R^{2}m_{\rm p}} \approx
    \begin{cases}
          1\,{\rm pc\,cm^{-3}}\left(\frac{\dot{M}_{\rm w}}{10^{3}\dot{M}_{\rm Edd}}\right)v_{\rm w,9}^{-2}\left(\frac{t}{42\,{\rm yr}}\right)^{-1} & (t < t_{\rm free}) \\
          1\,{\rm pc\,cm^{-3}}\left(\frac{L_{\rm w,39}}{n_{1}}\right)^{1/5}\left(\frac{t}{42\,{\rm yr}}\right)^{3/5} & (t > t_{\rm free}),
    \end{cases}
\end{equation}
where we take the mass in the expanding shell to be $M_{\rm sh}\sim\dot{M}_{\rm w}t$, and $m_{\rm p}=1.67\times10^{-24}\,g$ is proton's mass.

The host contribution to the DM for FRB\,20181030A is estimated to be DM$_{\rm host}\sim10\,{\rm pc\,cm^{-3}}$ \citep{BhardwajR42021}. The hypernebula shell will contribute this value of DM when the hypernebula's age is $t_{\rm DM}\sim2000$\,yr. The aforementioned three timescales are indicated by grey dotted ($t_{\rm free}$), dashed ($t_{\rm DM}$), and dash-dotted ($t_{\rm active}$) vertical lines in the left panel of Fig.~\ref{fig:hypernebula}. We compute the radio synchrotron spectrum at time $t_{\rm DM}$, which is shown as a brown curve in the right panel of Fig.~\ref{fig:hypernebula}. We note that the hypernebula model spectrum is consistent with the observations (red markers). Furthermore, in this model, the expected X-ray flux from the accretion disk---during the active period of FRB emissions when the accreting cone is aligned with the observer---is expected to be $<2\times10^{-11}\,{\rm erg\,s^{-1}\,cm^{-2}}$ (taking a distance of 20\,Mpc and X-ray luminosity of $\lesssim10^{42}\,{\rm erg\,s^{-2}}$). This is also consistent with the X-ray flare upper limits of $\lesssim10^{46}\,{\rm erg\,s^{-1}}$ \citep{Bhardwaj_2021}.

We next consider 20190417A-S1. A peak radio burst luminosity of $10^{41-42}\,{\rm erg\,s^{-1}}$ from FRB\,20190417A (adopting FRB fluxes from \citealt{Fonseca2020} and a redshift of $z=0.12817$ corresponding to the PRS candidate 20190417A-S1 presented here) would require an accreting engine transferring matter at $\dot{M}_\bullet\gtrsim10^{4-5}\,\dot{M}_{\rm Edd}$. A hypernebula powered by such an engine could contribute to a large DM$_{\rm host}\lesssim1250\,{\rm pc\,cm^{-3}}$ at the age of $t\sim7$\,yr (Eq.~\ref{eq:DM}). This is consistent with the observed upper limit on DM$_{\rm host}$ of FRB\,20190417A that one can obtain by roughly subtracting the Milky Way and host galaxy contributions \citep{Fonseca2020}. At this age, the hypernebula shell would still be in the free-expansion phase (Eq.~\ref{eq:t_free}), and the nebula could contribute to a large maximum rotation measure (RM$\lesssim10^7\,{\rm rad\,m^{-2}}$; see Eq.~50 of \citealt{Sridhar&Metzger_22}) which may be consistent with the current measurements of RM$\sim4681\,{\rm rad\,m^{-2}}$ \citep{Mckinven2023}. These properties resemble that of FRBs\,20121102A and 20190520B, which were also explained using the hypernebula model \citep{Sridhar&Metzger_22, Bhandari2023}.
However, due to the uncertain value of the DM$_{\rm host}$ associated with FRB\,20190417A \citep{Fonseca2020}, and a lack of a robust distance/host galaxy association between FRB\,20190417A and PRS\,20190417A-S1, we defer from performing a detailed spectral analysis for PRS\,20190417A-S1 (in a way that is self-consistent with the properties of the FRB).

\begin{figure}[t]
\centering
\includegraphics[width=.45\textwidth]{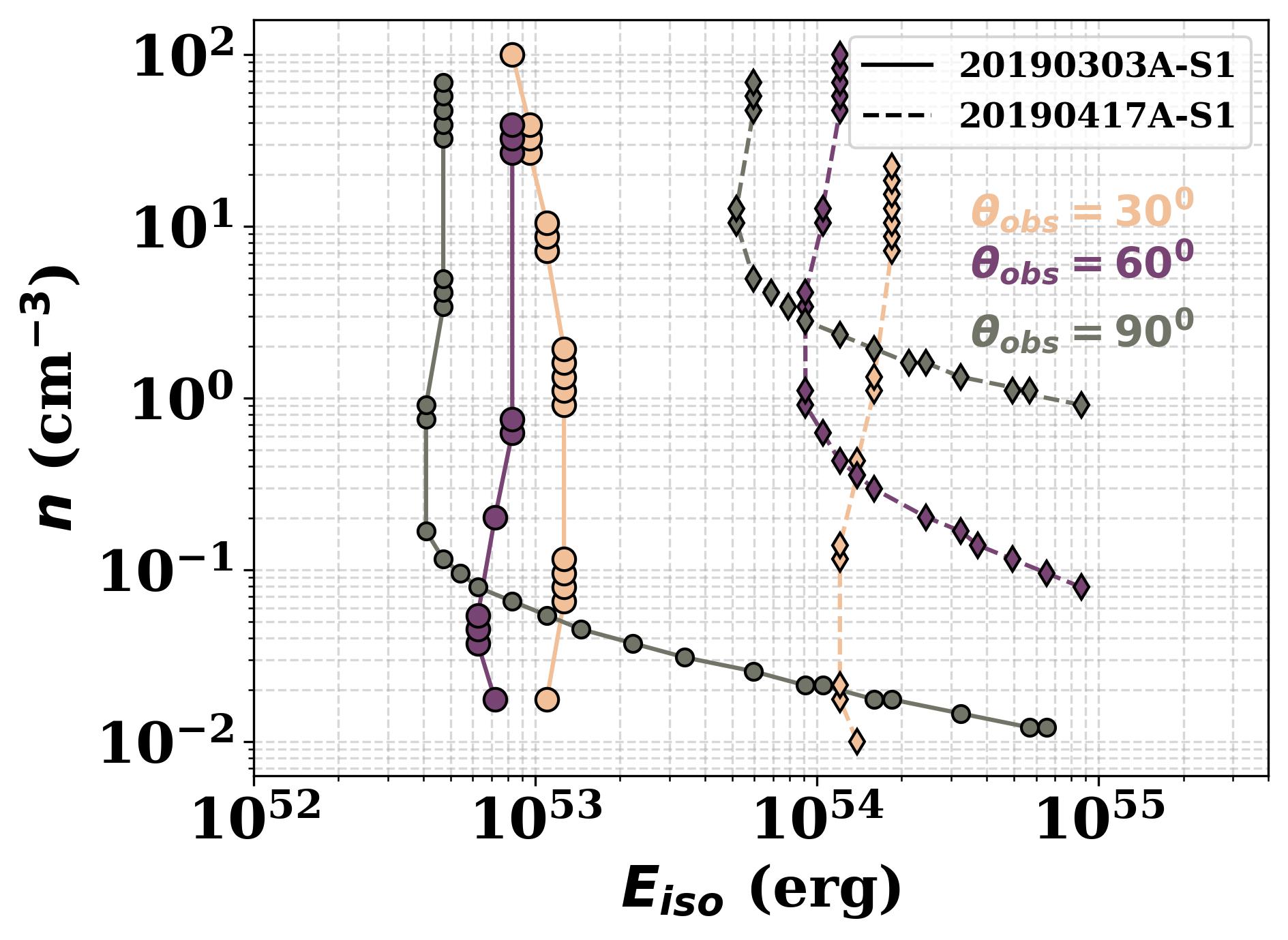}
\caption{Some allowable parameter space for the properties of 20181030A-S1 (circles) and 20190417A-S1 (diamonds) if they are each a radio afterglow from an off-axis jet of a long gamma-ray burst or a superluminous supernova associated with their FRBs. The allowed range of energies for 20190417A-S1 is consistent with higher values when compared to that of 20181030A-S1 for the same range of ISM densities. } 
\label{fig:grb-afterglow}
\end{figure}

\begin{figure*}[t]
\centering
\includegraphics[width=.45\textwidth]{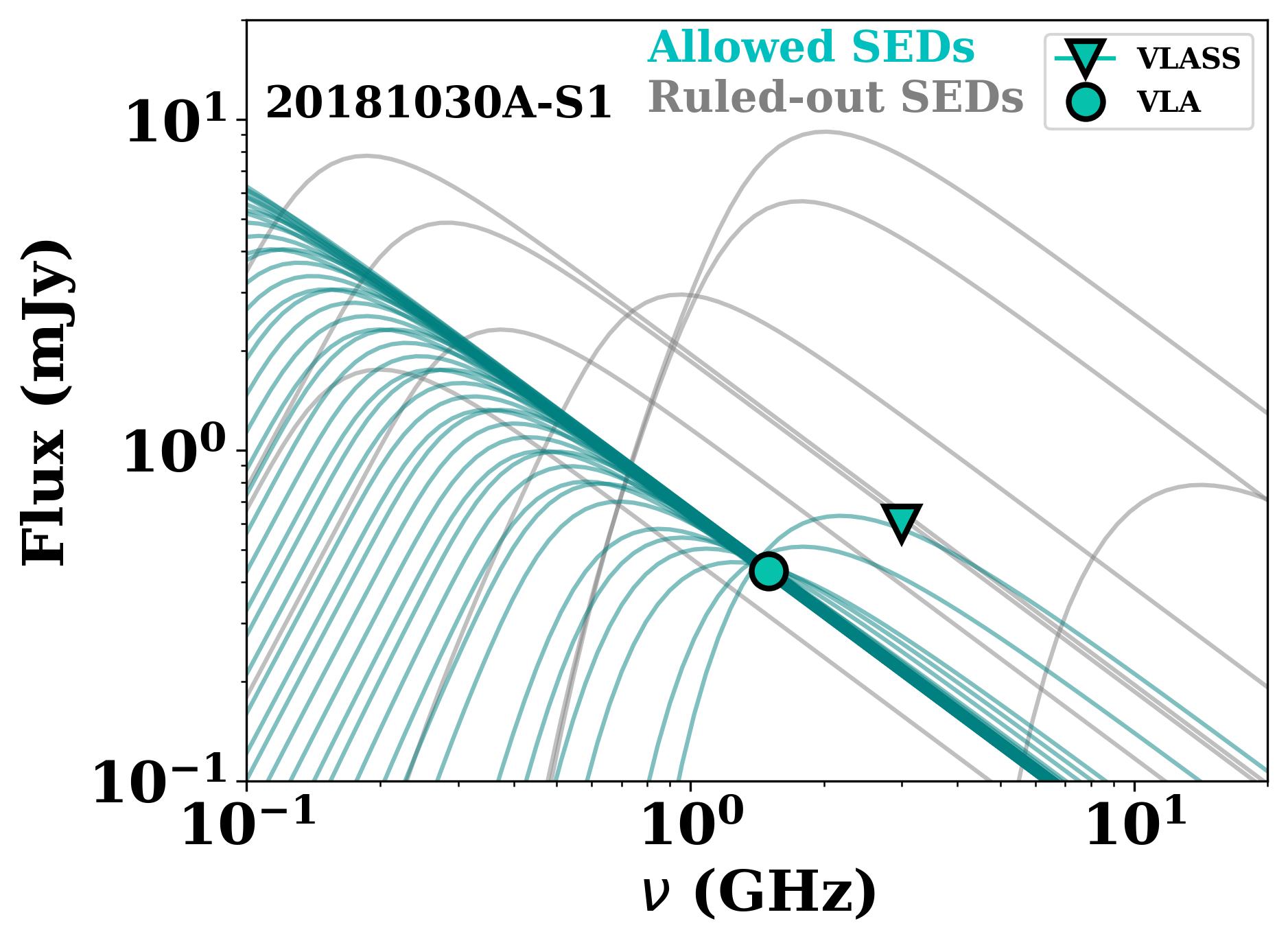}
\includegraphics[width=.42\textwidth]{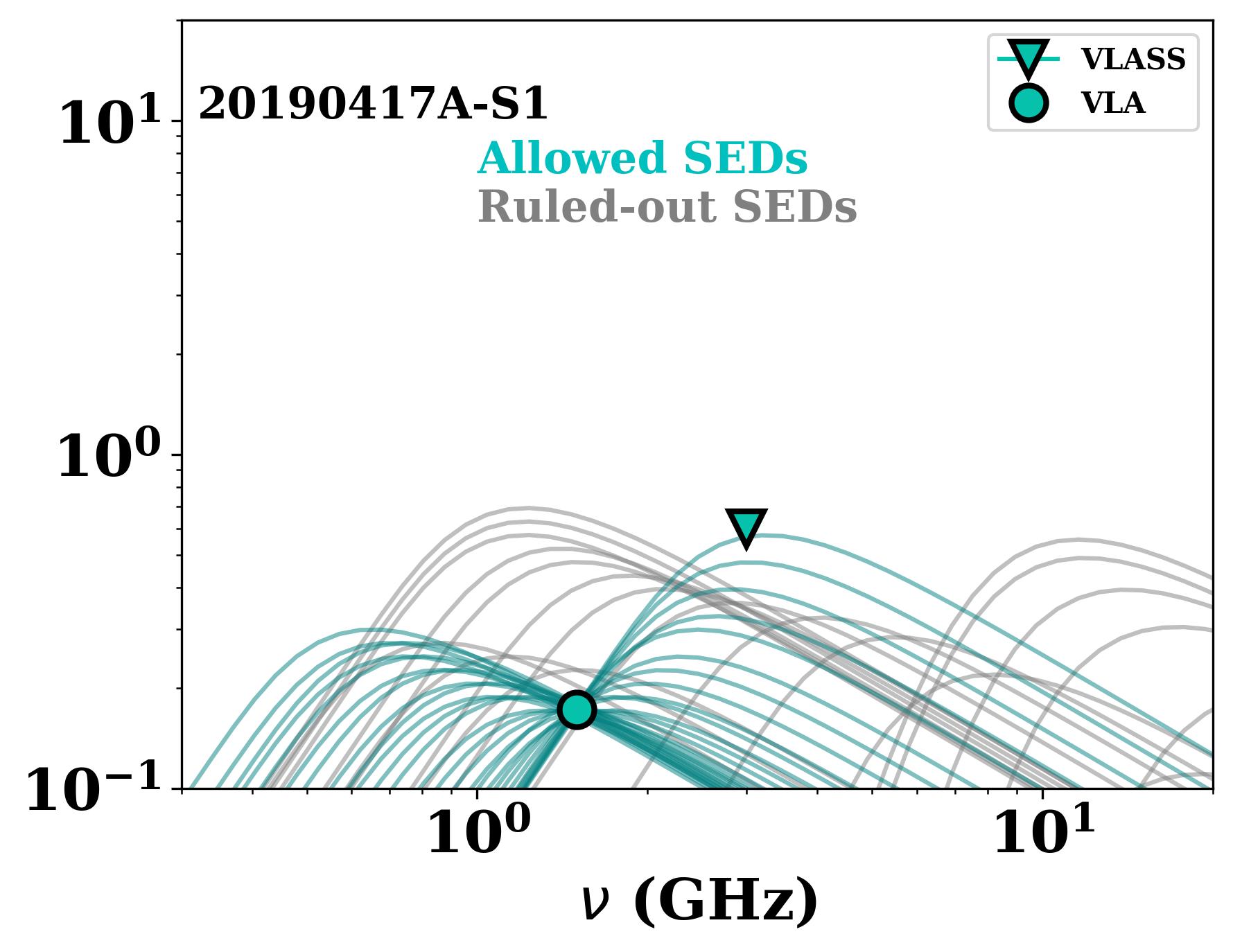}
\caption{Model SEDs of the radio source, 20181030A-S1 (left panel) and 20190417A-S1 (right panel) assuming an expanding SN ejecta colliding into external CSM. The cyan-colored lines are the allowed SEDs given the model. We report the corresponding allowed parameter space for the properties of the explosion in \S ~\ref{subsec:sn-ejecta} which are consistent with a moderately luminous supernova.} 
\label{fig:sn-ejecta}
\end{figure*}

 \subsection{Off-axis jet afterglow model from LGRB/SLSN} \label{subsec:grb-afterglow}

Next, we explore the off-axis jet afterglow model as a possible alternative source of PRS emission. Specifically, we examine whether it is possible to explain the radio sources 20181030A-S1 and 20190417A-S1 with parameters broadly consistent with those observed for other GRBs.
We used the open-source Python package \texttt{afterglowpy} \citep{Ryan2020afterglow}---software that uses numerical models to calculate structured jet afterglows synthetic light curves and spectra from any viewing angle. We run a dense grid of isotropic equivalent jet energies, $E_\mathrm{iso}$ and interstellar material (ISM) densities, $n$, through and calculate radio light curves for each pair at three viewing angles, $\theta_\mathrm{obs} = 30\arcdeg, 60\arcdeg, 90\arcdeg$.

Following previous GRB studies \citep[e.g.][]{Alexander2017grb, Laskar2019grb, Eftekhari2021}, we assume a Gaussian jet type, a $10\arcdeg$ half-opening angle, a truncation angle that is 5 times the opening angle, an electron energy distribution index of 2.6, and fractional values for post-shock energy contained in the relativistic electrons ($\epsilon_e$) and amplified magnetic fields ($\epsilon_b$) set at 0.005 and 0.01, respectively. Additionally, we fix the fraction of accelerated electrons ($X_n$) at 0.8. 

For each model in our grid, we generate light curves at 1.5 GHz and compare them with our radio detections. We consider a given model to be ``allowed'' if it passes through the observed radio flux within its 1$\sigma$ uncertainty and ruled out if it does not. We adopt ages of 864 and 682 days for 20181030A-S1 and 20190417A-S1, respectively. These correspond to lower limits on their true age, as described above. In Figure ~\ref{fig:grb-afterglow} we plot the allowed values of $E_\mathrm{iso}$ and $n$ at each viewing angle considered for both events.

For 20181030A-S1, we find that the radio constraints mostly allow the possibility of jets with energies $7\times10^{52}$ erg $\leq$ $E_\mathrm{iso}$ $\leq$ $2\times 10^{53}$ erg at $n \sim 10^{-2} - 10^{2}$\,cm$^{-3}$ for viewing angles, $\theta_\mathrm{obs} = 30\arcdeg$ and $60\arcdeg$. However, for the same ISM densities, we are unable to rule out the jet energies $E_\mathrm{iso}$ $\geq$ $4\times 10^{52}$ erg for $\theta_\mathrm{obs} = 90\arcdeg$. For 20190417A-S1, we have a tight range of allowable energies from (1.39 $-$ 1.84) $\times$ 10$^{54}$ erg for $\theta_\mathrm{obs} = 30\arcdeg$. However, we are unable to rule out jet energies $\geq$ $9\times 10^{53}$ erg and $\geq$ $5\times 10^{53}$ erg for $\theta_\mathrm{obs} = 60\arcdeg$ and $\theta_\mathrm{obs} = 90\arcdeg$ respectively. 

Our results for 20181030A-S1 are consistent with previously reported GRB afterglows at similar frequency \citep[e.g.][]{Berger2003,Resmi2005,Frail2005,Cenko2010}. However, the resultant jet energies for 20190417A-S1 are higher than previously reported radio afterglow limits except for the ultra-long GRB 111209A \citep{Stratta2013}. However, we emphasize that these constraints can be more robust with a well-defined light curve or SED and are highly sensitive to the adopted $t_\mathrm{age}$.

\subsection{Supernova Ejecta Model } \label{subsec:sn-ejecta}
Finally, we consider a radio supernova or a young supernova remnant as the cause of the radio emission. Radio emission originating from interacting supernovae predominantly arises from synchrotron radiation at the forward shock, formed by the fastest-moving ejecta, as the supernova shockwave collides with the circumstellar medium (CSM) \citep[e.g. PTF10hgi;][]{Eftekhari2019}.

We employ a modified version of the framework outlined by \cite{Chevalier1998ApJ} and \cite{Chevalier_2006} to model the radio spectral energy distribution (SED) of the radio source, focusing on synchrotron emission (SE) from interacting supernovae. Specifically, our model accounts for the influence of both synchrotron self-absorption (SSA) and free-free absorption (FFA), as described in \cite{Ibik2024}. Throughout our modeling, we assume baseline values for constants and free parameters. In particular, we assume the equipartition of energy ($\alpha$ $=$ 1), the radio filling factor $f=0.5$, and the fraction of post-shock energy contained in amplified magnetic fields of $\epsilon_b$ $=$ 0.1.

To investigate the implications of the VLA radio detection and VLASS upper limit\footnote{We note that use the flux density limit from the VLASS second epoch of observation. While these are not contemporaneous with the VLA observations, they were observed within 5--6 months, and old radio SN are expected to be relatively slowly evolving.} for the 20181030A-S1 and 20190417A-S1, we generate a dense grid of SSA+FFA SEDs for a wide range of peak fluxes (0.001$-$10\,mJy) and frequencies (0.05$-$50\,GHz). When constructing the SED grid, we apply values for the free-free optical depth that assume a wind-like medium for the external CSM (see \citealt{Ibik2024}). For each SED, we then determine whether it is consistent with, or ruled out by VLA radio detection. The left panel of Figure ~\ref{fig:sn-ejecta} shows examples of allowed (cyan) and ruled-out (grey) SEDs for 20181030A-S1.

For each allowed SED, we can then take the peak frequency, and peak flux and infer a radius for the radio-emitting region. When further coupled with an assumption for the age of the transient, we can also constrain the SN shock velocity and the density of the circumstellar material.  Given that we have only a single detection, the peak of the SED could be at arbitrarily low frequencies (see Figure ~\ref{fig:sn-ejecta})---as a result, our models are unconstrained at the high radius/velocity side.  However, if we 
adopt an age of 864 days (which is formally a lower limit; see above) and restrict ourselves to solutions with velocities $<$ 0.3c (as is typical for SN), then we would infer a radius for the radio-emitting region of $\sim$(2$-$70)$\times 10^{16}$\,cm and CSM density of $\sim$3$\times 10^{-25}$ $-$9$\times 10^{-20}$\,g cm$^{-3}$. In this scenario, allowed shock velocities range from $\sim$3500\,km s$^{-1}$ to our imposed upper limit. Broadly, these parameters are consistent with those observed for all types of core-collapse SN.

We apply a similar approach to the VLA detection and VLASS upper limit for 20190417A-S1 and show a set of example allowed and ruled out SED in the right panel of Figure ~\ref{fig:sn-ejecta}. As above, the radii and velocities are unconstrained on the high end. However, adopting an age of 682 days (a lower limit) and restricting ourselves to solutions with shock velocities $<$ 0.3c, we would infer a radius of the radio-emitting region of $\sim$(1$-$5)$\times 10^{17}$\,cm and CSM density of $\sim$10$\times 10^{-24}$ $-$6$\times 10^{-21}$\,cm$^{-3}$. In this scenario, the allowed shock velocities range from $\sim$2\,$\times$\,10$^{4}$\,km s$^{-1}$ up to our imposed upper limit. Such high shock velocities are not expected for normal core-collapse SN, but are instead closer to those found for some Type Ibc SN \citep[e.g. SN2008D][]{Malesani2009} including the 'broad-lined' subgroup. However, given that the explosion epoch used here is a lower limit on the age of the radio source, the inferred shock velocities could be lower (if the transient were older). Thus, overall, we find that given the uncertainty in both the age and spectral shape of 20181030A-S1 and 20190417A-S1, either could be consistent with emission from a SN.

\begin{figure}[t]
\centering
\includegraphics[width=.45\textwidth]{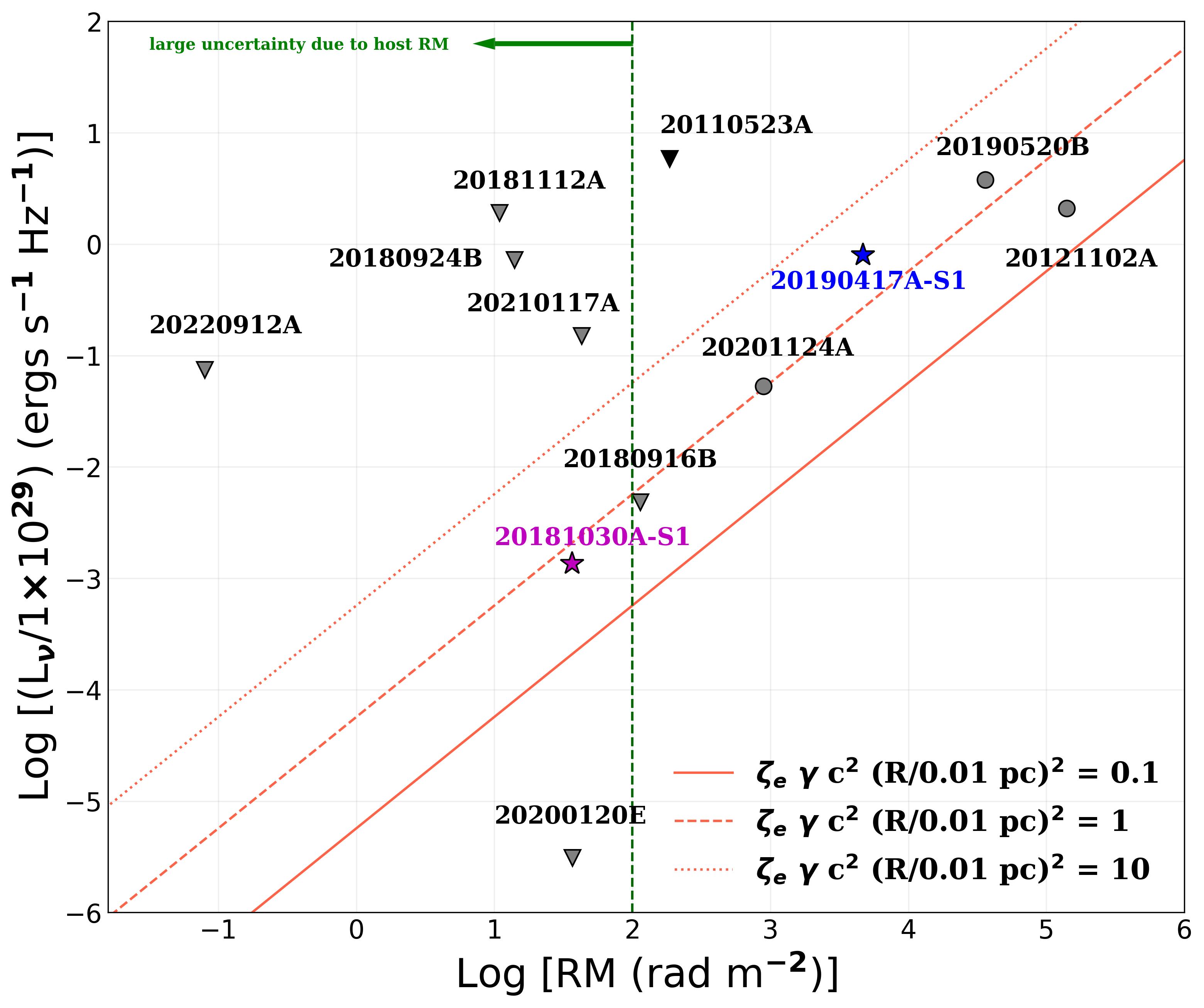}
\caption{The proposed relation between specific radio luminosity of PRS and the
FRB rotation measure adapted from \citealt{Bruni2023}. The colored stars and inverted triangles are potential PRSs (20181030A-S1 and 20190417A-S1) and upper limits from our sample (20190117A-S and 20190208A-S). The pink dotted dashed, and solid lines are predicted relations of values 0.1, 1, and 10 from the nebula model framework. Gray circles are values for the 3 known PRSs while the inverted gray triangles are PRS upper limits for those FRBs.} 
\label{fig:nebular-model}
\end{figure}

\subsection{Summary of models for 20181030A-S1 and 20190417A-S1} \label{subsec:summary-of-models}

In the sections above, we considered four theoretical models for PRS emission: (i) a pulsar wind or ion-electron wind nebula, (ii) a hypernebula, (iii) an off-axis jet GRB afterglow, and (iv) supernova ejecta-CSM interaction. Broadly, we find that---within current constraints---any of these models could explain the luminosity of the radio sources 20181030A-S1 and 20190417A-S1. However, we note that some of these results are highly model-dependent. For example, in the magnetar model, the size of the nebula would be strongly dependent on the source age ($t_\mathrm{age}$), for which we only have lower limits.  
In addition, for both the GRB and SN models, the observed radio emission can be explained using parameters similar to those seen in some supernovae and gamma-ray bursts. However, this explanation is highly contingent on several uncertain factors, particularly the true age ($t_\mathrm{age}$) of the event and the spectral shape. Depending on these variables, the models may become incompatible, but we cannot dismiss them as a potential source for the emission at this stage.

However, it is notable that several properties of both the FRBs and the candidates PRSs presented in this work are consistent with expectations for a nebular origin for PRSs. In particular, in Figure~\ref{fig:nebular-model} we plot the spectral luminosities and RMs of the 2 potential PRSs from this work alongside other previously published PRSs and upper limits. We see that both 20181030A-S1 and 20190417A-S1 align well with the nebula model framework as predicted and shown by \cite{Yang2020} and \cite{Bruni2023}. If 20181030A-S1 and 20190417A-S1 are confirmed to be associated with the FRBs, future work examining both FRB properties (e.g. activity\footnote{We note that despite the relatively young predicted ages for these sources within the magnetar and hypernebulae models, FRBs 20181030A and 20190417A are not amongst the most active FRB repeaters discovered by CHIME (with 9 and 19 bursts detected, respectively). However, this does not preclude them from being young and active FRBs. For instance, we highlight that only one burst from FRB\,20121102A has been detected by CHIME to date.}) and PRS properties (e.g. size) can further assess their consistency with this picture.

\subsection{Physical Implications for non-detections of PRSs} \label{subsec:implication-nondetect}
In Figure \ref{fig:vlasslim}, we presented the upper limits for VLASS non-detections in the majority of the searched FRB repeater fields. In Section~\ref{subsec:global-implication}, we used these limits to conclude that PRSs with high luminosities ($\sim$10$^{40}$ erg s$^{-1}$, which is $\sim$25 times brighter than the PRS associated with FRB\,20121102) must be rare.
This high radio luminsity exceeds what has been observed from SLSNe and LGRBs (at times $>$100 days) to date \citep{Ibik2024, Eftekhari2021}. We would therefore not expect to see such bright radio emission, except in very extreme cases, if these sources are the origin of the PRS emission.
However, in the context of nebular models, it is theoretically possible to achieve these high luminosities in environments with extremely strong magnetic fields and super-Eddington accretion rates \citep{Margalit2018, Sridhar_2021, Sridhar+24}, although no such environments have been observed.

We also established deep PRS flux limits for a subset of these FRBs. Notably, for 5 FRBs with known redshifts (all below $z < 0.1$), these limits are 2$-$6 orders of magnitude fainter than the PRS of FRB\,20121102A. This suggests that these FRBs either lack associated PRSs or have PRSs that are significantly less luminous than that of FRB\,20121102A. This has important implications for the nature of FRBs that should have PRSs. Although the small sample size of known PRSs limits our understanding, we discuss the following factors that may influence whether we can detect a PRS if associated with a repeating FRB \citep[see also discussions offered by][]{Marcote2020, kirsten2021, Chibueze2022}:
\begin{enumerate}
    \item \emph{Age:} Aging PRS sources may enter a quiescent phase, causing their emission to weaken and become undetectable. This is due to the shutdown of energy injection from a central engine \citep[e.g.,][]{Margalit2018}.
    
    \item \emph{Timing of observation:} For periodic and randomly variable sources, observing a PRS during a quiescent period may result in non-detection \citep[for example, this has been seen in pulsars;][]{Lorimer&Kramer2004}. 

    \item \emph{Beaming:} Another possibility is that the PRS emission is beamed, and if we are not observing from the optimal angle, the PRS may remain undetected \citep{Rybicki1986}. 
    
    \item \emph{Burst rate:} Certain FRB-PRS models, including the synchrotron maser and hypernebula models, suggest a positive correlation between burst rate and PRS luminosity. This correlation is thought to arise from an increased electron-ion injection rate, fueling maser emission \citep{metzger2019}, or an increased accretion rate fueling synchrotron emission \citep{Sridhar_2021}. According to these models, FRBs with low burst rates would be expected to have faint PRSs, potentially evading detection due to instrumental sensitivity limitations. The possibility of such a correlation will be explored in an upcoming work (Bhardwaj et al., in prep).   
    
    \item \emph{Environment:} Local environmental conditions, such as magnetic field strength, relativistic electron density, electron-ion injection rate, and accretion rate are crucial for synchrotron emission. Generally, even with a strong magnetic field, insufficient relativistic electrons can prevent radio emission. In the nebular scenario, the energy density in the surrounding environment is key, as it influences the correlation between FRB RM and PRS luminosity, as predicted by \cite{Yang2020} and observed in some FRBs by \cite{Bruni2023} (see \S \ref{subsec:summary-of-models} and Figure \ref{fig:nebular-model}). This suggests that FRBs with low RM may not exhibit a PRS due to a weaker magneto-ionic medium, hindering emission \citep{Bruni2023}.
    
\end{enumerate}

Some of these factors described above, specifically the burst/repetition rate and RM of FRBs, could inform future follow-up observations of FRB repeater fields and searches for their PRSs.

\section{Summary and Conclusions}\label{sec:conclusion}

We have conducted a comprehensive search for and characterization of radio sources within the $\sim1\prime$ baseband localization regions of 37 repeating FRBs discovered by CHIME/FRB. This is part of an ongoing effort to identify additional PRSs similar to those that have been found at the location of two confirmed repeating FRBs to date. We search archival radios surveys for all 37 FRBs and supplement this with deeper, targeted, VLA observations of the localization region for a subset of 8 events.  In total, we identify 25 radio sources (13 in archival surveys and 12 from deep VLA observations) within the fields of 14 of the FRBs. Of these, 10 radio sources were unresolved (as would be expected for true PRSs) while the rest were extended/resolved. We summarize the rest of our findings as follows: 

\emph{Multiwavelength Characterization:} We perform multiwavelength characterization of the 10 unresolved radio sources identified in our search---with an emphasis on 6 sources with detected optical counterparts/host associations. We examine properties such as host redshift, radio variability, and spectral shape, host galaxy offset, probability of chance alignment, optical-to-radio ratio, infrared colors, and host star formation rates. Our goal was to identify any radio sources that (i) appear to be inconsistent with expectations for either star formation or an AGN in the host galaxies, and (ii) have properties broadly comparable to the previously identified PRSs. 

\emph{Candidate PRSs:} After conducting our multi-wavelength analysis, we identify two radio sources of particular interest for which we disfavor either a star formation or host AGN origin. 20181030A-S1 overlaps with the spiral arm of NGC\,3252, which was previously identified as the most likely host of FRB\,20181030A \citep{BhardwajR42021}. If located in the galaxy, it has a radio luminosity of approximately 3 orders of magnitude fainter than the PRS of FRB\,20121102A.  In contrast, 20190417A-S1 is associated with a star-forming galaxy at a redshift of $z=0.128$ and has a luminosity broadly similar to that of the FRB\,20121102A PRS. Both sources have non-thermal spectral indices. 

\emph{Deep Limits:} We present deep radio limits on the presence of a PRS for the 8 FRBs with targeted VLA imaging. We include the fields of the PRS candidates in the case that these radio sources are not truly associated with the FRBs. These limits are generally 2--4 orders of magnitude fainter than the PRS associated with FRB\,20121102.

\emph{Implication of Global Search:} We also consider the global implications of our search of 37 CHIME/FRB localization regions on the prevalence of PRSs. VLASS was the deepest radio survey at GHz frequencies but is only sensitive to PRSs with similar luminosities to FRB\,20121102A out to a redshift of $z\sim0.1$. Hence, deeper targeted observations are necessary to probe such objects. However, VLASS would be sensitive to PRSs with radio luminosities of $\sim$10$^{40}$ erg s$^{-1}$ out to a redshift of $z\sim0.5$. Based on the lack of detected VLASS sources in fields with $z_{\rm{max}}$ less than this, we conclude such luminous PRSs must be rare ($<$5\% of repeaters).

\emph{Comparison to FRB-PRS Models:} Finally, we explored the implications for the two candidate PRSs identified here within the framework for four FRB-PRS models. Broadly, we found that it was possible to explain the radio detections with reasonable parameters in any of the (i) PWN/Magnetar ion-electron wind nebulae, (ii) hypernebulae, (iii) GRB afterglow or (iv) SN-ejecta-CSM interaction models. However, we note that both objects are consistent with the expected PRS luminosity versus RM relationship expected with the model nebulae framework \citep{Yang2020,Bruni2023} and the hypernebulae model can also broadly explain the amount of excess DM inferred for both events if they are associated with the FRB. Future observations with more constraining size estimates and SEDs for both sources would be able to further refine and potentially exclude a subset of models.

Overall the analyses in this manuscript are clues to finding PRSs in the coming era of CHIME/Outriggers \citep{Leung2021,Mena-Parra2022,Cassanelli2022,Lanman2024}, DSA-110 \citep{Ravi2022DS110}, and MeerTrap \citep{Rajwade2022meerkat}. We emphasize that we do not claim any associations between the potential PRSs with their FRBs. We recommend following up on these FRBs to confirm or rule out any of these potential PRSs, particularly by localizing subsequent bursts to $\sim$arcsecond precision.

\facilities{CHIME, VLA, Gemini North}

\software{\texttt{FRUITBAT} \citep{Batten2019},
\texttt{PHOTUTILS} \citep{larry_bradley_2023},
\texttt{SExtractor}  \citep{Bertin1996},
 \texttt{IRAF} \citep{Blondin2012},  \texttt{PYRAF} \citep{pyraf2012}, \texttt{PypeIt} \citep{Prochaska2020} \texttt{CASA} \citep{casa2007}, \texttt{pwkit} \citep{peterwilliams2017}, 
 \texttt{AEGEAN}  \citep{Hancock2018} 
 Astropy \citep{Astropy2013, Astropy2018, astropy2022}, Matplotlib \citep{matplotlib}, NumPy \citep{numpy}, SAOImage DS9\citep{ds92003}}

\section*{Acknowledgements}

We acknowledge Gabriele Bruni and Yuan-Pei Yang for their useful conversations.

We acknowledge that CHIME is located in the traditional, ancestral, and unceded territory of the
Syilx/Okanagan people. We are grateful to the staff
of the Dominion Radio Astrophysical Observatory,
which is operated by the National Research Council
of Canada. CHIME is funded by a grant from the
Canada Foundation for Innovation (CFI) 2012 Leading
Edge Fund (Project 31170) and by contributions from
the provinces of British Columbia, Quebec, and Ontario.
The CHIME/FRB Project is funded by a grant from the
CFI 2015 Innovation Fund (Project 33213) and by contributions from the provinces of British Columbia and Quebec, and by the Dunlap Institute for Astronomy and Astrophysics at the University of Toronto. Additional support was provided by the Canadian Institute for Advanced Research (CIFAR), McGill University, and the McGill Space Institute thanks to the Trottier Family Foundation and the University of British Columbia. The CHIME/FRB baseband
system is funded in part by a Canada Foundation for
Innovation John R. Evans Leaders Fund award to IHS. 

The National Radio Astronomy Observatory (NRAO) is a facility of the National Science Foundation operated under a cooperative agreement by Associated Universities, Inc. We would like to acknowledge NRAO for telescope time awarded through the Karl G. Jansky Very Large Array (VLA) interferometer for program numbers
20B-280, 20A-469, 21B-176, and 21A-387. 
We also thank the NRAO staff for their help in the preparation of observations. 

The Dunlap Institute is funded through an endowment established by the David Dunlap family and the University of Toronto.

Some of this work is based on observations obtained at the Gemini Observatory, which is operated by the Association of Universities for Research in Astronomy, Inc., under a cooperative agreement with the NSF on behalf of the Gemini partnership: the National Science Foundation (United
States), the Science and Technology Facilities Council (United Kingdom), the National Research Council (Canada), CONICYT (Chile), the Australian Research Council (Australia), Ministério da Ciência e Tecnologia (Brazil), and SECYT (Argentina). We appreciate the Gemini team for granting us observing time for the program IDs: GN-2022A-Q-212, GN-2022B-Q-308, and GN-2022B-Q-116. 

Basic research in radio astronomy at the U. S. Naval Research Laboratory is supported by 6.1 Base funding. The construction and installation of VLITE was supported by the NRL Sustainment Restoration and Maintenance fund.

B.M.G. acknowledges the support of the Natural Sciences and Engineering Research Council of Canada (NSERC) through grant RGPIN-2022-03163 and of the Canada Research Chairs program. MRD acknowledges support from the NSERC through grant RGPIN-2019-06186, and of the Canada Research Chairs Program, and the Dunlap Institute at the University of Toronto. V.M.K. holds the Lorne Trottier Chair in Astrophysics \& Cosmology, a Distinguished James McGill Professorship, and receives support from an NSERC Discovery grant (RGPIN-228738-13) and from the FRQNT CRAQ. J.W.T.H. is a Canada Excellence Research Chair in Transient Astrophysics.

N.S. acknowledges the support from NASA (grant number 80NSSC22K0332), NASA FINESST (grant number 80NSSC22K1597), Columbia University Dean’s fellowship, and a grant from the Simons Foundation.
T. E. is supported by NASA through the NASA Hubble Fellowship grant HST-HF2-51504.001-A awarded by the Space Telescope Science Institute, which is operated by the Association of Universities for Research in Astronomy, Inc., for NASA, under contract NAS5-26555.
A.M.C. is funded by an NSERC Doctoral Postgraduate Scholarship. 
A.B.P. is a Banting Fellow, a McGill Space Institute~(MSI) Fellow, and a Fonds de Recherche du Quebec -- Nature et Technologies~(FRQNT) postdoctoral fellow.
K.W.M. holds the Adam J. Burgasser Chair in Astrophysics and is supported by NSF grants (2008031, 2018490).
S.P.T. is a CIFAR Azrieli Global Scholar in the Gravity and Extreme Universe Program.
Z.P. was a Dunlap Fellow and is supported by an NWO Veni fellowship (VI.Veni.222.295).
A.P. is funded by the NSERC Canada Graduate Scholarships -- Doctoral program.
K.N. is an MIT Kavli Fellow. 
F.~K. acknowledges support from Onsala Space Observatory for the provisioning of its facilities/observational support. The Onsala Space Observatory national research infrastructure is funded through Swedish Research Council grant No 2017-00648. M.B is a McWilliams fellow and an International Astronomical Union Gruber fellow. M.B. also receives support from the McWilliams seed grant.

\clearpage

\appendix

\section{Details of Individual Sources}\label{sec:individual-radio-sources}
Here, we provide an overview of each FRB field where a radio source was identified, detailing the relevant properties of the radio sources and their hosts. Table ~\ref{tab:radio-archive} presents the basic properties, while Table ~\ref{tab:radio-derived} displays the derived properties of each radio source.

\subsection{FRB\,20180814A field} \label{subsec:R2}

FRB\,20180814A is part of the first set of repeating FRBs to be discovered by CHIME/FRB and has a DM of 189.4(4)\,pc\,cm$^{-3}$ \citep{Michilli2022}. We used the baseband localization region and $z_{\rm{max}}$ $= 0.091$ published by \cite{Michilli2022} to search for radio sources in its field. 
 
 We conducted an archival search for radio sources in the field of FRB\,20180814A but found none. However, we found 5 radio sources from the VLA deep image at 1.5\,GHz of the field. Out of the 5 radio sources, 2 (20180814A-S2 and 20180814A-S3) could not be associated with a galaxy while another 2 (20180814A-S4 and 20180814A-S5) have likely host galaxy associations.

 One of the VLA radio sources (20180814A-S1) found in this field is extended and spatially coincident with the plausible host galaxy of the FRB. 20180814A-S2 and 20180814A-S3 are unresolved sources and thus likely PRS candidates. However, 20180814A-S1, 20180814A-S4, and 20180814A-S5 are extended radio sources, thus we disfavor them as potential PRSs.

 The likely host galaxy of the 20180814A-S1 is the same as the plausible host of the FRB which was discovered from the PanSTARRS survey to be PanSTARRS-DR1 J042256.01+733940.7 at a $z_{\rm{spec}}$ $=$ 0.06835(1) with an AB apparent magnitude of 17.15\,mag, as reported by \cite{Michilli2022}.

 In the case where the FRB is not associated with any radio source, we take the 5$\sigma$ r.m.s. value of 17.5$\mu$Jy as the upper limit on the flux density of the PRS at 1.5 GHz. 

\subsection{FRB\,20180916B field} \label{subsec:R3}

FRB\,20180916B is a well-studied repeating FRB source. Its relatively close proximity of approximately 150 Mpc despite having a DM of 348.77\,pc\,cm$^{-3}$ \citep{CHIME-Anderson2020}, combined with thorough investigations into the bursts, has unveiled valuable insights into its characteristics — notably, the presence of a 16.3-day periodicity in its activity \citep{Pleunis2021}. \cite{Marcote2020} achieved a subarcsecond localization of the source using VLBI with the European VLBI Network (EVN), linking the FRB to a star-forming region within a massive spiral galaxy at a redshift of 0.0337.

There have not been any radio sources found at the location of the FRB after various efforts \citep{Marcote2020}. Here, we report a similar deep limit on the presence of a PRS at the location of the FRB using our realfast/VLA image of the field (see Figure ~\ref{fig:VLA-upp}). We measure an upper limit on flux density above a 3-$\sigma$ r.m.s. noise level of 18$\mu$Jy beam$^{-1}$ at 1.5 GHz. This resulted in an upper limit on PRS luminosity of $<$4.8$\times$10$^{26}$ erg s$^{-1}$ Hz$^{-1}$ which is the same as the earlier constraint provided by \cite{Marcote2020}.

 \subsection{FRB\,20181030A field} \label{subsec:R4}
FRB\,20181030A is a repeater that is prominent for coming from a large, nearby bright galaxy. It was reported to have a DM of 103.5\,pc\,cm$^{-3}$ which is approximately 20 Mpc away according to \cite{BhardwajR42021}. The plausible host of FRB\,20181030A is a large spiral galaxy known as NGC 3252 ($z = 0.00385$) with an NVSS (NVSS J103422+734554) radio source close to the center of the galaxy reported by \cite{BhardwajR42021} and a VLA source--- 20181030A-S1 seen at the edge as described here. Considering the offset, lack of WISE emission, and star formation information, we consider 20181030A-S1 a candidate PRS even though we are unable to rule out background AGN.

In the case where the FRB is not associated with any radio source, we take the 5$\sigma$ r.m.s. value of 50$\mu$Jy as the upper limit on the flux density of the PRS at 1.5 GHz. This results in an upper limit on luminosity of $<$2.4$\times 10^{25}$\,erg\,s$^{-1}$\,Hz$^{-1}$.

\subsection{FRB\,20181119A field} \label{subsec:R6}
FRB\,20181119A is one of the second set of repeaters discovered by CHIME/FRB with a $z_{\rm{max}}$ of $0.43$ \citep{CHIME2023}. 
The radio source (ILTJ124151.73+650802.7, hereafter 20181119A-S) was found in the LoTSS survey but did not have an optical host association. This is the only source in our sample with a positive spectral index limit similar to that found for a third potential PRS---FRB\,20201124A \citep{Bruni2023}.

Additionally, there appears to be a probable second LoTSS source located at another edge of this FRB field, yet it was not documented in the LoTSS catalog.

\subsection{FRB\,20190110C field} \label{subsec:C64}
FRB\,20190110C is a part of the third set of repeating FRBs discovered by CHIME/FRB with 
$z_\mathrm{max} = 0.22$ \citep{CHIME2023}. There is a 92\% likelihood (as estimated by the Probabilistic Association of Transients to their Hosts software PATH) of the FRB being linked to a galaxy at $z_\mathrm{spec}$ $= 0.12244(6)$ \citep{Ibik2023}. Initial archival searches within the uncertainty region revealed a radio source (20190110C-S) in various catalogs, including VLASS, FIRST, and LoTSS as shown in Table ~\ref{tab:radio-archive}.

The radio source is likely associated with a faint galaxy close to the PATH-preferred FRB host. Since PATH does not necessarily exclude other galaxies in the field, we investigate the galaxy further. We found a single line $z_\mathrm{spec} > z_\mathrm{max}$ and this rules out this source. Even if we think this redshift is incorrect, the offset, RO ratio, and WISE color ratios are consistent with AGN.

\subsection{FRB\,20190117A field} \label{subsec:R16}
FRB\,20190117A belongs to the second set of repeating FRBs to be discovered by CHIME/FRB. We used the baseband localization region, and $z_{\rm{max}}$ $= 0.46$ published by \cite{Michilli2022} to search for radio sources in its field. No radio source was detected in an initial archival search within the localization uncertainty region of the burst but we found a VLA source (20190117A-S) just outside the field. The radio source is likely associated with an unresolved optical counterpart with a small offset consistent with AGN. 
In the case where the FRB is not associated with any radio source, we take the 5$\sigma$ r.m.s. value of 80$\mu$Jy as the upper limit on the flux density of the PRS at 1.5 GHz.

\subsection{FRB\,20190208A field} \label{subsec:R12}
FRB\,20190208A is part of the second set of repeating FRBs to be discovered by CHIME/FRB. We used the baseband localization region, 
and $z_{\rm{max}}$ $= 0.68$ published by \cite{Michilli2022} to search for radio sources in its field. No radio source was detected in an initial archival search within the localization uncertainty region of the burst but we found a VLA source (20190208A-S) just outside the field. The radio source is likely associated with an unresolved optical counterpart with a small offset consistent with AGN. 

In the case where the FRB is not associated with any radio source, we take the 5$\sigma$ r.m.s. value of 25$\mu$Jy as the upper limit on the flux density of the PRS at 1.5 GHz. 

\subsection{FRB\,20190303A field} \label{subsec:R17}
FRB\,20190303A is among the second set of repeating FRBs whose baseband localizations were published by \cite{Michilli2022}, with $z_{\rm{max}}$ $ = 0.22$. The plausible host of FRB\,20190303A is two merging galaxies known as SDSS J135159.17+480729.0 and SDSS J135159.87+480714.2 with $z_{\rm{spec}}$$=$0.06437(1) and $z_{\rm{spec}}$ $=$ 0.06386(1) respectively as reported by \citep{Michilli2022}. We discovered an NVSS radio source spatially coincident with SDSS J135159.17+480729.0 (P$_{\mathrm{cc,rad}}$ $\sim$0.1673), a FIRST source spatially coincident with SDSS J135159.87+480714.2 (P$_{\mathrm{cc,rad}}$ $\sim$0.2268) and a LoTSS source overlapping the two galaxies (P$_{\mathrm{cc,rad}}$ $\sim$0.1283).

Since the LoTSS radio source is quite offset from both galaxies, we know that it is not an AGN but its blended nature makes it ambiguous and thus we are unable to associate it with any of the galaxies. A deeper observation of the field revealed two distinct extended radio sources (20190303A-S1 and 20190303A-S2) that nicely traced the shape of the two merging galaxies. 
However, we are not able to detangle any point source even if embedded in the radio source, hence we report the measured radio emissions and upper limits. The morphology of these emissions is consistent with star formation activities of their host galaxies or probably a consequence of the merging of the two galaxies, hence are ruled out as PRSs. 

In the case where the FRB is not associated with any radio source, we take the 5$\sigma$ r.m.s. value of 25$\mu$Jy as the upper limit on the flux density of the PRS at 1.5 GHz. This gives a radio luminosity of $<$2.6$\times$10$^{27}$\,erg\,s$^{-1}$\,Hz$^{-1}$ for the FRB.

\subsection{FRB\,20190417A field} \label{subsec:R18}

FRB\,20190417A is one of the very high DM repeating FRBs reported by CHIME/FRB among the second set of repeaters. We used the baseband localization region, and $z_{\rm{max}}$ $= 1.2$ published by \cite{Michilli2022} to search for radio sources in its field. No radio source was detected in an initial archival search within the localization uncertainty region of the burst.  

VLA imaging of the FRB localization uncertainty region showed two radio sources in the field. The first source (20190417A-S1)is likely a PRS when considered in the context of its optical host ($z_\mathrm{spec} = 0.12817(2)$) imaged with Gemini North.

Using the line fluxes from Gemini North spectra, we construct and place this galaxy on the updated ``Baldwin, Phillips \& Terlevich'' (BPT) \citep{BPT1-1981} diagram according to \cite{BPT22011}. The line ratio location is consistent with a star-forming galaxy. Using the extinction-corrected spectra, we estimate the $H_\mathrm{\alpha}$ luminosity to be 3.7$\times 10^{40}$\,erg\,s$^{-1}$. In addition, we also check for intrinsic host galaxy extinction by calculating the Balmer decrement from the H$_\mathrm{\alpha}$ and H$_\mathrm{\beta}$ line fluxes. Compared to the theoretical Case-B recombination line ratio of H$_\mathrm{\alpha}$/H$_\mathrm{\beta}$ $=$ $-$3.3, we do not find evidence for significant additional extinction given that we measure a Balmer decrement of 1.76. We then calculate the SFR$_\mathrm{H_\alpha}$ to be 0.1964\,M$_\mathrm{\odot}$\,yr$^{-1}$ as described in \S ~\ref{subsec:SFR}. We estimate a metallicity of $12 + \log(O/H) = 8.91$ using Equation 1 from 
\citet{Pettini2004}: $12 + \log(O/H)= 8.90 + 0.57 \times$ N2, where N2 $=$ 
[NII]$\lambda$6583/H$_\mathrm{\alpha}$. This value is approximately similar to solar metallicity assuming $12 + \log(O/H)_{\rm{solar}} = 8.69$. All these results insinuate that this unresolved optical source is a star-forming galaxy. Considering the physical offset, lack of WISE emission, and star formation information, we consider 20190417A-S1 a potential PRS.
The second source (20190417A-S2) is resolved and has no optical counterpart.

In the case where the FRB is not associated with any radio source, we take the 5$\sigma$ r.m.s. value of 42.5$\mu$Jy as the upper limit on the flux density of the PRS at 1.5 GHz.

\subsection{FRB\,20190604A field} \label{subsec:R13}
FRB\,20190604A is part of the second set of repeating CHIME/FRBs whose baseband localizations were published by \cite{Michilli2022}, and has $z_{\rm{max}}$ $= 0.7$ from its DM.
In the field of FRB\,20190604A, there are two resolved LoTSS sources; ILTJ143448.87+531731.2 (20190604A-S1) and ILTJ143448.97+531833.3 (20190604A-S2). 

Since these two radio sources are resolved, we did not discuss them further in the context of this study. 

\subsection{FRB\,20191106C field} \label{subsec:C27} 
FRB\,20191106C is one of the third set of repeaters discovered by CHIME/FRB with a $z_{\rm{max}}$ $= 0.36$ \citep{CHIME2023, Ibik2023}. We found a LoTSS source (ILTJ131819.22+425958.9; hereafter 20191106C-S)  spatially associated with the plausible host of the FRB with $z_{\rm{spec}}$$=$0.10775(1) as reported by \cite{Ibik2023} at an offset of 0.11" from the center of the galaxy. The radio source is resolved and is likely consistent with AGN given the offset.

\subsection{FRB\,20191114A field} \label{subsec:C30} 

FRB\,20191114A is among the third set of FRB repeaters discovered and positions published by CHIME/FRB with a $z_{\rm{max}}$ of $0.52$ \citep{CHIME2023}. 
We found a NVSS (NVSS J181426+194749), TGSS (J181426.2+194749), and RACS (RACS 1819+18A) radio source at the edge of the FRB region. A VLASS source (VLASS1QLCIR J181426.18+194748.2) was found at the same location but offset from the FRB region and referred to as 20191114A-S. There is no optical host association for the radio source, hence could not characterize the source further.

\subsection{FRB\,20200120E field} \label{subsec:R51}

FRB\,20200120E is the only repeating FRB that is associated with a globular cluster in the M81 spiral galaxy \citep{Bhardwaj_2021, kirsten2021}. Its low DM of 87.82 pc cm$^{-3}$ \citep{Bhardwaj_2021} is consistent with the distance to M81 of 3.6 Mpc. \cite{kirsten2021} reported a VLBI localization of the source with the EVN. To date, no coincident radio sources have been identified at this location. Here, we present an upper limit on the presence of a PRS at the FRB's location using realfast/VLA image of the field (see Figure ~\ref{fig:VLA-upp}).  Our analysis yields an upper limit on the flux density, exceeding a 3$\sigma$ r.m.s. noise level of 27$\mu$Jy beam$^{-1}$ at 1.5 GHz. Consequently, we establish an upper limit on the PRS luminosity of $<$4.2$\times$10$^{23}$ erg s$^{-1}$ Hz$^{-1}$ which is slightly higher by an order of magnitude than the earlier constraint provided by \cite{kirsten2021}.

\subsection{FRB\,20200223B field} \label{subsec:C35} 
FRB\,20200223B is one of the third set of repeaters discovered by CHIME/FRB with a DM of 201.8(4)\,pc\,cm$^{-3}$ corresponding to a $z_{\rm{max}}$ $= 0.19$ \citep{CHIME2023, Ibik2023}.
We found an extended LoTSS source (ILTJ003304.67+284952.4) in the field of FRB\,20200223B, spatially associated with the plausible host of the FRB ($z_{\rm{spec}}$$=$0.06024(2)) at an offset of 0.17" as reported by \cite{Ibik2023}.

\subsection{FRB\,20200619A field} \label{subsec:C37} 
FRB\,20200619A is one of the third set of repeaters discovered by CHIME/FRB with a $z_{\rm{max}}$ $= 0.45$ \citep{CHIME2023}. An archival search of the localization region revealed a NVSS (NVSS J181016+553724) and a TGSS ADR1 (J181015.3+553736) source at the edge of the CHIME error region for the FRB. The source looks like 3 individual sources in the VLASS image. However, the source catalog only identified the one close to the center of the emission VLASS1QLCIR J181017.37+553715.0 (which is a bit outside the FRB field). The radio source is likely associated with an unresolved optical galaxy.

\subsection{FRB\,20200929C field} \label{subsec:C43} 
FRB\,20200929C is a repeating FRB identified by CHIME/FRB among the third set of repeaters with a $z_{\rm{max}}$ of 0.44 \citep{CHIME2023}. We found a LoTSS source (ILTJ010811.77+182829.7) within the error region.

While we still do not know the host galaxy of the FRB, the radio source is spatially coincident with an SDSS galaxy (SDSSJ010811.69+182830.8) at $z_{\rm{ph}} = $0.40$\pm$0.09.


\begin{longrotatetable}
\movetabledown=22mm
\begin{deluxetable*}{l|cc|lccc|cc|ccc}
\tabletypesize{\small}
\tablecaption{Summary of Archival, VLA and VLITE Radio Results \label{tab:radio-archive}}
\tablehead{\colhead{FRB} & \colhead{Reference for FRB} & \colhead{Radio source} & \colhead{S$_\mathrm{\nu,V}$$^a$} & \colhead{S$_\mathrm{\nu,N}$$^b$} & \colhead{S$_\mathrm{\nu,F}$$^c$} & \colhead{S$_\mathrm{\nu,L}$$^d$} & \colhead{RA} & \colhead{DEC} &  \colhead{P$_\mathrm{cc,rad}$$^e$} & \colhead{Nature$^m$} \\
\colhead{Name} & \colhead{region used$^{r}$ (90\% cl)} & \colhead{} & \colhead{(mJy)} & \colhead{(mJy)} & \colhead{(mJy)} & \colhead{(mJy)} & \colhead{(J2000)} & \colhead{(J2000)}  & \colhead{} & \colhead{} 
}
\startdata
FRB\,20181030A & 1 & 20181030A-S2$^f$ & $<$0.6 & 3.8$\pm$0.5 & $<$0.75 & $<$0.42 & 
10:34:22.74(1.48s) & $+$73:45:54.9(4.5") &  0.0954 & R \\ 
FRB\,20181119A & 2 & 20181119A-S & $<$0.6 & $<$12.5 & $<$0.75 & 0.495$\pm$0.095 & 
12:41:51.72(0.48s)	& $+$65:08:02.76(0.23") &  0.7121 & U\\
FRB\,20190604A & 2 & 20190604A-S1 & $<$0.6 & $<$12.5 & $<$0.75 & 1.803$\pm$0.188 & 
14:34:48.86(0.21s) & $+$53:17:31.56(0.21") &  0.2209 & R  \\ 
FRB\,20190604A & 2 & 20190604A-S2 & $<$0.6 & $<$12.5 & $<$0.75 & 0.971$\pm$0.175& 
14:34:48.97(0.30s)	& $+$53:18:33.37(0.42") & 0.4186 & R  \\ 
FRB\,20190303A & 2 & 20190303A-S1 & $<$0.6 & 2.2$\pm$0.4  & $<$0.75 & $<$0.42 & 
13:51:59.91(0.69s) & $+$48:07:19.0(6.9") & 0.1673 & E \\ 	
FRB\,20190303A & 2 & 20190303A-S2 & $<$0.6 &  $<$12.5 & 1.25$\pm$0.13 &  $<$0.42 &
13:51:59.10	& $+$48:07:29.3 &  0.2267 & E  \\ 
FRB\,20190303A & 2 &  20190303A-S3 & $<$0.6 & $<$12.5 & $<$0.75 & 16.92$\pm$0.39 & 
13:51:59.518(0.25s)	& $+$48:07:21.22(0.16")	&  $-$ & B, E  \\ 
FRB\,20190110C & 3 & 20190110C-S & 0.77$\pm$0.19 & $<$12.5 & 0.91$\pm$0.15 & 10.05$\pm$0.29 & 
16:37:17.84(0.14s)	& $+$41:26:33.96(0.14") &  0.0515$^{l}$ & U  \\ 
FRB\,20191106C & 3 & 20191106C-S & $<$0.6 & $<$12.5 & $<$0.75 & 3.29$\pm$0.14 & 
13:18:19.22(0.09s) &	$+$42:59:58.9(0.08") & 0.1302 & R  \\ 
FRB\,20191114A & 3 & 20191114A-S & 4.46$\pm$0.26 & 9.1$\pm$0.5 & $<$0.75 & $<$0.42 & 
18:14:26.19(0.06s)	& $+$19:47:48.3(0.03") & 0.2913$^{i}$  & U  \\ 		
FRB\,20200223B & 3 & 20200223B-S & $<$0.6 & $<$12.5 & $<$0.75 & 2.964$\pm$0.588 & 
00:33:04.68(1.22s) & $+$28:49:52.5(1.17") & 0.0863 & E \\
FRB\,20200619A & 3 & 20200619A-S & 1.89$\pm$0.28 & 6.3$\pm$0.5 & $<$0.75 & $<$0.42 & 
18:10:17.37(0.1s) & $+$55:37:15.1(0.12") &  0.9065$^{k}$ & U  \\ 
FRB\,20200929C & 3 & 20200929C-S & $<$0.6 & $<$12.5 & $<$0.75 & 1.705$\pm$0.313 & 
01:08:11.77(0.27s)	& 18:28:29.8(1.4") &  0.5516 & R  \\	
\hline
&  &  & S$_\mathrm{p,VLA}$$^p$ & S$_\mathrm{tot,VLA}$$^q$ & S$_\mathrm{p,VLITE}$$^s$& S$_\mathrm{tot,VLITE}$$^s$ &  &  &  \\
 & & & (mJy/beam) & (mJy) & (mJy/beam) & (mJy) & &  & & & \\
\hline
 FRB\,20180814A & 2 & 20180814A-S1 & 0.018$\pm$0.003	 & 0.0673$\pm$0.0128 &  & $<$1.6 & 
 04:22:56.11(0.23s) &	$+$73:39:40.3(0.30") & 0.5162 & E  \\ 
 FRB\,20180814A & 2 & 20180814A-S2 & 0.020$\pm$0.004 & 0.0197$\pm$0.0038 &  & $<$1.6 & 
 04:22:51.82(0.15s)	 & $+$73:39:53.68(0.15") &  0.0192 & U  \\ 
 FRB\,20180814A & 2 & 20180814A-S3  & 0.022$\pm$0.004 & 0.026$\pm$0.0044 &  & $<$1.6 & 
04:22:38.8(0.15s)	 & $+$73:40:15.21(0.14") &  0.0129 & U \\ 
FRB\,20180814A & 2 & 20180814A-S4 & 0.029$\pm$0.003 & 0.065$\pm$0.007 &  & $<$1.6 & 
04:22:41.3(0.14s)	 & $+$73:40:21.04(0.12") &  $-$ & E  \\ 
FRB\,20180814A & 2 & 20180814A-S5 & 0.033$\pm$0.002 & 0.20$\pm$0.01 &  & $<$1.6 & 
04:22:46.3(0.13s)	& $+$73:40:20.41(0.13") & $-$ & E  \\ 
 FRB\,20181030A & 1 &    20181030A-S1 & 0.40$\pm$0.02 & 0.43$\pm$0.02 &  & $<$2.2 & 
     10:34:14.25(0.02s) &	$+$73:45:04.1(0.02") & 0.3718 & U  \\ 
  FRB\,20190208A & 2 &  20190208A-S & 0.07$\pm$0.02 & 0.0791$\pm$0.0055 &  & $<$1.7 & 
     18:54:07.11(0.05s)	& $+$46:55:51.9(0.03") & 0.8217 & U \\ 
  FRB\,20190117A & 2 &   20190117A-S & 0.18$\pm$0.02  & $0.2078\pm0.0194$ &  & $<$4.8 & 
    22:06:36.95(0.15s)	& $+$17:22:25.7(0.08") & 0.6982 & U  \\ 
  FRB\,20190303A & 2 &   20190303A-S1$^{n}$ & 0.06 & 0.2677 & 1.4$\pm$0.3 & 7.7$\pm$1.9 & 
     13:51:59.9	& $+$48:07:17.55 &  0.6521 & E \\ 
  FRB\,20190303A & 2 &   20190303A-S2$^{o}$ & 0.07 & 0.9516 & 1.9$\pm$0.3  & 10.0$\pm$2.1 & 
     13:51:59.01(0.02s)	& $+$48:07:24.3(0.02") &  0.2700 & E \\
    FRB\,20190417A & 2 &  20190417A-S1 & 0.164$\pm$0.008 & 0.1727$\pm$0.0089 &  & $<$1.1 & 
     19:39:05.89(0.03s)	& $+$59:19:36.8(0.03") &  0.6705 & U  \\ 
    FRB\,20190417A & 2 &  20190417A-S2 & 0.043$\pm$0.008 & 0.057$\pm$0.012 &  & $<$1.1  & 
     19:39:00.44(0.18s)	& $+$59:20:02.4(0.12") & 0.9914 & R \\ 
 \enddata
\tablecomments{This Table contains all radio (archival and targeted VLA) sources including extended and blended ones. The topmost set is archival data followed by the VLA data after the dividing line and all upper limits quoted for each survey are the 5-$\sigma$ rms value.  }
$^a$ V means VLASS, 
$^b$ N means NVSS,
$^c$ F means FIRST,
$^d$ L means LoTSS\\
$^e$ P$_\mathrm{cc,rad}$ is the probability of finding a radio flux as bright as S$_\nu$ within each FRB error region (see \S ~\ref{subsec:pcc-fr}). \\
$^f$ Flux density values and other information related to this source are taken from \citet{BhardwajR42021} \\
$^{i}$ The P$_\mathrm{cc,rad}$ value shown in the table is for VLASS.  P$_\mathrm{cc,rad}$ for the NVSS source is 0.7154. 
There is also radio emission at the location of 20191114A-S from RACS with a flux density of 15.49$\pm$0.72\,mJy at P$_\mathrm{cc,rad}$ of 0.2347 and emission from TGSS with a flux density of 33.9$\pm$6.3\,mJy at P$_\mathrm{cc,rad}$ of 0.3181. \\
$^{k}$ The P$_\mathrm{cc,rad}$ value shown in the table is for VLASS. The P$_\mathrm{cc,rad}$ for NVSS is 0.7707. There is also a TGSS source at the location of 20200619A-S with a flux density of 28.6$\pm$5.6\,mJy at P$_\mathrm{cc,rad}$ of 0.785. \\
$^{l}$ The P$_\mathrm{cc,rad}$ value shown in the table is for the VLASS source. The P$_\mathrm{cc,rad}$ value is 0.1012 for the FIRST source while the LoTSS source is blended.\\
$^m$ Nature here refers to the size of the radio source. U means that the radio source is unresolved, R means that the source is resolved (radio source size is more than the beam size), B means that the source is blended with another radio source and E means that the radio source is extended (see \S ~\ref{subsec:size} for details.). We refrained from computing P$_\mathrm{cc,rad}$ values for the blended sources due to the ambiguity in determining their flux densities.\\
$^{n}$ This VLA source is multi-component but at the location of the NVSS source.\\
$^{o}$ This VLA source is at the location of the FIRST source.\\
$^p$ S$_\mathrm{p,VLA}$ is the peak flux density for the VLA source.\\
$^q$ S$_\mathrm{tot,VLA}$ is the integrated flux density for the VLA source.\\
$^r$ [1]----\cite{BhardwajR42021}, [2]---\cite{Michilli2022}, [3]---\cite{CHIME2023}\\
$^s$ Peak and integrated fluxes and limits from VLITE (0.34 GHz). The limits are 5-$\sigma$ upper limits used to constrain the spectral indices of the VLA sources.
\end{deluxetable*}
\end{longrotatetable}

\begin{deluxetable*}{l|cc|ccccc|ccc}
\tabletypesize{\small}
\tablecaption{Derived Properties of the Radio sources \label{tab:radio-derived}}
\tablehead{\colhead{Radio source} & \colhead{$z_{\rm{max}}$} & \colhead{ $z_{\rm{spec}}$} & \colhead{L$_{\mathrm{\nu,V}}$} & \colhead{L$_{\mathrm{\nu,N}}$} & \colhead{L$_{\mathrm{\nu,F}}$} & \colhead{L$_{\mathrm{\nu,L}}$} & L$_{\mathrm{\nu,VLA}}$$^t$ & \colhead{$\alpha_{in}$$^d$} & \colhead{$\alpha$$^d$}  & \colhead{Host$^n$} \\
 \colhead{} & \colhead{} & \colhead{} & \colhead{($\times 10^{30}$)} & \colhead{($\times 10^{30}$)} & \colhead{($\times 10^{30}$)} & \colhead{($\times 10^{30}$)} & \colhead{($\times 10^{28}$)}  & \colhead{}  & \colhead{} \\
 \colhead{} & \colhead{} & \colhead{} & \multicolumn{5}{c}{(erg\,s$^{-1}$\,Hz$^{-1}$)}  & \colhead{} & \colhead{}  & \colhead{} 
}
\startdata
20181030A-S2$^f$ & $<$0.05$^f$ & 0.00385$^f$ & $<$0.0002 & 1.57(27)& $<$0.0003 &  $<$0.0001 & $-$ &  & $>-$0.43 & Y, E \\ 
 20181119A-S & $<$0.43$^{g}$  & $-$ & $<$4.27 & $<$89.01 & $<$5.34 & 3.526 & $-$ &  & $<-$0.063 & N \\
20190604A-S1 & $<$0.7$^{g}$ & $-$ & $<$13.80 & $<$287.57 & $<$17.25 & 41.48 & $-$ &  & $<-$0.359 & N \\ 
 20190604A-S2 &  $<$0.7$^{g}$ & $-$ & $<$13.80 & $<$287.57 & $<$17.25 & 22.33 & $-$ &  & $<-$0.157 & Y, E \\ 
20190303A-S1 & $<$$0.22$$^{g}$ & 0.06437(1)$^{g}$ & $<$0.063 & 0.24(4) & $-$ & $-$ & $-$ &  & $<-$1.70 & Y, E \\ 
 20190303A-S2 & $<$$0.22$$^{g}$ & 0.06386(1)$^{g}$ & $<$0.063 & $-$ & 0.13(1) & $-$ & $-$ &  & $<-$0.96 & Y, E \\ 
20190303A-S3 & $<$$0.22$$^{g}$ & 0.06386(1)$^{gh}$ & $<$0.063 & $-$ & $-$ & 1.78(4) & $-$ &  & $<-$1.09 & Y, E \\ 
 20190110C-S & $<$0.22$^{j}$ & $-$ & 1.17 & $<$1.89 & 1.38 & 15.25$^{l}$ & $-$ &  & $-$0.09(1)$^{p}$ & Y, E \\
 20191106C-S & $<$$0.36$$^{j}$ &0.10775(1)$^{j}$ & $<$0.19 & $<$3.97 & $<$0.24 & 1.05(4) & $-$ &  & $<-$0.55 & Y, E \\
 20191114A-S & $<$0.52 & $-$ & 49.80 & 101.64 & $<$8.40 & $<$4.71 & $-$ &  & $-$2.9(4)$^{o}$ & N \\
 20200223B-S & $<$0.19$^{j}$ & 0.06024(2)$^{j}$ & $<$0.056 & $<$1.164 & $<$0.069 & 0.28(5) & $-$ & & $<-$0.52  & Y, E\\
 20200619A-S & $<$0.45 & $-$ & 14.97 & 49.97 & 5.95 & 3.332 & $-$ &  & $-$2.8(4)$^{o}$ & Y, U \\
 20200929C-S & $<$0.44 & $-$ & $<$4.513 & $<$94.03 & $<$5.641 & 12.83 & $-$ &  & $<-$0.34 & Y, E \\	
\hline
  20180814A-S1 & $<$0.091$^{g}$ & 0.06835(1)$^{g}$ &  & & & & 0.82(15) 
  & $-$1.8(1.2) & $>-$3.02$^q$ & Y, E \\
  20180814A-S2 & $<$0.091$^{g}$ & $-$ &  & & & & 0.44(8) 
& $-$2.7(1.1) & $>-$2.95$^q$ & N \\
  20180814A-S3 & $<$0.091$^{g}$ & $-$ &  & & & & 0.58(9) 
 & $-$1.3(9) & $>-$2.88$^q$ & N \\
 20180814A-S4 & $<$0.091$^{g}$ & 0.376(1)$^{g}$ &  & & & & 1.44(16) 
& 0.1(6) & $>-$2.70$^q$ & Y, E \\
20180814A-S5 & $<$0.091$^{g}$ & 0.237(1)$^{g}$ &  & & & & 4.43(22) 
& 0.2(6) & $>-$2.61$^q$ & Y, E \\
      20181030A-S1 & $<$0.05$^{f}$ & 0.00385(2)$^{f}$ &  & & & & 0.02(1) 
     & $-$0.77(2) & $>-$1.13$^q$ & Y, E \\
     20190208A-S & $<$0.68$^{g}$ & $-$ &  & & & & 16.9(1.2) 
    & $-$1.3(5) & $>-$2.15$^q$ & Y, U \\
      20190117A-S & $<$0.46$^{g}$ & $-$ &  & & & & 174(16) 
     & $-$0.8(6) & $>-$2.21$^q$ & A$^r$ \\
      20190303A-S1 & $<$0.22$^{g}$ & 0.06437(1)$^{g}$ &  & & & & 2.86$^{u}$  
      & $-$ & $-$1.1$\pm$0.3$^q$ & Y, E \\
      20190303A-S2 & $<$0.22$^{g}$ & 0.063861(1)$^{g}$ & & & & & 0.1$^{u}$ 
      & $-$ & $-$1.6$\pm$0.3$^q$ & Y, E \\
     20190417A-S1 & $<$1.2$^g$ & 0.12817(2) &  & & & & 7.97(41) 
    & $-$1.2(4) & $>-$1.25$^q$ & A$^s$ \\
     20190417A-S2 & $<$1.2$^g$ & $-$ &  & & & & 493(104) 
      & $-$4.2(1.4) & $>-$2.15$^q$ & N \\
\enddata 
\tablecomments{The table comprises properties of all radio sources obtained from archival and targeted VLA observations. Luminosities were calculated utilizing the available z$_\mathrm{spec}$ (spectroscopic redshift) or z$_\mathrm{max}$ (maximum redshift) in cases where z$_\mathrm{spec}$ is unavailable. Luminosities derived from z$_\mathrm{spec}$ are treated as absolute values, whereas those derived from z$_\mathrm{max}$ serve as upper limits within the FRB framework. In instances of non-detections, the symbol $<$ denotes upper limits.}
V means VLASS at 3 GHz, N means NVSS at 1.4 GHz, \\
F means FIRST at 1.4 GHz, L means LoTSS at 0.14 GHz,\\
$^d$ The spectral index ($\alpha$) characterizes the radio source based on flux densities obtained from at least two frequencies, while $\alpha_{in}$ is the in-band spectral index, ($\alpha$ $=$ $I_{1}/I_{0}$) (see \S ~\ref{subsec:variability} for details). Upper and lower limit values of $\alpha$ were determined utilizing VLASS non-detection data and detections of each individual radio source.\\
$^f$ Values taken from \citet{BhardwajR42021} \\
$^{g}$ Values taken from \citet{Michilli2022} \\
$^{h}$ Two merging galaxies have been reported as the host of FRB\,20190303A. The LoTSS source associated with this FRB is in between the two merging galaxies,\\
so we adopted the galaxy redshift of $z = 0.064(1)$ for estimating its luminosity. \\
$^{j}$ Values are taken from \citet{Ibik2023}.\\
$^n$ Y means yes, there is an optical source that is likely the host galaxy associated with the radio source, E means that there is an extended optical source at the position of the radio source, U means that the optical source at the position of the radio source is unresolved, R means that the optical source at the position of the radio source is resolved and N means no, there is no optical source that is likely a host galaxy of the radio source. \\
$^{o}$ These spectral indices were estimated using the VLASS and the NVSS sources. \\
$^{p}$ The spectral index for this source was calculated using the VLASS and the FIRST sources. \\
$^q$ Contemporaneous spectral index taken from VLITE (0.34 GHz) and VLA (1.5 GHz).\\
$^r$ This source is classified by DESI to be a point source with a very low probability of being a star. This may be an unresolved galaxy. \\
$^s$ This source was not detected in any archival catalog but was found using targeted Gemini observation in 'grz' bands (See \S ~\ref{subsec:R18} for details). This source is classified as unresolved.\\
$^t$ L$_\mathrm{\nu,VLA}$ is obtained using the integrated flux density of each VLA source.\\
$^{u}$ The source finder could not find flux density errors for these sources.
\end{deluxetable*}


\end{document}